\documentclass[twocolumn,preprintnumbers,amsmath,amssymb,physrev,superscriptaddress]{revtex4-1}
\usepackage{graphicx}
\usepackage{epstopdf}
\usepackage{dcolumn}
\usepackage{bm}
\usepackage{amsmath}
\usepackage{amssymb}
\usepackage{mathrsfs}
\usepackage{amsfonts}
\usepackage{times}
\usepackage{subfigure}
\usepackage{amstext}
\usepackage[colorlinks=true, citecolor=blue]{hyperref}
\usepackage[dvipsnames]{xcolor}
\newcommand{\beqa}{\begin{eqnarray}}
\newcommand{\eeqa}{\end{eqnarray}}

\newcommand{\bra}[1]{\langle #1|}
\newcommand{\ket}[1]{|#1\rangle}
\newcommand{\braket}[2]{\langle#1|#2\rangle}

\newcommand{\ann}[2][]{\hat #2^{#1}}
\newcommand{\crea}[2][]{\hat #2^{\dagger #1}}

\begin{document}
\title{Theoretical methods for ultrastrong light-matter interactions}
\author{Alexandre Le Boit\'e}
\email{alexandre.leboite@univ-paris-diderot.fr}
\affiliation{
Universit\'e de Paris, Laboratoire Mat\'eriaux et Ph\'enom\`enes Quantiques, CNRS, F-75013 Paris, France}

\begin{abstract}
This article reviews theoretical methods developed in the last decade to understand cavity quantum electrodynamics in the ultrastrong-coupling regime, where the strength of the light-matter interaction becomes comparable to the photon frequency. Along with profound modifications of fundamental quantum optical effects giving rise to a rich phenomenology,  this regime introduces significant theoretical challenges. One of the most important is the break-down of the rotating-wave approximation which neglects all non-resonant terms in light-matter interaction Hamiltonians. Consequently, a large part of the quantum optical theoretical framework has to be revisited in order to accurately account for all interaction terms in this regime. We give in this article a broad overview of the recent progress, ranging from analytical estimates of ground-state properties to proper derivations of master equations and computation of photodetection signals. For each aspect of the theory, the basic principles of the methods are illustrated on paradigmatic models such as quantum Rabi and spin-boson models.
In this spirit, most of the article is devoted to effective models, relevant for the various experimental platforms in which the ultrastrong coupling has been reached, such as semiconductor microcavities and superconducting circuits. The validity of these models is discussed in the last part of the article, where we address recent debates on fundamental issues related to gauge invariance in the ultrastrong-coupling regime.
\end{abstract}

\maketitle
\tableofcontents

\section{Introduction}

The search for a microscopic theory of light-matter interactions has played a central role in the development of quantum theory since its origin. A constant refinement of the theoretical framework has been fuelled by experimental progress and technological advances. While quantization of atomic degrees of freedom was the essential ingredient of early models of light-matter interactions in atomic physics, the advent of the laser made it crucial to build a more complete theory that accounted also for the quantum nature of light in optical experiments~\cite{Glauber:1963}. 
Decades after the pioneering works that led to the birth of quantum optics, both its theoretical and experimental aspects are extremely active research fields. Indeed, the control, at a microscopic level, of coherent interactions between light and matter is now at the heart of the blooming field of quantum technologies. In this context, the ability to increase the light-matter coupling strength has played a crucial role. To achieve this goal, experiments in cavity quantum electrodynamics (cavity QED) have proved to be extremely valuable tools~\cite{Haroche:2006}. 
 
Hallmarks of quantum coherence in cavity-QED setups, such as vacuum Rabi oscillations, are observed when the light-matter coupling strength becomes larger than any dissipation rate in the system. This so-called strong-coupling regime has now been demonstrated in various platforms, including atomic cavity QED~\cite{Rempe:1987}, semiconductor nanostructures~\cite{Reithmaier:2004, Peter:2005} and superconducting circuits~\cite{Wallraff:2004}. From a theoretical standpoint, this regime is remarkable in that the dynamics of the system must be understood in terms of hybrid light-matter eigenstates. The resulting notion of dressed state has been key to our understanding of quantum features in the output photon statistics in the strong-coupling regime, such as photon antibunching~\cite{Imamoglu:1997, Birnbaum:2005, Bozyigit:2010, Lang:2011, Hoffman:2011}. Historically, in the platforms mentioned above, an important feature of the strong-coupling regime was that the coupling strength remained much smaller than the frequency of the cavity mode. As a result, only the resonant terms in the interaction Hamiltonian (i.e. conserving the number of excitations), play a significant role. The remaining anti-resonant (or counter-rotating) terms, couple states that are much wider apart in energy and their contribution can be neglected. This so-called rotating-wave approximation (RWA) provides a simpler, intuitive and accurate picture of atom-photon interaction processes in this regime. The success of the exactly solvable Jaynes-Cummings model~\cite{Jaynes:1963}, describing the interaction between a two-level atom and a single cavity mode, in various contexts is a prominent example.

In the last decade, we entered in a new era of cavity QED with the achievement~\cite{Anappara:2009, Todorov:2010,Niemczyk:2010,Forn-Diaz:2010,Forn-Diaz:2017,Yoshihara:2017}  of the ultrastrong coupling (USC) regime, where the coupling strength becomes comparable or even larger than the cavity frequency~\cite{Ciuti:2005, Devoret:2007, Bourassa:2009}. The rich phenomenology of this new regime of cavity QED has been the focus of an intense research activity : the USC regime has indeed proved to induce profound modifications in a variety of fundamental quantum optical phenomena, ranging from vacuum radiation~\cite{DeLiberato:2009, Sanchez-Burillo:2019} to single-photon emission~\cite{Ridolfo:2012, LeBoite:2016}, scattering processes~\cite{Shi:2018} and transport properties~\cite{Felicetti:2014, Bartolo:2018}. We refer the reader to Refs.~\cite{Frisk-Kockum:2019, Forn-Diaz:2019} for a detailed presentation of the USC phenomenology and experimental setups. It was clear from the first theoretical predictions that the USC regime would give rise to counter-intuitive phenomena rooted in the break-down of the RWA and the resulting significant contribution of non-resonant terms. Equally clear was the necessity of developing new techniques to handle correctly all interaction terms, and checking the validity of widely used effective models in this regime. The need to provide a complete theoretical framework valid at arbitrary strong coupling strength was further motivated by recently developed quantum simulation techniques. Various experimental schemes have made it possible to explore~\cite{Langford:2017, Braumuller:2017,Markovic:2018, Lv:2018, Peterson:2019} ultrastrong-coupling physics even in systems that do not naturally achieve the required interaction strength. 

In this article, we review the different theoretical methods that have been designed in recent years to go beyond the RWA and treat ultrastrong light-matter interactions. We also discuss the recent debates on some fundamental issues and limitations of USC cavity QED.  
As the break-down of the RWA has dramatic consequences on nearly all aspects of the theory, both in open and closed systems, a wide range of questions have been revisited to face the challenges of the USC regime.  This includes  exact and approximate diagonalization methods, the treatment of dissipation and driving or the theory of photodetection. Our aim is to provide a pedagogical overview of these different topics. The term deep-strong coupling (DSC) regime has been introduced in the literature to designate more specifically the regime in which the coupling strength becomes larger than the photon frequency~\cite{Casanova:2010}. In the rest of this review we use the term USC in its most general sense, which includes the DSC regime. The article is structured as follows. Notations and effective models of cavity QED that are used throughout the paper are introduced in Sec.~\ref{sec:models}. Section~\ref{sec:spectrum} is devoted to approximation strategies and exact results for spectral properties of closed systems in the USC regime. The treatment of dissipation and external driving fields is the subject of Sec.~\ref{sec:open}. Theoretical methods specific to waveguide QED setups, such as scattering theory, are discussed in Sec.~\ref{sec:scattering}. In Sec.~\ref{sec:limitations}, we discuss gauge invariance issues and other fundamental limitations of effective models in the USC. We conclude in Sec.~\ref{sec:conclusion}.    

\section{Models}
\label{sec:models}
In this section, we introduce the effective models of light-matter interaction that will be used in the rest of the article to illustrate the different methods presented. As pointed out in the introduction there are a variety of experimental platforms in which the USC regime has been reached. Nevertheless, the essential features of the USC regime are for the most part captured by effective models that share platform-independent characteristics. As the main focus of this review is on general methods and tools, we will restrict ourselves to a set of models that best exemplify the theoretical challenges of USC cavity QED. All effective models of light-matter considered in what follows are based on a general non-relativistic formulation of quantum electrodynamics~\cite{Cohen-Tannoudji:1997} valid within the long-wavelength approximation. The question of their range of validity is discussed in Sec. \ref{sec:limitations}. 

Among the set of models, a distinction can be drawn between two classes, based on the nature of the matter degrees of freedom that they describe. In many setups, the latter can be well approximated by two-level systems (TLS). Hence the first paradigmatic family of models are spin-boson Hamiltonians. In its most general form it describes the interaction between an ensemble of TLS with several modes of the electromagnetic field. Its simplest, single-spin and single-boson version is the quantum Rabi model~\cite{Rabi:1937}
\begin{align}\label{HamQRM}
H_{\mathrm{QRM}} = \omega \crea{a}\ann{a} + \frac{\Omega}{2}\hat{\sigma}_z + g\hat{\sigma}_x(\crea{a}  + \ann{a}),
\end{align}
where $\ann{a}$ is the annihilation operator of the cavity mode of frequency $\omega$. The operators $\hat{\sigma}_z$, $\hat{\sigma}_x$ are Pauli matrices $\Omega$ denotes the energy of the TLS  and $g$ is the light-matter coupling strength (In all that follows we have set $\hbar = 1$). Equation~(\ref{HamQRM}) has been widely used as model of cavity QED systems. Therefore, many of the methods presented below were first applied or originally taylored for this Hamiltonian. Several extension of the Rabi model have been studied, including two-photon versions~\cite{Travenec:2012}, where the interaction term also contains $\ann[2]{a}$ and $\crea[2]{a}$ operators, $N$-level extensions~\cite{Albert:2012}, replacing the TLS by more complex level structures, or models including chiral light-matter interaction~\cite{Mahmoodian:2019}.
In this family of spin-boson model, the multimode version has been widely used to describe of a single spin strongly coupled to its environment~\cite{Leggett:1987}
\begin{equation}\label{HamSBM}
H_{\mathrm{SB}} = \frac{\Omega}{2}\hat{\sigma}_z + \sum_k g_k \hat{\sigma}_x(\crea{a}_k + \ann{a}_k) + 
\sum_k \omega_k \crea{a}_k\ann{a}_k.
\end{equation}
It is particularly relevant in the context of waveguide QED, where a single emitter is coupled to a continuum of modes. 

A second class of models for light-matter interaction are purely bosonic. Originally introduced by Hopfield \cite{Hopfield:1958}, such models are relevant when the matter degrees of freedom behave as bosonic (quasi)-particles~\cite{Ciuti:2005,Ciuti:2006, Bamba:2012}. It is the case in the first experimental platforms in which a signature of ultrastrong coupling was reported, where quantum well excitons are coupled to an intersubband transition~\cite{Todorov:2010}. A two-mode example of the Hopfield Hamiltonian is the following
\begin{align}\label{HamBosonic}
H = \omega_c \crea{a}\ann{a} + \omega_X \crea{b}\ann{b} + ig(\crea{a} + \ann{a})(\crea{b} -\ann{b}) + D(\crea{a} + \ann{a})^2,
\end{align}
where $\omega_X$ denotes the frequency of the matter degree of freedom. The last term, quantified by $D$, is a diamagnetic term. A multimode generalization to handle translation-invariant planar structures is straightforward.


\section{Spectral properties}
\label{sec:spectrum}
In the regime where the rotating-wave approximation is valid, the task of finding the ground state and other spectral properties of cavity QED systems is greatly simplified. Indeed, the RWA-version of the Rabi model, the Jaynes-Cummings Hamiltonian,
\begin{equation}\label{HamJC}
H_{\mathrm{JC}} = \omega \crea{a}\ann{a} + \frac{\Omega}{2}\hat{\sigma}_z + g(\hat{\sigma}_+\ann{a}  +\hat{\sigma}_-\crea{a}),
\end{equation}
conserves the total number of excitations $\crea{a}\ann{a} + \hat{\sigma}_+\hat{\sigma}_-$ and is exactly solvable. It is no longer the case when all the terms of Eq.~(\ref{HamQRM}) are included. 
Finding spectral properties of ultrastrongly coupled systems becomes highly nontrivial. This section surveys the different analytical tools that have been developed for this purpose. Note that although the advent of USC cavity QED has motivated most of the works reported here, some of the idea presented in the following where introduced much before the first proposals of USC experiments~\cite{Benivegna:1987}. We begin this section by reviewing different approximation schemes, based on perturbation theory, generalized RWA and variational approaches. The last part is devoted to exact results, such as the one leading to an analytical solution for the spectrum of the quantum Rabi model~\cite{Braak:2011}.   

\subsection{Perturbative approach}

A first approach is to handle non-resonant terms as a perturbation of a Hamiltonian $H_0$, 
containing both the free terms and the resonant part of the interaction.

\paragraph{Effective Hamiltonian }
This choice for the ''unperturbed'' Hamiltonian is motivated by the following feature.  The spectrum of $H_0$ is structured around subspaces $\mathcal{E}_N$ spanned by eigenstates that are close in energy within one of theses subspaces but far from any other eigenstate belonging to another subspace $\mathcal{E}_{N'}$. Under such circumstances, the dominant effect of the perturbation is to affect the dynamics within each subspace, while the coupling between different subspaces can be neglected at lowest order. As explained below, the label $N$ is related to the symmetry of the Hamiltonian and refers to a quantity that is conserved by the resonant part of the Hamiltonian. For example, for the quantum Rabi model, $H_0$ is given by  the Jaynes-Cummings Hamiltonian $H_{\mathrm{JC}}$ and  $N$ corresponds to the total number of excitations. The space $\mathcal{E}_N$ and $\mathcal{E}_{N+1}$ are separated by an energy of order $\omega$, while the level spacing within one subspace is of order $g$.

A precise formulation of these ideas consists therefore in finding a effective Hamiltonian accounting for the effect of the perturbation within each subspace $\mathcal{E}_N$ \cite{Cohen-Tannoudji:1998}. Formulated by Schrieffer and Wolff~\cite{Schrieffer:1966} in the context of condensed-matter physics, the method is quite general and has found many applications~\cite{Bravyi:2011}. In quantum optics, these ideas were applied to strongly driven systems in the form of quantum averaging techniques~\cite{Jauslin:2000}. Making use of symmetry properties and of the underlying Lie Algebra structure of light-matter interaction models, Klimov et al. also gave a systematic algebraic formulation known as the ``small rotation method''~\cite{Klimov:2000,Klimov:2002a, Klimov:2009}. More recently, this approach was exploited to obtain effective low-energy approximations to finite-component Hamiltonians exhibiting a superradiant phase transition~\cite{Hwang:2015, Zhang:2019}. A mathematical expression for the general perturbation scheme is the following. Consider the general Hamiltonian~\cite{Amniat-Talab:2005}
\begin{equation}
H = H_{0} + \epsilon V,
\end{equation}
and a conserved quantity $\hat{N}$, such that $[H,\hat{N}] = 0$ that defines the subspaces $\mathcal{E}_N$. One looks for a unitary transformation $e^{\epsilon W}$ such that in the transformed Hamiltonian $\tilde H = e^{-\epsilon W}He^{\epsilon W}$, the effect of the perturbation outside of $\mathcal{E}_N$ is of second order:
\begin{equation}\label{pUSC}
\tilde H = H_0 + \epsilon \hat{D} + \epsilon^2\hat{V}_2,
\end{equation}
with $[\hat{N},\hat{D}] = 0$. In many light-matter interaction Hamiltonians, $W$ is constrained by the underlying algebraic structure of $H$ \cite{Klimov:2009}. In particular, the perturbative expansion can be derived in a systematic way  when $V = X_+  + X_-$, such that
\begin{equation}\label{algBS}
[H_0, X_{\pm}] = \pm X_{\pm} \quad \text{and} \quad [X_+,X_-] = P(H_0),
\end{equation}
with $P$ a polynomial function. Under these assumptions the operator $W$ takes the form 
\begin{equation}
W \propto X_+ - X_-.
\end{equation}
\paragraph{Bloch-Siegert corrections and multiphotonic resonances} Applied to the QRM Hamiltonian, this perturbation scheme gives the so-called Bloch-Siegert corrections to the energy spectrum~\cite{Bloch:1940}. In this case, the counter-rotating terms define the operators $X_+ = \crea{a}\hat{\sigma}_+$ and $X_- = (X_+)^{\dagger}$.  
Applying first $U_1 = \mathrm{exp}\left[\frac{g}{\omega + \Omega}(\ann{a}\hat{\sigma}_- -\crea{a}\hat{\sigma}_+)\right]$ yields at first order
\begin{align}
\tilde{H}_1 = U_1^{\dagger}HU_1 =& H_{\mathrm{JC}}  + \frac{g^2}{\omega + \Omega}[\hat{\sigma}_z(\crea{a}\ann{a} + \frac{1}{2}) -\frac{1}{2}] \nonumber \\
&+ \frac{g^2}{\omega + \Omega}[(\crea[2]{a} + \ann[2]{a})\hat{\sigma}_z]. 
\end{align}
The last term does not commute with $\hat{N}$. This comes from the fact that the algebra of Eq.~(\ref{algBS}) is only obtained if $H_0$ is the non-interacting Hamiltonian. This last term is generated by the resonant interaction terms and can be eliminated by a second small rotation $U_2 = \mathrm{exp}\left[\frac{g^2}{2\omega(\omega + \Omega)}\hat{\sigma}_z(\ann[2]{a} - \crea[2]{a})\right]$ that produces no additional first order terms. Hence the Bloch-Siegert effective Hamiltonian is given by
\begin{equation}
H_{\mathrm{BS}} = H_{\mathrm{JC}} + \frac{g^2}{\omega + \Omega}[\hat{\sigma}_z(\crea{a}\ann{a} + \frac{1}{2}) -\frac{1}{2}]. 
\end{equation}
Note that the unitary transformations $U_1$ and $U_2$ also give the first correction to the eigenstates. 

The validity of this approximation for the QRM has been extensively studied~\cite{Irish:2005, Irish:2007, Rossatto:2017}. In the resonant case, the first Juddian points~\cite{Judd:1979}, defined as the first energy-level crossings in the spectrum, were proposed as a boundary for the perturbative regime~\cite{Rossatto:2017}. This perturbative approach is consistent with the first experimental observations involving an ultrastrongly coupled qubit-oscillator system in circuit QED~\cite{Forn-Diaz:2010}. 

Perturbation theory also applies to the study of implicit resonances induced by counter-rotating terms. For certains values of the detuning between the atom and the cavity, some multiphotonic processes involving intermediate states connected through counter-rotating terms may become resonant. An effective Hamiltonian that captures the dynamics of these processes can be derived within the general algebraic framework mentioned above~\cite{Klimov:2003, Klimov:2009}. Other derivations have been obtained by adiabatic elimination of the fastest dynamical variables in the relevant truncated Hilbert space~\cite{Ma:2015, Garziano:2015}. Several proposals have been made in recent years to exploit this feature of the USC regime and engineer various nonlinear optical analogs~\cite{Garziano:2016, Frisk-Kockum:2017, Stassi:2017,Munoz:2019}. 

To go beyond the perturbative USC regime, other approximation schemes have been developed. We first discuss the generalized rotating wave approximation.  
 
 \subsection{Generalized RWA}
\begin{figure}[]
\begin{center}
\includegraphics[width=0.98\columnwidth]{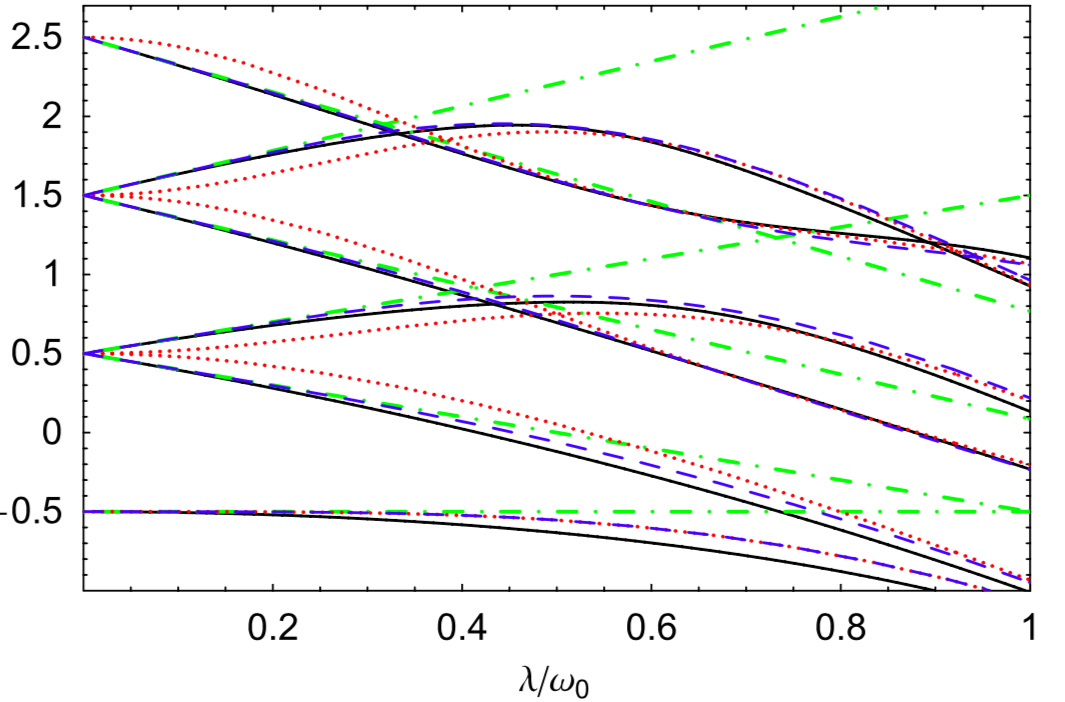}
\caption{\label{fig:Irish} Generalized rotating wave approximation applied to the Rabi Hamiltonian $H_{\mathrm{QRM}}$. Comparison of RWA (dotted-dashed line), limit $\Omega = 0$ (dotted line), and GRWA (dashed line) with the exact spectrum (solid line). Reproduced with permission~\cite{Irish:2007}. Copyright 2007, American Physical Society.
}
\end{center}
\end{figure}
  
 In $H_0$ defined above, the free Hamiltonian is taken as the Hamiltonian at $g = 0$. The subsequent separation into resonant and non-resonant interaction terms, as well as the perturbative treatment of non-resonant terms is relative to this reference point. It turns out that more accurate results at large $g$ can be obtained by starting instead from the Hamiltonian at $g/\Omega \to +\infty$ (or equivalently $\Omega = 0$). 
 More precisely, the Generalized RWA introduced by Irish \cite{Irish:2007} combines the RWA selection of resonant contributions with a unitary transformation yielding a free Hamiltonian in the limit $g/\Omega \to +\infty$. Hence one changes the noninteracting Hamiltonian with respect to which the RWA is applied. Note that in contrast to the perturbation method described above, the unitary transformation need not be a small rotation. The implementation of the generalized RWA for the quantum Rabi model is the following~\cite{Irish:2007}. The new exactly solvable Hamiltonian that is considered reads 
 \begin{equation}
 H' = \omega \crea{a}\ann{a} + g\hat{\sigma}_x(\crea{a} + \ann{a}).
 \end{equation}
 The eigenstates of this Hamiltonian are factorized and doubly degenerate. They are of the form $\ket{\pm}\ket{\pm\alpha, N}$, where $\ket{\pm}$ are the eigenstates of the operator $\hat{\sigma}_x$ and $\ket{\pm\alpha, N}$ is the $N^{th}$ Fock state, displaced by the operator $D[\alpha] = \mathrm{exp}(\alpha \crea{a}- \alpha^* \ann{a})$. It is actually more convenient to introduce the basis $\ket{\Psi_{\pm,N}} = 1/\sqrt{2}(\ket{+}\ket{\alpha,N} \pm \ket{-}\ket{-\alpha,N})$, the operator $\hat{\sigma}_x$ being diagonal in the subspace spanned by $\{\ket{\Psi_{+,N}}, \ket{\Psi_{-,N}}\}$. Note that the states $\ket{\Psi_{\pm,N}}$ have a defined parity of the number of excitations.
 The change of basis outlined above is given by the following unitary transformation
 \begin{equation}\label{UGRWA}
 \ket{\Psi_{-,N}} = \hat{U} |\downarrow,N\rangle = e^{\frac{g}{\omega}\hat{\sigma}_x(\ann{a}-\crea{a})}|\downarrow,N\rangle,
 \end{equation}
 with the parameter $\alpha = -g/\omega$. In this new basis the QRM Hamiltonian of Eq.~(\ref{HamQRM}) is expressed as 
 \begin{equation}\label{QRM2}
 \tilde{H}_{\mathrm{QRM}} = \omega \crea{a}\ann{a} + \frac{\Omega}{2}\hat{\sigma}_z \mathrm{exp}
\left[ -\frac{2g}{\omega} \hat{\sigma}_x(\crea{a} - \ann{a}) \right].
 \end{equation}
The GRWA consists in keeping in $\tilde{H}_{\mathrm{QRM}}$ only the resonant terms. It can indeed be shown that in matrix form, $\tilde{H}_{\mathrm{QRM}}$ has the same structure as the original Hamiltoninan. Hence the GRWA yields a block-diagonal Hamiltonian, whose blocks are spanned by the eigenstates $\{\ket{\Psi_{N,-}}, \ket{\Psi_{N-1,+}} \}$. Indeed  the resulting Hamiltonian can be expressed as
\begin{equation}\label{HamGRWA}
\tilde H_{\mathrm{GRWA}} = \tilde \omega \crea{a}\ann{a} + \frac{\tilde \Omega}{2} \hat{\sigma}_z  + \tilde g[f(\crea{a}\ann{a})\crea{a}\hat{\sigma}_- + f^*(\crea{a}\ann{a})\ann{a}\hat{\sigma}_+],
\end{equation}
 where the coupling constants are in general renormalized with respect to the original Hamiltonian and $f$ is an analytical function coming from the series expansion of Eq.~(\ref{QRM2}). Note that resonant terms in $\tilde H_{\mathrm{GRWA}}$ do not conserve the number of excitations. Indeed, in the transformed basis, the spin degree of freedom represents the parity of the number of excitation and not the original spin. 
The GRWA spectrum, obtained after diagonalization of $\tilde H_{\mathrm{GRWA}}$ within each $\{\ket{\Psi_{N,-}}, \ket{\Psi_{N-1,+}} \}$-subspace was also derived earlier by other methods~\cite{Feranchuk:1996, Amniat-Talab:2005}. Interestingly, whereas the RWA for the Rabi model breaks down at the first level-crossing, the GRWA gives more accurate results for a wider range of parameters (see Fig.~\ref{fig:Irish}). The GRWA approach has been successfully applied to other light-matter interaction models related to the QRM, namely to the two-qubit Rabi model~\cite{Zhang:2015} and to a biased version of the QRM~\cite{Zhang:2013}. A so-called symmetric implementation the GRWA, that exploit more efficiently the symmetries of the model has also been proposed for the single- and two-photon models~\cite{Albert:2011}. A possible explanation for the wider range of validity of the GRWA is the following. In the RWA, the degeneracy in the energy spectrum associated with the resonant terms is exact only when $\Omega = \omega$ and $g= 0$. The resonant terms in the GRWA are related to a degeneracy that is exact for $\Omega = 0$ (or $g \to \infty$), which represents a larger part of the parameter space.
 
 An extension of these ideas was put forward by Zhang \cite{Zhang:2016}, who suggested to add to the displacement defined in Eq.~(\ref{UGRWA}), a squeezing operation. The starting point for the RWA now becomes 
 \begin{equation}
 \tilde H = \hat{V}^{\dagger}\hat{U}^{\dagger} H \hat{U}\hat{V},
 \end{equation}
 with
 \begin{equation}
 \hat{U} = e^{\beta\hat{\sigma}_x(\ann{a}-\crea{a})} \quad \text{and} \quad \hat{V} = e^{\lambda(\crea[2]{a} - \ann[2]{a})}.
 \end{equation}
In contrast to the transformation of Eq.~(\ref{UGRWA}), the displacement and squeezing parameters $\beta$ and $\lambda$, are not determined from the diagonalization of a new free Hamiltonian, but are computed variationally by minimizing the ground state energy.  After this change of basis, the Hamiltonian is simplified in the same way as before. After the transformation, the Hamiltonian still takes the general form of Eq.~(\ref{HamGRWA}), with the additional subtlety that the renormalized atomic frequency is also a function of $\crea{a}\ann{a}$. Nevertheless, the problem is reduced to the diagonalization of $2\times2$ matrices.

\begin{figure}[]
\begin{center}
\includegraphics[width=0.89\columnwidth]{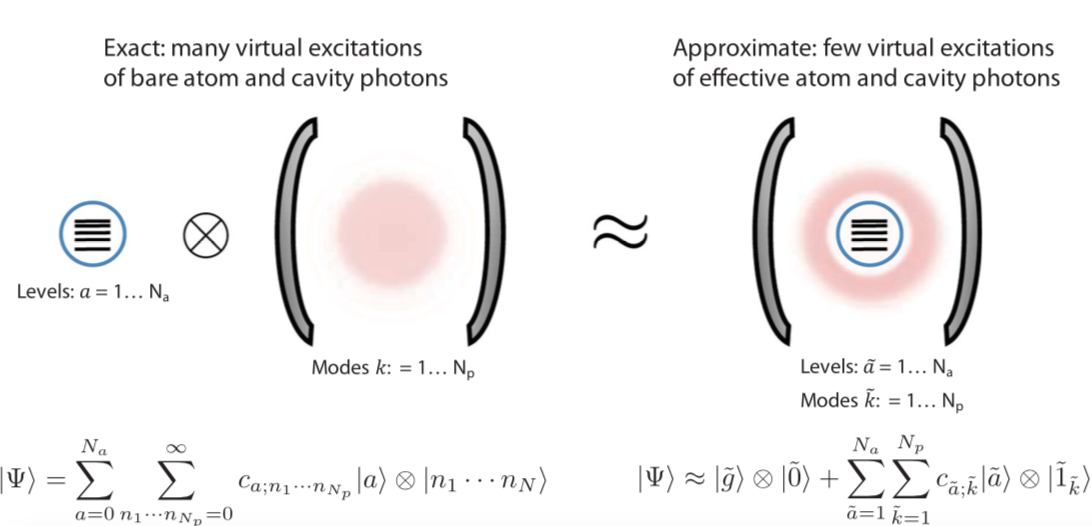}
\caption{\label{fig:Rivera}Physical picture undelying the ab initio approach described in Eqs.~(\ref{abinitio1}) and (\ref{abinitio2}). (left) Bare description of the coupled light-matter ground state in terms of many virtual excitations of the emitter state and the bare cavity photons. (right) Quasiparticle description of the coupled system as a factorizable state of an effective emitter in its ground state and the vacuum of an effective photonic degree of freedom. Reproduced with permission~\cite{Rivera:2019}. Copyright 2019, American Physical Society.
}
\end{center}
\end{figure}

\subsection{Variational methods}

In the previous section, we have seen that GSRWA method combined the generalized rotating wave approximation with a variational determination of additional squeezing paramaters. More generally, the ground-state properties of various models have been investigated using a variational method.

\paragraph{Polaron picture} In the case of spin-boson models, multi-polaron test functions and their generalization have proved to be efficient \textit{Ans\"atze}. For the spin-boson model of Eq.~(\ref{HamSBM}), a multi-polaron wave function is defined as~\cite{Bera:2014}
\begin{equation}\label{polaronWF}
\ket{\Psi}  = \sum_{n = 1}^{N_{\mathrm{pol}}} C_n [\ket{+, \alpha^{(n)}} - \ket{-,-\alpha^{(n)}}],
\end{equation}
where $\ket{\alpha^{(n)}} = \ket{\alpha_1,\alpha_2 \dots}$  is a multimode coherent state, $\ket{\alpha} = \mathrm{exp}(\sum_n \alpha_n \crea{a}_n - \alpha_n^*\ann{a}_n)\ket{0}$.

In the case of the quantum Rabi model, the state $\ket{\Psi_{-,0}}$ introduced in Eq.~(\ref{UGRWA}) is an example of a single polaron wave-function. As mentioned above, this state is the exact ground state in the limit $\Omega = 0$. The relevance of polaron wave functions as trial functions for the ground state of the QRM was actually recognized in early studies of the model, long before the advent of USC cavity QED~\cite{Benivegna:1987}. The reasoning is based on the parity symmetry of the QRM. As the parity of the number of excitations, given by the operator $P = \hat{\sigma}_z e^{i\pi\crea{a}\ann{a}}$, is conserved, the problem is simplified by considering separately the two subspaces of states with odd or even parity.  Restricted to such subspaces, the Hamiltonian becomes purely bosonic and can be expressed as \cite{Benivegna:1987, Hwang:2010}
\begin{equation}\label{QRMbose}
H_{\pm} = \omega \crea{b}\ann{b} + g(\crea{b} + \ann{b}) \pm \frac{\Omega}{2}\cos(\pi\crea{b}\ann{b}),
\end{equation}
where the $\pm$ signs refer to the odd and even parity subspaces. A coherent state in this representation of the Hamiltonian correspond to a polaron state in the original basis. 

Such an ansatz was later improved by considering squeezed coherent states~\cite{Hwang:2010}, and deformed, frequency renormalized polarons~\cite{Ying:2015, Cong:2017}.
This approach was also applied to the two-qubit~\cite{Mao:2019}  and two-photon~\cite{Cong:2019} Rabi models. Unlike the GRWA or the effective Hamiltonians presented above, these polaron-based methods have been used mostly to extract ground-state properties, although possible extensions to excited states have been proposed~\cite{Ying:2015}. Other applications of polaron transformations in the context of QED are presented in Sec. \ref{DynPolaron}.

\paragraph{Ab initio approaches} Variational methods have also been developed to tackle more complex QED systems in which it is essential to take into account complex electronic configurations and  multiple cavity modes. Inspired by density functional theory and its recent extension to quantum electrodynamics~\cite{Tokatly:2013, Flick:2018}, such methods take as a starting point the general light-matter Hamiltonian in the Coulomb gauge. 
 In this context, a variational principle able to tackle ultrastrongly coupled systems was recently proposed by Rivera et al. \cite{Rivera:2019}. The ground-state is assumed to  be a product of a Fermi sea of quasi-particules parametrized by one-particle wavefunctions $\psi_i(\mathbf{r})$, and of the vacuum of effective photonic degrees of freedom (see Fig.~\ref{fig:Rivera}). The photonic variables are the cavity mode functions $F_i(\mathbf{r})$, defined such that the vector potential $\hat{\mathbf{A}}(\mathbf{r})$ is expressed as $\hat{\mathbf{A}}(\mathbf{r})= \sum_i [F_i(\mathbf{r})\crea{a}_i +  F^*_i(\mathbf{r})\ann{a}_i]$. The set of coupled equations for the dressed one-particle wave function $\psi_i(\mathbf{r})$, effective photonic mode functions $F_i(\mathbf{r})$ and mode frequency $\omega_i$ are obtained by minimizing the ground state energy, under normalization constraints. The set of equations takes the general form
 \begin{align}
 (\frac{\mathbf{p}^2}{2m} + v_{\mathrm{ext}}(\mathbf{r}))\psi_i(\mathbf{r}) + F[\{\psi\}] & \nonumber\\
+ \frac{\hbar e^2}{4m\epsilon_0}\left(\sum_n \frac{1}{\omega_n}F^2_n(\mathbf{r})\right)\psi_i(\mathbf{r}) &= E_i \psi_i(\mathbf{r}),\label{abinitio1}\\
 \left[ \nabla \times \nabla \times \frac{\omega^2_i}{c^2}(1-\frac{\omega^2_p(\mathbf{r})}{\omega_i^2})\right]\mathbf{F}_i(\mathbf{r}) &=0,\label{abinitio2}
 \end{align}
where $\omega^2_p(\mathbf{r}) = \frac{e^2}{m\epsilon_0}\sum_n |\psi^2_n(\mathbf{r})|$ defines a position-dependent plasma frequency. Given the form of the Ansatz, the interaction term proportional to $\mathbf{A}\cdot \mathbf{p}$ in the Hamiltonian does not enter into the above equations. The effect of this term is taken into account self-consistently by means of second-order perturbation theory.
So far, a proof of principle has been given for a single emitter placed in a 1D cavity. Comparison with exact numerical calculation for the ground state and one excited state shows that the self-consistent corrections to the dressed wave function and photonic modes make the scheme accurate also in the ultrastrong coupling regime.
  
  A type of variational computation based on quantum algorithms has been recently adapted for cavity QED in the USC regime~\cite{DiPaolo:2019}. However, algorithms relying on quantum hardware lie outside of the scope of this review.

\subsection{Exact results}
\begin{figure}[]
\begin{center}
\includegraphics[width=0.85\columnwidth]{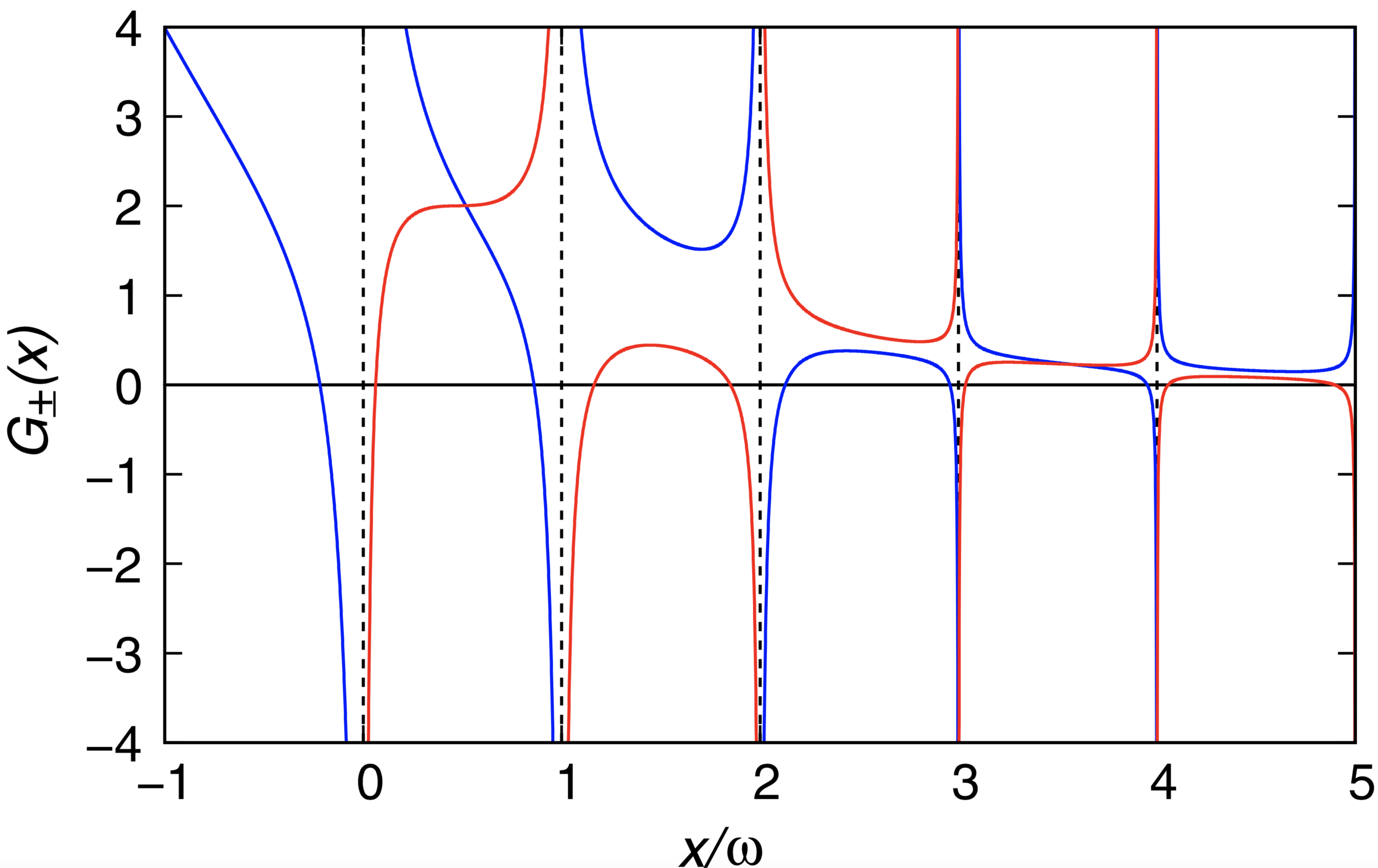}
\caption{\label{fig:Braak}. Transcendental functions $G_{+}(x)$[red] and $G_{-}(x)$ [blue] whose zeros give the energy levels of the Rabi Hamiltonian of Eq.~(\ref{HamQRM}). The parameters are $\omega = 1$, $g = 0.7$ and $\Omega = 0.4$. Reproduced with permission~\cite{Braak:2011}. Copyright 2011, American Physical Society.
}
\end{center}
\end{figure}
 
Although for most cavity QED problems one has to rely on approximation methods, such as the one presented in the preceding sections, there exist exact analytical solutions for the eigenvalue problem of a class of light-matter interaction models, including the Rabi model. The exact solution of the Rabi model found by Braak~\cite{Braak:2011} has triggered an important research activity bringing together both theoretical physicists and mathematicians~\cite{Braak:2016}. 

\paragraph{Bargmann space representation} Braak's solution to the eigenvalue problem of the quantum Rabi model is based on the Bargmann space representation of bosonic creation and annihilation operators~\cite{Bargmann:1961}. The first application of Bargmann space methods to quantum optical models dates back to the late 60s \cite{Schweber:1967} and was used to establish several analytical results~\cite{Reik:1986, Kus:1986} prior to the proof of the exact solution. 
In this representation, the Hilbert space of physical states is that of analytical functions of a complex variable $z$, on which the creation and annihilation operators act in the following way
\begin{equation}\label{Bargmann}
\crea{a} \to \frac{\partial}{\partial z} \quad \text{and}\quad \ann{a} \to z.
\end{equation}
The inner product in this Hilbert space is defined as 
\begin{equation}
\braket{\psi}{\phi} = \frac{1}{\pi}\int dz dz^* e^{-zz^*} \psi^*(z)\phi(z),
\end{equation}
which also specifies the normalization requirement for the wave functions $\braket{\psi}{\psi} < \infty$.
The Schr\"odinger equation is therefore mapped to a differential equation in the complex plane in which the energy $E$ enters as a parameter. The requirement that the function $\psi(z)$ be analytical in the whole complex plane imposes some constraints on the admissible values of $E$. These constraints, along with all the symmetry properties of the model can be exploited to extract  an exact (but implicit) expression for the spectrum of the Hamiltonian.

In the case of the quantum Rabi model, due to the spin degree of freedom, the total wave function in Bragmann space has two components $(\phi_1(z), \phi_2(z))$. After a $\pi/2$-rotation of the spin, the Hamiltonian reads
\begin{equation}
H = 
\begin{pmatrix}
\omega z \partial_z + g(z + \partial_z)&\frac{\Omega}{2}\\
\frac{\Omega}{2}&\omega z \partial_z - g(z + \partial_z)
\end{pmatrix},
\end{equation}

which yield the following Schr\"odinger equation
\begin{align}
(z+g)\partial_z\phi_1(z) + (gz-E)\phi_1(z) + \frac{\Omega}{2}\phi_2(z) = 0, \label{BargODE1}\\
(z-g)\partial_z\phi_2(z) - (gz+E)\phi_2(z) + \frac{\Omega}{2}\phi_1(z) = 0. \label{BargODE2}
\end{align}

From Eqs. (\ref{BargODE1}) and (\ref{BargODE2}), the spectrum can be extracted in different equivalent forms. Due to the $\mathbb{Z}_2$ symmetry of the model, the problem can be solved separately in the odd and even parity subspace. Braak's solution exploits this feature by working directly on the Bargmann representation of Eq. (\ref{QRMbose}). It is then shown that $E$ belongs to the spectrum if and only if $x = E+g^2$ belongs to the set of zeros of some transcendental functions $G_{\pm}(x)$, where the index $\pm$ indicates the parity subspace (see Fig.~\ref{fig:Braak}). The functions $G_{\pm}(x)$ are expressed as power series whose coefficients can be computed recursively through the Taylor expansion of the solutions to Eqs.~(\ref{BargODE1}) and (\ref{BargODE2}).
Maciejewski et al.~\cite{Maciejewski:2014} gave an equivalent solution for the energy spectrum by relating Eqs.~(\ref{BargODE1}) and (\ref{BargODE2}) to the general theory of Heun differential equations~\cite{Ronveaux:1995}. By mapping the original problem to a second order Heun confluent equations, the results can be expressed in terms of confluent Heun functions. For a detailed presentation of these special functions we refer the reader to Ref.~\cite{Slavyanov:2000}.
Note that applications of the Bargmann representation and the theory of complex differential equation is not limited to parity-symmetric models. Solutions to the more general class of anisotropic Rabi models, where the symmetry is explicitly broken by an additional $\hat{\sigma}_x$ term and in which resonant and anti-resonant interaction terms depend on two-different coupling constants, were later derived~\cite{Xie:2014, Tomka:2014}. Exact solutions have also been found for other generalizations of the Rabi model such as the two-photon model~\cite{Travenec:2012}. A more complete presentation of the solution for these models can be found in Ref.~\cite{Xie:2017}.

\paragraph{Generalized coherent states and Bogoliubov operators}
Within the family of spin-boson models with a finite number of bosonic modes, the insight provided by the ``polaron picture'' -- displaced states of the oscillators conditioned by the $\hat{\sigma}_x$ projection of the TLS -- has been used to derive analytical and numerical exact results. The generalized coherent state approach can be viewed as an algebraic implementation of this idea, in which displacements are introduced in the form of Bogoliubov transformations on the bosonic operators~\cite{Chen:2008}. In contrast to the GRWA or variational approaches, this method is not restricted to the low-lying energy states but was implemented to obtain exact results on the full spectrum of several light-matter interaction models. We illustrate the principle of this method in the specific case of the Rabi model, for which it provides an alternate derivation of the exact results obtained trough the Bargmann space representation of the wave function. 
Given the role played by the $\sigma_x$ (and its relation to the symmetry of the model), the first step is to apply a $\pi/2$ rotation of the spin and write the Hamiltonian in matrix form as~\cite{Chen:2012}
\begin{equation}
H  =  
\begin{pmatrix}
\crea{a}\ann{a} + g(\crea{a} + \ann{a}) & \frac{\Omega}{2}\\
\frac{\Omega}{2} & \crea{a}\ann{a} - g(\crea{a} + \ann{a})
\end{pmatrix}.
\end{equation}
Displacements of the oscillator are subsequently introduced in the form of two Bogoliubov transformations $\ann{a}  \to \ann{A} = A + g$ and $\ann{a} \to \ann{B} = \ann{a} -g$. Applying each one of these of this transformation yields a Hamiltonian of the form
\begin{equation}
\begin{pmatrix}
\crea{A}\ann{A} - g^2 & \frac{\Omega}{2}\\
\frac{\Omega}{2} & \crea{A}\ann{A} - 2g(\crea{A} + \ann{A}) + 3g^2
\end{pmatrix}.
\end{equation}
The Schr\"odinger equation can now be written in the displaced Fock basis $\{\ket{n}_A = \frac{(\crea{A})^n}{\sqrt{n!}}\ket{0}_A\}$, or equivalently using the basis $\{\ket{n}_B= \frac{(\crea{B})^n}{\sqrt{n!}}\ket{0}_B\}$, where the states $\ket{0}_B$ and $\ket{0}_A$ are the coherent states such that $\ann{B}\ket{0}_B = 0$ and $\ann{A}\ket{0}_A = 0$. Writing the wave function as
\begin{equation}
\ket{\Psi} =
\begin{pmatrix}
\sum_{n = 0}^{\infty}\sqrt{n!}e_n\ket{n}_A\\
\sum_{n = 0}^{\infty}\sqrt{n!}f_n\ket{n}_A
\end{pmatrix},
\end{equation}
the Schr\"odinger equation translates into recursion relations for $e_n$ and $f_n$ that depend on the energy $E$. Applying the same reasoning to the second Bogoliubov transformation yields a second set of recursion relation relative to the basis  $\{\ket{n}_B\}$. As the two representations correspond to a unique state, an implicit equation for $E$ can be extracted, equivalent to the solution of Ref.~\cite{Braak:2011}. 
The method has first been applied to the finite-size Dicke model \cite{Chen:2008}, for which it provided an efficient way of computing ground-state observables for a large number of atoms and with arbitrary precision. More recently it has also been applied to the two-mode Rabi model~\cite{Duan:2015, Duan:2016, Cui:2017} and the quantum Rabi-Stark model~\cite{Xie:2019}.


\section{Open systems}
\label{sec:open}

\begin{figure}[]
\begin{center}
\includegraphics[width=0.99\columnwidth]{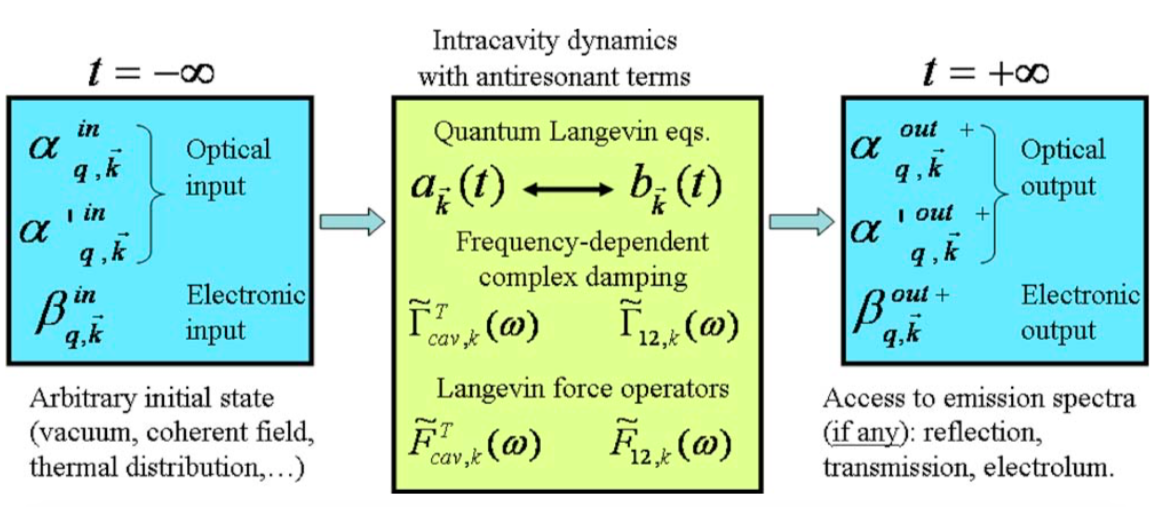}\\
\includegraphics[width=0.99\columnwidth]{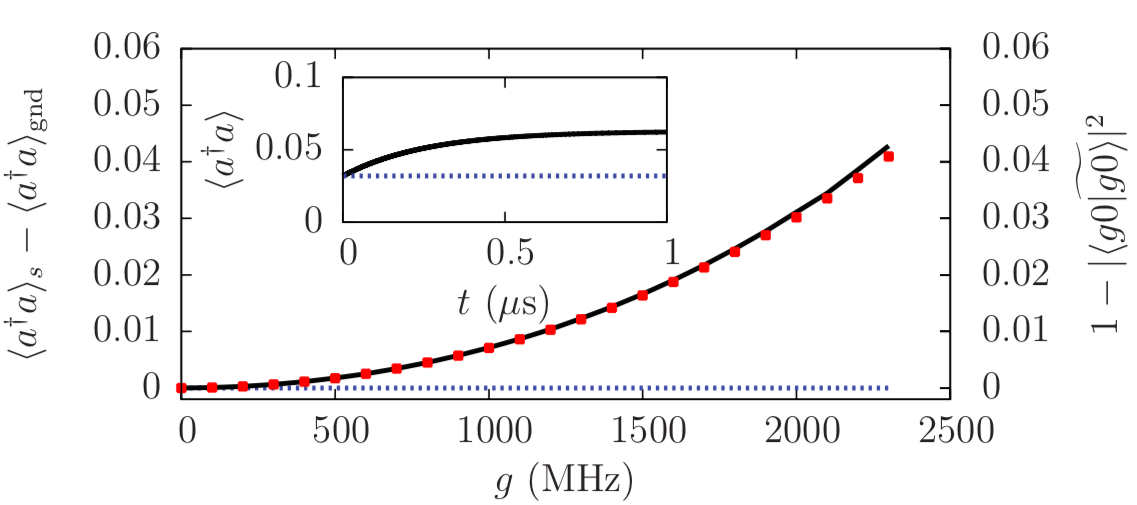}
\caption{\label{fig:dissipation}(Top) A  sketch of the input-output theory in the form of Langevin equations as originally presented in Ref.~\cite{Ciuti:2006}. Labels $k$ and $q$ correspond to in-plane and orthogonal wave vector respectively. Reproduced with permission~\cite{Ciuti:2006}. Copyright 2006, American Physical Society. 
(Bottom) Example of failure of the phenomenological master equation Eq.~(\ref{MESC}) in the form of an excess of photon (black line) in the seady-state $|g,0\rangle$ with respect to the true ground state $\tilde{|g,0\rangle}$. Simulations are performed for the Rabi Hamiltonian with $\Omega/2\pi = \omega/2\pi$ = 6GHz and $\gamma/2\pi = \kappa/2\pi 0.1$ GHz.  Red dots correspond to one minus the fidelity. Dotted lines are the result computed with the correct master equation. Reproduced with permission~\cite{Beaudoin:2011}. Copyright 2011, American Physical Society.
}
\end{center}
\end{figure}

The methods and results presented in the previous section dealt only with closed systems, setting aside the issue of the experimental signature of the various spectral features in realistic setups~\cite{DeLiberato:2007,Lolli:2015, Felicetti:2015}. It proved particularly fruitful to address this question within the more general framework of open quantum systems, where dissipative processes stemming from the coupling of the system to its environment play a crucial role. In this context, it was recognized that without proper modifications, the usual approach would lead to unphysical predictions such as the emission of photons by a system in its ground state~\cite{Werlang:2008, DeLiberato:2009, Beaudoin:2011}. 
The theory described in this section is structured around three elements. The first one is the master equation, governing the dynamics of the internal degrees of freedom of the system. The second one is the input-output theory, relating the signal observed via a given detection scheme to the internal degrees of freedom and to possible input field. The last aspect that is discussed is the specific case of open systems driven by an external periodic field. In this context, Floquet theory can be employed to efficiently treat the time dependency in an exact way. 

\subsection{Master equation}


Master equations form an important branch of the theory of open quantum systems which has found numerous applications in quantum optics~\cite{Breuer:2002}. It has successfully been applied to the paradigmatic model of cavity QED, i.e. a system composed of a TLS and a cavity mode. At temperature $T= 0$, the usual quantum optical master equation for the density matrix $\rho$ of such a system takes the canonical Lindblad form and reads
\begin{align}
\frac{d\rho}{dt} &= -i[H,\rho] + \frac{\gamma}{2}[2\ann{a} \rho \crea{a} - \crea{a}\ann{a}\rho - \rho\crea{a}\ann{a}] \nonumber \\
&+ \frac{\kappa}{2}[2\hat{\sigma}_-\rho \hat{\sigma}_+ - \hat{\sigma}_+\hat{\sigma}_-\rho - \rho \hat{\sigma}_+\hat{\sigma}_-], \label{MESC}
\end{align}
where $\kappa$ and $\gamma$ are damping rates associated with the cavity and the atom respectively. The non-unitary terms in this equation translate the fact that both the cavity and the atom are coupled to their environment. It remains partly phenomenological, as dissipative processes involving the atom and the cavity are treated independently. The total dissipator in Eq.~(\ref{MESC}) is indeed the sum of the dissipators obtained if the cavity and the atom were not coupled. 
In the absence of pumping terms in the Hamiltonian, such a master equation will drive the system to d the state $\ket{g}\ket{0}$. 
It is clear that Eq.~(\ref{MESC}) cannot correctly account for the dynamics in the USC since the state $\ket{g}\ket{0}$ is no longer the ground state of the system. Such a master equation would lead in particular to emission of photons in the ground state (see Fig.~\ref{fig:dissipation})~\cite{Werlang:2008, Beaudoin:2011}. 

A first approach to handle corrections to the phenomenological equation is to go back to the microscopic derivation of Markovian master equations~\cite{Breuer:2002}.  First applied to the Jaynes-Cummings~\cite{Rau:2004, Scala:2007a, Scala:2007b}, this method was later extended to the Rabi model~\cite{Beaudoin:2011}, providing a realistic description of dissipative processes in the USC. In particular, it was shown that unphysical predictions arise even when the effect of counter-rotating terms in the atom-cavity Hamiltonian can be treated perturbatively.
In order to highlight the correction to Eq.~(\ref{MESC}), we first recall the basic principles on which the microscopic derivation is based, as outlined, e. g., in Ref. \cite{Breuer:2002}.

\paragraph{Setup and notations}
Let us consider the typical system - bath Hamiltonian
\begin{equation}\label{HamBath}
H = H_S + H_B + \sqrt{\gamma}A\otimes B, 
\end{equation}
where $H_S$ and $H_B$ are respectively the system and bath Hamiltonians. In the interaction term, $A$ and $B$ are Hermitian operators acting only on the system and bath Hilbert spaces respectively. The bath is typically assumed to be an infinite collection of harmonic oscillators. Generally speaking, the effect of the bath on the system dynamics is to induce transitions between eigenstates of the system Hamiltonian $H_S$. Transitions at frequency $\omega$ are determined by the following jump operator 
%
\begin{equation}\label{Aomega}
A(\omega) = \sum_{\omega_{ki} = \omega} |i\rangle\langle i| A|k\rangle\langle k|,
\end{equation}  
where the $|i\rangle$ are eigenstates of $H_S$ and $\omega_{ki}$ are the corresponding Bohr frequencies. As we will see in the following, it is essential in the USC regime to write Eq.~(\ref{Aomega}) in the dressed-state basis of the full system.  Note that the operators $A(\omega)$ introduce a decomposition of $A$, such that $A = \sum_\omega A(\omega)$. The aim of the master equation is to obtain an approximate equation of motion for the reduced density matrix $\rho$ of the system by tracing out the bath degrees of freedom. Instead of keeping track of all its microscopic degrees of freedom, the relevant information on the bath is  encoded in the functions
\begin{equation}
\Gamma(\omega) = \gamma \int_{0}^{+\infty} d\tau e^{i\omega \tau}\langle B(\tau)B(0)\rangle. 
\end{equation}

\paragraph{Born-Markov approximation}
The derivation of the master equation relies on two main assumptions (Born-Markov approximation): a perturbative treatment, up to second-order, of the system-bath interaction and a fast decay of the bath correlation functions on the time scale of the system's internal dynamics. These assumptions are not \textit{a priori} incompatible with the USC regime, as the latter only quantify the strength of the interactions within subparts of the system. This approximation scheme, along with the trace over bath degrees of freedom, it is most conveniently expressed in the interaction picture, where the time evolution of $A(\omega)$ is straightforwardly given by $A(\omega,t) = e^{-i\omega t}A(\omega)$. Within this framework, the equation of motion for the reduced density matrix reads
\begin{align}\label{MEi}
\frac{d\rho}{dt} &= \sum_{\omega, \omega'}  e^{i(\omega'-\omega)t} \Gamma(\omega)(A(\omega)\rho A^{\dagger}(\omega')- A^{\dagger}(\omega')A(\omega)\rho)\nonumber\\
& + \sum_{\omega, \omega'} e^{i(\omega-\omega')t} \Gamma^*(\omega)(A(\omega')\rho A^{\dagger}(\omega)- \rho A^{\dagger}(\omega)A(\omega')),
\end{align}
 which yields the following equation in the Schr\"odinger picture
\begin{equation}
\frac{d\rho}{dt} = -i[H_S,\rho] + \sum_{\omega, \omega'} \Gamma(\omega)(A(\omega)\rho A^{\dagger}(\omega')- A^{\dagger}(\omega')A(\omega)\rho) + \text{h.c}.
\end{equation}
The particularity of this equation is that it does not guarantee complete positivity of the density matrix. This requirement is fulfilled by assuming an additional simplification, the secular approximation.

\paragraph{Secular approximation}
If $|\omega - \omega'| \gg \gamma$ for $\omega \neq \omega'$, one can perform a rotating-wave approximation in Eq. (\ref{MEi}) and keep only the terms for which $\omega = \omega'$. This gives a Lindblad equation
\begin{equation}\label{MEsec}
\frac{d\rho}{dt} = -i[H_S,\rho] + \sum_{\omega} \Gamma(\omega)(A(\omega)\rho A^{\dagger}(\omega)- A^{\dagger}(\omega)A(\omega)\rho) + \text{h.c.},
\end{equation}
which guarantees that the dynamical map it defines is complete positive and trace preserving.  For the physical interpretation of the different terms, it is useful to write $\Gamma(\omega) = \gamma(\omega) + iD(\omega)$. With these notations the general expression of the master equation at $T = 0$ reads
\begin{equation}\label{microME}
\frac{d\rho}{dt} = -i[H_S + H_L,\rho] + \mathcal{D}[\rho],
\end{equation}
with
\begin{align}
H_L &= \sum_{\omega>0}D(\omega)A^{\dagger}(\omega)A(\omega), \\
\mathcal{D}[\rho] & = \sum_{\omega>0} \gamma(\omega)(2A(\omega)\rho A^{\dagger}(\omega)- \{A^{\dagger}(\omega)A(\omega),\rho\}). 
\end{align}
The operator $H_L$ gives a Lamb shift that is usually absorbed in $H_S$ by redefining the system Hamiltonian, while the dissipator $\mathcal{D}[\rho])$ governs the non-unitary part of the dynamics. In the particular case introduced at the beginning of this section, where both the atom and the cavity are coupled to a bath, there are two operators, $A_1 = \crea{a} + \ann{a}$ and $A_2 = \sigma_+ + \sigma_-$, coupling the system to its environment. 
In light of the above considerations, the essential corrections to Eq.~(\ref{MESC}) imposed by the USC regime can be formulated in the following way. As the coupling strength becomes comparable to the cavity frequency, one cannot assume that $\gamma(\omega) \approx \gamma(\omega_{cav})$ for all transitions. Moreover, as the structure of the energy levels in the USC regime is expected to differ drastically from the uncoupled system, the jump operators can be defined consistently only in the dressed-state basis. In the case of the Rabi model, it was shown that for realistic experimental implementation in circuit QED and for a large range of coupling strengths, the correct dissipator in the USC regime includes individual jump operators for all possible transition~\cite{Beaudoin:2011, Ridolfo:2012, LeBoite:2016}. In other words all transitions have to be considered non-degenerate, in sharp contrast to Eq.~(\ref{MESC}).

\paragraph{quasi-degenerate spectrum}
For extreme values of the coupling strength, such that it dominates all other energy scales, a wide of class of systems exhibit quasi-degenerate eigenstates~\cite{Felicetti:2020}. In driven-dissipative setups, such degeneracies can be lead to appearance of long-lived metastable states~\cite{LeBoite:2017}. Under such circumstances, the assumption that $|\omega - \omega'| \gg \gamma$ for $\omega \neq \omega'$, underlying the secular approximation breaks down. It is however possible to recover a Lindblad equation such as Eq.~(\ref{MEsec}) by extending the definition of this approximation. Suppose that approximate equivalent classes of frequencies can be defined as $\mathcal{P(\bar\omega)} = \{\omega :  \gamma(\omega) \simeq \gamma(\bar\omega)\}$, and let us write
$\omega_1 \equiv \omega_2 \quad \text{if} \quad \mathcal{P}(\omega_1) = \mathcal{P}(\omega_2)$.
In the general experssion of the dissipator in the interaction pictutre given in Eq. (\ref{MEi}), terms involving frequencies $\omega$ and $\omega'$ such that $ \omega \equiv \omega' $ factorize. 
Expressing the secular approximation as $|\bar{\omega}_1-\bar{\omega}_2| \gg \gamma$ for $\bar{\omega}_1 \not \equiv \bar{\omega}_2$, one recovers a dissipator in the Lindblad form, which reads 
\begin{align}
\mathcal{D}[\rho] = \sum_{\bar\omega >0} \gamma(\bar\omega)(2\tilde A(\bar\omega)\rho \tilde A^{\dagger}(\bar\omega)- \{\tilde A^{\dagger}(\bar\omega)\tilde A(\bar\omega),\rho\}),
\end{align}
where the sum is now performed over inequivalent frequencies only and the jump operators are defined as $\tilde A(\bar\omega) = \sum_{\omega \in \mathcal{P}(\bar\omega)} A(\omega)$
In this formulation of the secular approximation, problems arise when $\bar{\omega}_1 \not \equiv \bar{\omega}_2$ and $|\bar{\omega}_1-\bar{\omega}_2| \sim \gamma$. This can be a serious issue in the case of strongly driven systems. To go beyond the secular approximation in the USC regime, a more general formalism was introduced by Settineri \textit{et al.} in the context of hybrid quantum systems~\cite{Settineri:2018}.

\paragraph{Non-Markovian effects}
To account for non-Markovian effects, an exact master equation for a general open quantum system can be written following the Nakajima-Zwanzig projection method~\cite{Breuer:2002}. Without further approximation, such an equation is integro-differential and contains an involved time convolution with a memory kernel, but approximation strategies relying on a perturbative expansion of the coupling to the bath are  available to make the master equation local in time while going beyond the Markov approximation. Before the study of Beaudoin et al.~\cite{Beaudoin:2011}, De Liberato et al. \cite{DeLiberato:2009} have followed such an approach, called the second-order time-convolutionless operator approach, to correct the master equation in the USC regime. It was also implemented by Nataf et al. in a proposal for protected superconducting qubits in the USC regime~\cite{Nataf:2011}. The form of the master equation is 
\begin{align}
\frac{d\rho}{dt} = -i[H, \rho] + \sum_{j}(\hat{U}_j\rho\hat{S}_j + \hat{S}_{j}\rho\hat{U}^{\dagger}_j  - \hat{S}_j\hat{U}_j\rho - \rho\hat{U}_j^\dagger \hat{S}_j),
\end{align}
where the $\hat{S}_j$ are system-bath coupling operators. The operators $\hat{U}_j$ are time-dependent and given by
\begin{align}
\hat{U}_j(t) = \int_0^\infty v_j (\tau)e^{-iH(t) \tau} \hat{S}_je^{iH(t) \tau} d\tau,
\end{align}
with $\nu(\tau) = \langle B(\tau)B(0)\rangle$ at $T = 0$.
When the Born-Markov approximation is justified, the two approaches yield the same master equation. 
\paragraph{Counter-rotating terms in the system-bath Hamiltonian} In the approach of Beaudoin et al.  \cite{Beaudoin:2011}, a rotating-wave approximation is still carried out in the system-bath Hamiltonian. Bamba et al \cite{Bamba:2012} performed a detailed study of corrections to the master equation when no RWA is performed at that stage. These authors focused on a system where the matter part is composed of (quasi-bosonic) excitons. As a result, the system Hamiltonian is bosonic and quadratic, which from a methodological point of view offers additional possibilities. In particular, by performing an exact diagonalization of the full system-bath Hamiltonian, they show that the reduced density matrix of the bath in the true ground state is not a vacuum state. As a result, considering the bath to be in a vacuum state when deriving the master equation induce a non-physical excitation of the system resulting in a non-vanishing polariton population (in the absence of driving). The solution to this paradox is to estimate the correlations induced in the reservoir by its coupling to the system when the latter is in its ground state and the former in the vacuum state. A consistent master equation is then obtained when injecting the corrected bath correlations functions in the microscopic derivation. 
Corrections to the dissipative dynamics beyond the Born-Markov approximation due to counter-rotating terms in the system-bath Hamiltonian have also been recently studied for spin-boson models by means of a dynamical polaron ansatz~\cite{Zueco:2019} (see Sec.\ref{sec:scattering}).
Note that in all this section we have considered the correction to the standard master equation of a given system-bath Hamiltonian. A discussion on meaningful ways of deriving such Hamiltonians is presented in Ref.~\cite{Bamba:2014}.

%
\subsection{Input-output theory}

While the master equation describes the dynamics of the internal degrees of freedom of the system, the input-output theory relates these internal variables to external ones at the origin of input and output signals. Another aspect of the input-output formalism is that it allows to formulate the dynamics of the system in term of Langevin equations. This approach is particularly fruitful when dealing with quadratic bosonic Hamiltonians, in which case the associated Langevin equations are linear \cite{Ciuti:2006}.
As the input-output theory gives access to quantities such as transmission or fluorescence spectra and correlation functions, it is well suited to tackle questions related to the experimental signatures of the USC. For example, in the case of a two-dimensional electron gas in multiple-quantum-well structures, a signature of the USC was identified in the form of an asymmetric anticrossing of polariton modes visible in the optical spectra~\cite{Todorov:2010}. Within this theoretical framework, several works have studied the output photon statistics for systems described by the Rabi model and its generalizations. In particular essential modification of the phenomenology of the photon blockade effect have been reported in the USC regime of single~\cite{Ridolfo:2012, LeBoite:2016} and two-photon models~\cite{Felicetti:2018}

\paragraph{Input-output relation for nonlinear systems} The general setting is the same as that of Eq.~(\ref{HamBath}). In this section we explicitly write the interaction part of the system-bath Hamiltonian in the following form :
\begin{equation}
H_{\mathrm{I}} = \hat{X} \otimes \sum_k \gamma_k i(\crea{a}_k - \ann{a}_k). 
\end{equation}
where $\hat{X}$ is a generic Hermitian operator acting on the Hilbert space of the system and   the operators $\ann{a}_k$ are bosonic annihilation operators defining the bath modes. The bath Hamiltonian is given by$H_{\mathrm{B}} = \sum_k \omega_k \crea{a}_k\ann{a}_k$. 
The input-output relation is derived from the Heisenberg equations of motion which for the bath modes read
\begin{equation}\label{HeisenBath}
\frac{d\ann{a}_k}{dt} = -i\omega_k \ann{a}_k+ \hat{X}(t),
\end{equation}
which yields
\begin{equation}
\ann{a}_k(t) = e^{-i\omega_k(t-t_0)}\ann{a}_k(t_0) + \gamma_k e^{-i\omega_k t}\int_{t_0}^t e^{i\omega_k \tau}\hat{X}(\tau) d\tau,
\end{equation}
for an arbitrary initial time $t_0$. With the definition of the input and output field as  $\ann{a}_{\mathrm{out}(\mathrm{in})}(t) = \lim_{t_0\to \pm\infty} \frac{1}{\sqrt{2\pi}}\int d\omega e^{i\omega(t-t_0)}\ann{a}(t_0)$ and combining the above equation for $t_0 \to + \infty$ and $t_0 \to -\infty$ we obtain
\begin{equation}
\ann{a}_{\mathrm{out}}(t) - \ann{a}_{\mathrm{in}}(t) = \sum_k \gamma_k e^{-i\omega_k t} \hat{X}(\omega_k),
\end{equation}
where $\hat{X}(\omega_k)$ denotes the Fourier transform of $\hat{X}(t)$. In the continuous limit, where $\sum_{k}\gamma_k \to \int_0^{+\infty}d\omega \gamma(\omega)$, the r.h.s. becomes the inverse Fourier transform of the quantity $\gamma(\omega)\Theta(\omega)\hat{X}(\omega)$, where $\Theta(\omega)$ is the Heaviside step function.  Note that the operator $\Theta(\omega)\hat{X}(\omega)$ is the Fourier transform of the positive-frequency part of $\hat{X}$, which in principle should be defined with respect to the full system-bath Hamiltonian. However, when the coupling to the bath is weak, one can define to a good approximation, the positive-frequency part from the eigenstates of $H_{\mathrm{S}}$ alone. Hence in the frequency domain, the general input-output relation for a weak system-bath coupling reads
\begin{equation}\label{genIO}
\ann{a}_{\mathrm{out}}(\omega) - \ann{a}_{\mathrm{in}}(\omega) = \sqrt{2\pi}\gamma(\omega)\hat{X}^+(\omega),
\end{equation}
with
\begin{equation}
\hat{X}^+ = \sum_{\omega_i<\omega_j}X_{ij}\ket{i}\bra{j},
\end{equation}
and $X_{ij} = \langle i|\hat{X}|j\rangle$. It is clear from the above expressions that we recover the key elements specific to the USC regime that appeared in the derivation of the master equation:  the ``white noise'' assumption $\gamma(\omega) \approx \gamma $ is not legitimate in the general case~\cite{Ciuti:2006}. In addition, $\hat{X}^+$ differs from $\ann{a}$ when the cavity mode is ultrastrongly coupled to the quantum emitter~\cite{Ridolfo:2012}. 

The input-output relation in Eq.~(\ref{genIO}) allows to compute various correlation functions of the output field by combining Eq.~(\ref{genIO}) with the master equation approach. Indeed, the key quantities are now correlation functions of the field $\hat{X}^+(t)$, which may in turn be calculated via the quantum regression theorem. However, such a scheme requires to find an explicit time-domain expression for Eq.~(\ref{genIO}), which depends on microscopic details of the model, such as the bath spectral density $\gamma(\omega)$. For specific setups in which the frequency dependence of $\gamma(\omega)$ can be neglected even in the USC, this expression takes the simple form $\ann{a}_{\mathrm{out}} - \ann{a}_{\mathrm{in}} \propto \hat{X}^+$~\cite{Garziano:2013, DiStefano:2017}. Note that alternative definitions for the input and output fields may be considered, depending on the actual measurement scheme under consideration. For example, considering a circuit QED model in which the resonator is coupled to a waveguide, Ridolfo et al. \cite{Ridolfo:2012} defined the output field as $\lim_{t_0\to +\infty} \int d\omega \sqrt{\omega}e^{i\omega(t-t_0)}\ann{a}(t_0)$, whose correlation functions are directly proportional to photodection signals from the electric field. For this setup, a relevant choice for the spectral density is $\gamma(\omega)~\propto \sqrt{\omega}$~\cite{Ridolfo:2012, Lalumiere:2013}, leading to an input-output relation of the form $\ann{a}_{\mathrm{out}} - \ann{a}_{\mathrm{in}} \propto \frac{d}{dt}\hat{X}^+$. In the bosonic model considered in \cite{Bamba:2012} to describe intersubband polaritons,  the output field is defined as the field operator that couple in to the upper and lower polariton branches. 
\paragraph{Langevin equations}
When the system Hamiltonian is quadratic, the coupled Heisenberg equations of motion defining the exact system-bath dynamics can be cast into analytically solvable Langevin equations. This approach was applied to a bosonic model such as the one of Eq.~(\ref{HamBosonic}) \cite{Ciuti:2006, Huppert:2016}. To the equation on the bath variables, that defined the input-output relation, must be added the equation on the internal degrees of freedom (see Fig.~\ref{fig:dissipation}). The underlying physical system is often a planar structure, where quasiparticles are labeled by their in-plane momentum $\mathbf{k}$. As we only sketch the method here, we omit this label here. Denoting by $\ann{a}$ the cavity modes, the general form of the equations for the photonic field is 
\begin{equation}\label{Langevin}
\frac{da}{dt} = -\frac{i}{\hbar}[a, H_{sys}] - \int_{-\infty}^{\infty}dt' \Gamma(t-t')a(t') + F(t),
\end{equation}
where the memory kernel $\Gamma$ involves the spectrum of the photonic bath and is responsible for the complex energy shifts of the photonic mode. The force $F$ can be written as a function of the input or output field depending of what initial time is chosen. A similar equation is derived for the matter degrees of freedom.
Due to the linearity of the system, one can obtain a algebraic expression relating input and output fields in the frequency domain. The final results take the general form.
\begin{equation}
\begin{pmatrix}
\alpha^{\mathrm{out}}(\omega)\\
\beta^{\mathrm{out}}(\omega)
\end{pmatrix}
= \mathcal{U}(\omega) 
\begin{pmatrix}
\alpha^{\mathrm{in}}(\omega)\\
\beta^{\mathrm{in}}(\omega)
\end{pmatrix}, 
\end{equation}
where this relation combines the input-output relation, as given e.g. by Eq.~(\ref{genIO}) and the algebraic representation of the Langevin equations of Eq.~(\ref{Langevin}) expressed in Fourier space. 
Assuming that the latter is written formally as
\begin{equation}
\mathcal{M}(\omega) 
\begin{pmatrix}
\ann{a}(\omega)\\
\ann{b}(\omega)\\
\crea{a}(-\omega)\\
\crea{b}(-\omega)
\end{pmatrix}
+i
\begin{pmatrix}
F_{c}(\omega)\\
F_{e}(\omega)\\
F_c^{\dagger}(-\omega)\\
F_e^{\dagger}(-\omega)
\end{pmatrix}.
\end{equation}
with the Fourier transform of the Langevin forces $F(\omega)$ directly proportional to the input fields, the key quantity entering in the expression for $\mathcal{U}(\omega)$ is the Green function $\mathcal{G}(\omega) = -i\mathcal{M}^{-1}(\omega)$. In the matrix $\mathcal{M}(\omega)$ appear the complex frequency dependent damping rates $\Gamma(\omega)$ that give rise to damping terms and Lamb shifts, as in the derivation of the master equation.
    
\paragraph{Photodection}
The considerations underlying the derivation of the input-output relations are also relevant for the theory of photodetection, whose aim is to  determine the relevant observable of the output field that one need to compute to reproduce photodetection signals. The basic principle of Glauber's original theory~\cite{Glauber:1963} are also valid in the USC regime. However, as pointed out by Di Stefano et al \cite{DiStefano:2018b}, its application requires the same kind of adjustments that lead to Eq.~(\ref{genIO}). In this spirit,  the theory can be established for a device coupled to a generic light-matter system. Hence, the operator involved in the system-detector coupling Hamiltonian is not limited to the electric field. In return, the frequency dependence of the system-detector coupling coefficient has to be taken into account. The general setting is formally very similar to was was presented above. In particular,  the system-detector coupling Hamiltonian may be written as $ H_{sd} = \sum g_n (\crea{c}_n + \ann{c}_n)\otimes \hat{X }$, where the operators $\ann{c}_n$ are annihilation operators for the $n^{th}$ mode of the detector and need not be bosonic. $\hat{X}$ is an operator acting only on the system. Within this framework, the equivalent of Glauber's formula, giving the expression of the detector probability of being excited, is obtained through the Fermi golden rule and reads
\begin{equation}
\frac{dW}{dt} = \langle \hat{O}^-\hat{O}^+\rangle,
\end{equation}
where the operator $\hat{O}^+$ is similar to the quantity that appear in the r.h.s. of Eq. (\ref{genIO}). It is such that its Fourier transform is given by
\begin{equation}
\hat{O}(\omega) = \sqrt{2\pi}g(\omega)\hat{X}^+,
\end{equation}
where the positive-frequency part is defined relative to the system Hamiltonian, assuming weak coupling to the detector. The notation $g(\omega)$ refers to the continuous limit of $g_n$.

\subsection{Driven systems }

A typical way to study, e.g., the output photon statistics of a cavity QED device is to couple it to an external coherent field. Such driving mechanism is accounted for in the theoretical description by adding a term proportional to $F\cos(\omega_dt + \phi)$ in the Hamiltonian of the system, where $\omega_d$ denotes the frequency of the field and $F$ its amplitude. In a regime where all counter-rotating terms can be safely neglected, the time-dependence induced by the driving term is subsequently removed by expressing all quantities in a frame rotating at the driving frequency $\omega_d$. In the ultrastrong coupling regime the Hamiltonian is still time-dependent in the rotating frame but other strategies, relying on Floquet theory, are available to handle the driving term. The use of Floquet theory in quantum mechanics is not restricted to open systems and was first employed to treat strong driving in closed systems~\cite{Shirley:1965, Grifoni:1998, Hausinger:2011}. We restrict ourselves here to its application to the master equation, although other approaches have also been proposed~\cite{Hausinger:2008, Restrepo:2016}.

\paragraph{Floquet-Liouville approach}

A rigorous and general derivation of the master equation for a time-dependent Hamiltonian is not a trivial task~\cite{Rivas:2010}. However when the driving is weak, one can assume that there is no ``dressing of the dressed-state'' by the external field and that the dissipator is left unchanged. A possible strategy, designated as the Floquet-Liouville approach~\cite{Ho:1986, Chu:2004}, is then to apply Floquet theory to the resulting time-periodic master equation, $\partial_t \rho = \mathscr{L}(t) \rho$, where $\mathscr{L}(t+T)=\mathscr{L}(t)$, with $T = 2\pi/\omega_d$. Here $\mathscr{L}(t)$ denotes the Liouvillian superoperator defined by Eq.~(\ref{microME}).
The Floquet theorem~\cite{Floquet:1883} states that there exist solutions of the master equation of the form
\begin{equation}\label{floquetSol}
\rho(t) = \sum_{\alpha}c_{\alpha}e^{\Omega_{\alpha}t}R_{\alpha}(t).
\end{equation}
Here, $R_{\alpha}(t)$ is a periodic function of period $T$ and $\Omega_{\alpha}$ is a complex number, which are eigenfunctions and eigenvalues, respectively, of the following operator
\begin{equation}\label{eqPeriodic_main}
(\mathscr{L}(t)-\partial_t)R_{\alpha}(t) = \Omega_{\alpha}R_{\alpha}(t).
\end{equation}
Practical implementations of the Floquet-Liouville approach amount to finding an algebraic representation of Eq.~(\ref{eqPeriodic_main}) that makes the problem time-independent. This is carried out by introducing the so-called Floquet Hilbert space $\mathcal{H}^2\otimes \mathcal{T}$, where $\mathcal{T}$ denotes the Hilbert space of $T$-periodic functions. A natural choice of basis for the space $\mathcal{T}$, is obviously the functions $\phi_n(t) = e^{-in\omega_dt}$. Following Refs.~\cite{Grifoni:1998,Hausinger:2010}, we denote $\phi_n$ by $|n)$. In this basis, the generic expression for an element $|A\rangle \rangle$ of $\mathcal{H}^2\otimes \mathcal{T}$, 
$|A\rangle \rangle =  \sum_{n= -\infty}^{+\infty} A^{(n)}\otimes|n)$ coincides with its Fourier series expansion $A(t) = \sum_{n= -\infty}^{+\infty} A^{(n)}e^{-in\omega_dt}$. Note that the scalar product on the Floquet space derives from the usual scalar product on $\mathcal{T}$, $(f|g) = \frac{1}{T}\int_{0}^T f^*(t)g(t)\mathrm{d}t$ and the scalar product on $\mathcal{H}^2$, $\langle A|B\rangle = \mathrm{Tr[A^{\dagger}B}]$. Within this framework, the quantities $R_\alpha(t)$ are represented as right-eigenvectors $|R_{\alpha}\rangle\rangle $ (corresponding to the eigenvalue $\Omega_\alpha$), of a non-Hermitian superoperator $\tilde{\mathscr{L}}$ in Floquet space. The matrix elements of this operator derive from the expression of Eq.~(\ref{eqPeriodic_main}) in Fourier space, which reads
\begin{equation}\label{floquetEig}
\sum_{m = -\infty}^{\infty} \mathscr{L}^{(n-m)}R^{(m)}_{\alpha,k} +in\omega_dR_{\alpha,k}^{(n)} = \Omega_{\alpha}R_{\alpha,k}^{(n)},
\end{equation}
Note that the range of the index $\alpha$ in Eq.~(\ref{floquetSol}) is equal to the dimension of the physical space of density matrices. However, given the dimension of the Floquet space it is necessary to label the eigenstates $|R_{\alpha,k}\rangle \rangle$ of $\tilde{\mathscr{L}}$ with an extra index $k\in\mathbb{Z}$. This apparent discrepancy reflects the fact that, similarly to Bloch functions in solid state physics, the matrices $R_{\alpha}(t)$ in Eq.~(\ref{floquetSol}) are not uniquely defined. Indeed, the equation is left invariant by the transformation $\{\Omega_{\alpha} \to \Omega_{\alpha} -ik\omega_d$, $R_{\alpha}(t) \to e^{ik\omega_d} R_{\alpha}(t)\}$. The full dynamics can be expressed as a function of eigenstates and eigenvectors of $\tilde{\mathscr{L}}$ leading to the Floquet space equivalent of Eq.~(\ref{floquetSol})
\begin{equation}
|\rho(t)\rangle\rangle = \sum_{\alpha,k}c_{\alpha,k}e^{\Omega_{\alpha,k}t}|R_{\alpha,k}\rangle \rangle,
\end{equation}
where $c_{\alpha,k} =  \langle\langle L_{\alpha,k}|\rho_0\rangle\rangle$, with $\langle\langle L_{\alpha,k}|$ the left-eigenvectors of $\tilde{\mathscr{L}}$. In this expression the periodic part of the time evolution is implicitly encoded in $|R_{\alpha,k}\rangle \rangle$ Note that for a given initial density matrix $\rho_0$, the choice of the $|\rho_0\rangle\rangle$ is not unique, but this arbitrariness has no influence on the dynamics. One possible choice is for example $|\rho_0\rangle\rangle =\rho_0\otimes|0)$. In addition, due to the degeneracy mentioned above, the sum over $k$ can always be suppressed and all quantities expressed as a functions of $\Omega_{\alpha,0}$ and $|R_{\alpha,0}\rangle\rangle$.
More generally the propagator for the master equation can be expressed as
\begin{align}
\rho(t+\tau) &= U(t+\tau,t)[\rho(t)] \nonumber \\
&= \sum_{\alpha,k} e^{-i\Omega_{\alpha,k}\tau}\langle\langle L_{\alpha,k}|\rho\rangle\rangle R_{\alpha,k}(t+\tau).
\end{align}
This algebraic formulation of the master equation in Floquet provides an efficient way of computing the dynamics for driven-dissipative systems with a small number of components, without numerically integrating a time-dependent master equation. In this respect, this approach was particularly useful to address the question of metastability in the driven-dissipative Rabi model~\cite{LeBoite:2017}. It also allowed to find semi-analytical expressions for the fluorescence spectrum of ultrastrongly coupled devices~\cite{Felicetti:2018}.
\paragraph{Floquet-Markov approach} 
A related application of Floquet theory to the master equation is the Floquet-Markov approach. 
 Originally conceived for quantum systems in strong driving fields~\cite{Blumel:1991, Breuer:1997, Breuer:2000} it consists in deriving the master equation directly in the Floquet basis associated with the periodic Hamiltonian. Let $|u_{\alpha}(t)\rangle$ be the Floquet eigenstate satisfying $|u_{\alpha}(t+T)\rangle = |u_{\alpha}(t)\rangle$ and
\begin{equation}
U_S(t,0)|u_{\alpha}(0)\rangle = e^{-i\epsilon_\alpha t}|u_{\alpha}(t)\rangle.
\end{equation}
In the Schr\"odinger picture, operators are defined in the basis $|u(0)_{\alpha}\rangle$ and the matrix elements of the density matrix as
\begin{equation}
\rho_{\alpha,\beta}(t) = \langle u_\alpha (t)| \rho(t)|u_\beta (t)\rangle.
\end{equation}
In the interaction picture relative to the Floquet basis, in which the master equation is derived, the matrix elements of an operator $A(t)$ are
\begin{align}
\langle u_\alpha(0) |A(t)|u_\beta(0)\rangle &= \langle u_\alpha(0)|U_S^\dagger(t,0) A U_s(t,0)|u_\beta(0)\rangle \nonumber\\
&= e^{i(\epsilon_\alpha - \epsilon_\beta)t} \langle u_\alpha(t) |A|u_\beta(t)\rangle \nonumber\\
& = \sum_{k = -\infty}^{+\infty}e^{-i(\epsilon_\beta - \epsilon_\alpha +k\omega_d)t}A_{\alpha\beta}^{(k)}.
\end{align}
Once a  meaningful interaction picture has been defined, a procedure similar to the one outlined at the beginning of this section applies. The relevant jump operators are now of the form
\begin{equation}
A(\omega) = \sum_{\epsilon_\beta - \epsilon_\alpha + k\omega_d = \omega} A^{(k)}_{\alpha \beta} |u_{\alpha}(0)\rangle \langle u_{\beta}(0)|,
\end{equation}
from which we recover  an equation that is formally equivalent to Eq.~(\ref{MEi}). Summing over indices $\alpha, \beta, k$ rather than frequencies $\omega$, the equation (without the secular approximation) reads~\cite{Blumel:1991}
\begin{align}
\frac{d\rho}{dt} = \sum_{\alpha,\beta,k, \alpha',\beta',k'}&\Big[ e^{i(\Omega_{\alpha'\beta'}(k') - \Omega_{\alpha\beta}(k))t}A_{\alpha \beta}^{(k)}A_{\alpha'\beta'}^{(k')*}\times\\
&\times\Gamma(\Omega_{\alpha,\beta}(k))[P_{\alpha\beta}, 
\rho P^{\dagger}_{\alpha'\beta'}]\Big] + \; \text{h.c.},
\end{align}
with $\Omega_{\alpha,\beta}(k) = \epsilon_\beta - \epsilon_\alpha + k\omega_d$ and $P_{\alpha\beta} =  |u_{\alpha}(0)\rangle \langle u_{\beta}(0)|$.
An autonomous equation in the interaction picture is obtained only when performing the secular approximation, i.e. the assumption $\omega \neq \omega' \implies |\omega-\omega'| \gg \gamma$. In this setting, the transitions frequencies not only involve the the quasienergies $\epsilon_{\alpha}$ but also all equivalent quasienergies obtain by adding a multiple of the frequency $\omega_d$. When the secular approximation understood in this way is valid, the final expression for the master equation (in the interaction picture) reads  
\begin{equation}
\frac{d\rho}{dt} = \sum_{\alpha,\beta,k}|A_{\alpha \beta}^{(k)}|^2\Gamma(\Omega_{\alpha,\beta}(k))[ P_{\alpha\beta}, \rho P^{\dagger}_{\alpha\beta}] + \quad \text{h.c.}
\end{equation}
%

\section{Waveguide QED}
\label{sec:scattering}

The methods presented in the previous sections where first and foremost tailored to solve cavity QED problems where a finite (and usually small) number of emitters interact ultrastongly with a finite number of cavity modes. As the result, even when the coupling to a continuum of modes was considered in Sec. \ref{sec:open} to model the environment, the system-bath coupling was considered small enough, so that the bath degrees of freedom could be effectively traced out. The possibility of reaching the USC regime in waveguide QED, where an atom is coupled to the continuum of electromagnetic modes propagating in a 1D waveguide, raises a new set of theoretical issues. While the paradigmatic model of Sections \ref{sec:spectrum} and \ref{sec:open} was the quantum Rabi model (Eq. (\ref{HamQRM})), this section focuses on the spin-boson model, presented in Eq.~(\ref{HamSBM}). 

\subsection{Ultrastrong coupling to a continuum}
In the context of waveguide QED, a typical situation is to consider a single quantum emitter of frequency $\Omega$ emitting light into the waveguide with a rate $\Gamma$. The system enters the strong coupling regime when the the emission rate $\Gamma$ becomes larger than the decoherence rate or any other dissipation rate into other channels, while  the relation $\Gamma \ll \Omega$ still holds. In analogy with cavity QED, the USC regime is reached when $\Gamma$ becomes a significant fraction of $\Omega$. Such a regime is achievable with superconducting architectures involving superconducting qubits coupled to a 1D transmission line~\cite{Bourassa:2009,Forn-Diaz:2017}.
In this settings, an experimental signature of the different coupling regime may be obtained by measuring the transmitted coherent scattering. For example, in an open setting involving a single atom and single-photon pulses, a hallmark of strong interaction between the atom and propagating photons is the extinction of the transmitted light, resulting from an interference process between the incoming light and the light emitted collinearly by the atom~\cite{Lalumiere:2013, Astafiev:2010}.
As in driven-disspative cavity QED scenarios, non-poissonian statistics of the transmitted and reflected field are also a manifestation of strong-light matter interaction~\cite{Hoi:2012, Hoi:2013, Pletyukhov:2015}. 
 From a theoretical perspective, the most important underlying model is the spin-boson Hamiltonian~\cite{Leggett:1987}. It is indeed the simplest model describing the coupling of an atom, assumed for simplicity to be a TLS, to a continuum of bosonic modes modelling the waveguide. The model, which is related to quantum impurity problem~\cite{Shi:2009, LeHur:2012} has numerous application outside of quantum optics and only a small fractions of the theoretical literature is touched upon here. 

Within this framework, several strategies are available to compute the output of scattering experiments. We note first, that in the strong-coupling regime, when the coupling to the waveguide modes is still much weaker than the atom frequency, a microscopic master equation approach is legitimate.  Scattering amplitudes and correlation functions of the output field are linked to the density matrix of the atom via standard input-output relations~\cite{Peropadre:2013a}. In the USC regime, it was shown that numerical schemes based on Matrix Product States~\cite{Orus:2014} could successfully be adapted to scattering problems~\cite{Peropadre:2013b, Sanchez-Burillo:2014}. They have in particular been used to benchmark two of the methods that are presented below: the dynamical polaron ansatz~\cite{Diaz-Camacho:2016} and the extension to the USC of field-theoretic scattering theory~\cite{Shi:2018}.

\subsection{Dynamical Ans\"atze}\label{DynPolaron}
The polaron states such as the one introduced in Eq.~(\ref{polaronWF}) proved also useful to tackle scattering problems, and more generally dynamical quantities relevant to waveguide QED setups.  In particular Diaz-Camacho et al.~\cite{Diaz-Camacho:2016} have developed a variational semi-analytical approach to the dynamics of the spin-boson model based on a dynamical polaron ansatz.
Note that this general framework is well suited to multi-spin configurations. The first step of the method is to find an optimized static polaron transformation that minimize the ground-state energy. 
The rationale behind  and the way it is implemented is similar to what was done for scattering theory: the static polaron transformation allows to disentangle spins and boson. More precisely, the general static ansatz is expressed through the polaron transformation
\begin{equation}
U_P[f_{ik}] = \bigotimes_{i,k}e^{\sigma_i^x(f^*_{ik}\crea{a}_k-f_{ik}\ann{a}_k)},
\end{equation}
where the $f_{ik}$ are the variational parameters of the transformation. For a multi-spin system there are additional variable parametrizing the ground state, namely the spin degrees of freedom defining the spin state,
\begin{equation}
|\psi_s[c_\sigma] \rangle  = \sum_{\sigma \in \{\uparrow, \downarrow \}^{N_s}} c_{\sigma}\bigotimes|\sigma_i\rangle.
\end{equation}
The optimal polaron state is therefore the state $|\Psi^P\rangle = U_P^\dagger[f_ik]|0\rangle|\psi_s[c_\sigma] \rangle$, for the values of $f_{ik}$ and $\sigma$ that minimize the energy.
Once the optimized polaron transformation is found, it defines a new basis, the polaron picture in which to express the dynamics of the system. The time evolution is then handled within a subspace with a defined number of excitation.  For example, in the one-excitation subspace the general state is parametrized as
\begin{equation}
U_P[f_{ik}]W[\alpha_s(t),\alpha_k(t)]\ket{0}\otimes\ket{\psi_{g.s.}},
\end{equation}
where
\begin{equation}
W[\alpha_s(t),\alpha_k(t)] = \sum_{s=1}^{N_s}\alpha_s(t)\ket{0}\otimes\ket{\psi^e_s}\bra{\psi_{g.s.}} + \sum_k \alpha_k(t)\crea{a}_k.
\end{equation}
The equations of motion for the coefficients $\alpha_{k,s}(t)$ that defines the dynamics take the form of Euler-Lagrange equations.The corresponding Lagrangian is derived from the energy functional for $\alpha_{k,s}$ associated with the Hamiltonian.

Starting similarly from a optimized static polaron wavefunction for the ground-state of the system, Gheeraert \textit{et al.}~\cite{Gheeraert:2017, Gheeraert:2018} have proposed an alternative dynamical ansatz, referred to as the``Multimode Coherent States ansatz''. Considering general superpositions of multimode coherent states they derived a formalism that proved efficient in predicting phenomena intrinsic to the USC regime such as frequency conversion processes arising in off-resonant inelastic scattering~\cite{Gheeraert:2018}.  

\subsection{Scattering theory}
Polaron transformations have also found application in the extension to the USC regime of scattering theory~\cite{Shi:2018}. The possibility of applying to quantum optics the theoretical apparatus developed in the context of high-energy physics to compute the $S$ matrix is not restricted to the USC regime. The formalism is indeed general and aims at extracting transmission rates and correlation between outgoing photons from the quantity~\cite{Shi:2009, Pletyukhov:2012}
\begin{equation}
S = T \mathrm{exp}[-i\int_{-\infty}^{+\infty} H_{\mathrm{int}}(t)dt],
\end{equation}
where $T$ denotes the time-ordering operation and $H_{\mathrm{int}}(t)$ is the atom-waveguide interaction Hamiltonian expressed in the interaction picture. from the $S$-matrix are extracted the scattering amplitudes
\begin{equation}
_{\mathrm{out}}\langle f|i\rangle_{\mathrm{in}} = _{\mathrm{in}}\langle f|S|i\rangle_{\mathrm{in}},
\end{equation}
where the input and output states are asymptotically free multiphoton states.
Several approaches are available to compute the $S$-matrix both for single and multi-photon scattering. Early results were obtained through integrability-based methods~\cite{Yudson:1985, Yudson:2008}, other approaches rely on the Lippmann-Schwinger formalism~\cite{Shen:2007a, Shen:2007b, Zheng:2010}. Path integral methods have also been developed to treat photonic scattering problems~\cite{Shi:2009}. They exploit the Lehmann-Symanzik-Zimmermann (LSZ) reduction, which relates the connected $T$-matrix to the photonic Green function. The Green function itself is then computed via a path-integral representation of its generating functional.

Shi et al. succeeded in extending the range of application of these computational techniques to the ultrastrong coupling by finding an effective low-energy particle-conserving Hamiltonian. In the spirit of the GRWA, they first applied an optimized polaron transformation to find the effective ground state, before applying the rotating-wave approximation. The model gave good results in the single photon regime and captures the renormalization of the spin frequency and strong Lamb shift characteristic of this regime. As the validity of the GRWA is not easy to prove in this case, the robustness of the approximation is established by comparing the results with MPS numerical simulations.
%


\section{Validity of effective models}
\label{sec:limitations} 

\begin{figure*}[]
\begin{center}
\includegraphics[width=1.8\columnwidth]{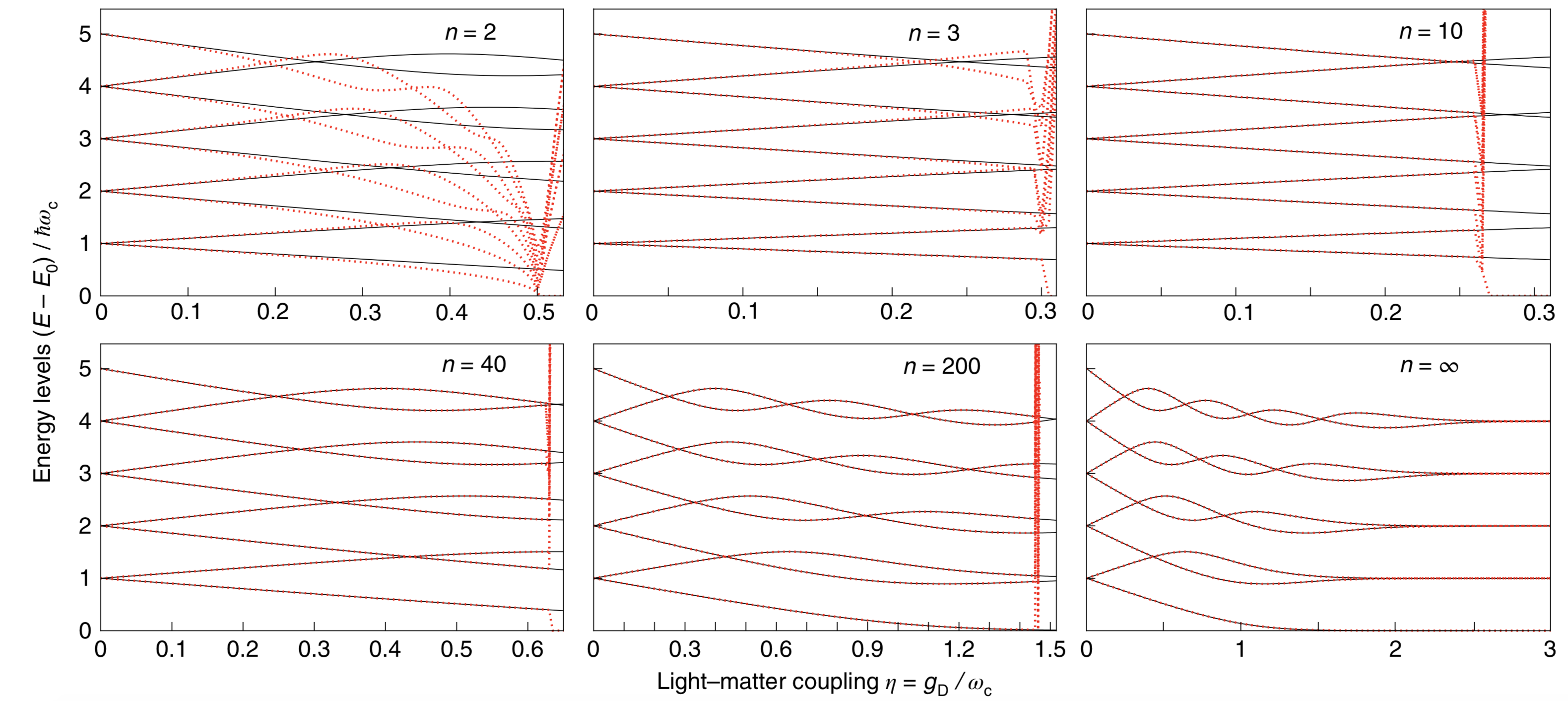}
\caption{\label{fig:gauge}Monitoring the break-down of gauge invariance. The energy spectra of the quantum Rabi Hamiltonian $H_\mathrm{C}$ in the Coulomb gauge (red dotted curves), derived from Eq.~(\ref{HamCoulomb})  is shown for various order of the Taylor expansion. Each panel also shows the exact spectra (black continuous curves). Reproduced with permission~\cite{DiStefano:2019}. Copyright 2019, Springer Nature.
}
\end{center}
\end{figure*}

In this last section we present some of the recent debates regarding fundamental limitations of effective models in the ultrastrong-coupling regime. While these models proved to be successful in predicting experimental results for currently achievable coupling strength, the prospect of reaching larger values of the interaction strength, where $g \gg \omega$, lead to question some of the approximations they are inevitably based on~\cite{Manucharyan:2017}. It was shown for example that the usually neglected diagmagnetic $A^2$ term can act as a potential barrier and lead to a decoupling of light and matter in the USC regime~\cite{DeLiberato:2014,Garcia-Ripoll:2015}. The role of the diamagnetic term has also been the focus of vivid debates in the context of the Dicke superradiant phase transition~\cite{Nataf:2010b, Viehmann:2011, Ciuti:2012, Lambert:2016, Jaako:2016, Bamba:2017, Kirton:2019, Andolina:2019}. We focus more specifically in the following on recent developments regarding the related question of gauge invariance.

Attempts in deriving a microscopic model for cavity QED setups share a common background: the theory is based on a non-relativistic formulation of quantum electrodynamics within the long-wavelength approximation. At the classical level, this formulation is conveniently expressed in the Coulomb gauge~\cite{Cohen-Tannoudji:1997}, from which quantization of the theory follows the canonical procedure. A general expression for the classical Hamiltonian of a system of charges interacting with the electromagnetic field is the following~\cite{Vukics:2014}
\begin{equation}\label{ClassHam1}
H = \sum_{\alpha} \frac{(\mathbf{p}_\alpha - q_{\alpha}\mathbf{A})^2}{2m_{\alpha}} + \int_{\mathcal{D}}dr^3(\nabla U)^2 + H_{\mathrm{field}},
\end{equation}
where $H_{\mathrm{field}}$ may be expressed as
\begin{equation}\label{ClassHam2}
\frac{\epsilon_0}{2} \int_{\mathcal{D}}dr^2\left[\left(\frac{\mathbf{\Pi}}{\epsilon_0}\right)^2 + c^2(\nabla\times \mathbf{A})^2\right],
\end{equation}
where $U$ is the scalar potential and $\mathbf{\Pi}$ the canonical conjugate momentum to $\mathbf{A}$.  In the context of cavity QED, the domain $\mathcal{D}$ in which the field lives is not the free space. Therefore, as pointed out by Vukics et al. \cite{Vukics:2014}, boundary conditions on the fields such as $U|_{\partial D} = 0$ and $\mathbf{A}\times \mathbf{n} = 0|_{\partial D}$, must be added to the Coulomb gauge condition $\nabla\cdot\mathbf{A} = 0$, in order to completely remove gauge ambiguities.
In view of a non-relativistic quantum treatment of the problem the Hamiltonian is further simplified by the long-wavelenth approximation. The charges are assumes to form well localized clusters of small radius, such that the position dependence of the field $\mathbf{A}(\mathbf{r})$ can be neglected at this scale. A difficulty inherent to such formulation of cavity QED problems, is that the notion of the charge clusters forming (natural or artificial) atoms are not gauge invariant concepts. Moreover, the effective models presented in the previous sections also rely on additional simplifications such as the two-level approximation for the atom or a single-mode description of the electromagnetic field. If not applied carefully, these approximations break the gauge invariance and may result in unphysical predictions. In the following we split the discussion into two parts, presenting first the debates focusing on the two-level approximations in various gauges. In a second part we review microscopic models that were build to describe many-particle systems, with emphasis on the existence of superradiant transitions and the role of the $A^2$ term.

\subsection{Gauge non-invariance of the two-level approximation}  

Several recent works~\cite{DeBernardis:2018b, Stokes:2019a, DiStefano:2019, Stokes:2019b, Garziano:2020} have identified important issues arising in the USC when performing the two-level approximation in different gauges.
A systematic study of the two-level approximation in the dipole and Coulomb gauge was performed by De Bernardis et al.~\cite{DeBernardis:2018b}. It is shown that already at the level of a single electric dipole coupled to a single cavity mode, serious discrepancies appear in two-level models resulting from different choice of gauges. It is exemplified in the computation of the matrix elements of the interaction Hamiltonian in the eigenbasis of the atom: while truncation of the particle Hilbert space gives consistent results for the terms $i\hat{x}(\hat{a} - \hat{a}^\dagger)$ present in the electric dipole gauge, the same truncation scheme is not justified for the operator $\hat{p}(\hat{a}+\hat{a}^{\dagger})$ stemming from the Coulomb gauge, The quality of the approximation depend also on the type of confining potential that is considered. Interestingly, comparison with exact diagonalization in the Coulomb gauge shows that the Rabi model is robust in the ultrastrong coupling regime when the charge confining potential is a double well, when derived in the electric dipole gauge. Conversely for a square-shape confining potential no Rabi Hamiltonian reproduces the exact result.
Stokes et al.~\cite{Stokes:2019a} tackle the validity of the two-level approximation in a more general setting by considering a family of gauge transformations parametrized by a real parameter $0 \leq \alpha \leq 1$. The relation between gauge-invariant variable and gauge-dependent canonical conjugate variables are given by
\begin{align}
m\dot{\mathbf{r}} &= \mathbf{p}_\alpha - q(1-\alpha)\mathbf{A}\\
\mathbf{E}_{T} &= -\Pi_{\alpha} - \alpha\frac{\mathbf{\epsilon}(\mathbf{d}\cdot \mathbf{\epsilon})}{V}M,   
\end{align}
where $\mathbf{A}$ is the transverse vector potential, $\mathbf{E}_T$ the transverse electric field, $\epsilon$ and $V$ are the cavity polarization vector and volume and $\mathbf{d}$ the matter dipole moment. The electric dipole and Coulomb gauges are recovered for  $\alpha = 1$ and $\alpha = 0$ respectively. The unitary gauge transformation going within this one-parameter family is $R_{\alpha\alpha'} = e^{i(\alpha - \alpha')\mathbf{d}\cdot \mathbf{A}}$. 
They show that, for each value $\alpha$, the TLA results in a generalized Rabi Hamiltonian with potentially asymmetric rotating and counter-rotating terms. Given the importance of counter-rotating terms in the phenomenology of the USC reigme, such arbitrariness may seem paradoxal. In particular it implies that there exists a value of $\alpha$ for with the TLA yields a Jaynes-Cumming Hamiltonian, whose validity does not rely on the RWA. Numerical simulations regarding the first two-level and eigenstates and energy show that the TLA in the Jaynes-Cummings gauge may give more accurate results than the other two gauges considered.  

Following these studies, a prescription for recovering consistent results in systems involving a charged particule in a confining potential interacting with a single mode of the electromagnetic field, was proposed by Di Stefano et al.~\cite{DiStefano:2019}. As in previous studies, the long-wavelength approximation is assumed to be valid and the electromagnetic field is uniform in space. The reason for the failure of the TLA in the Coulomb gauge presented above was identified as related to the non-local character of the potential after truncation of the Hilbert space. 
Hence, starting from the Rabi Hamiltonian in the dipole gauge
\begin{equation}
H_D = \omega \crea{a}\ann{a} + \frac{\Omega}{2}\hat{\sigma}_z + ig\hat{\sigma}_x(\crea{a}  - \ann{a}),
\end{equation}
which has proved to be in good agreement with the exact one, the change of gauge and subsequent truncation of the Hilbert space can be written as a true unitary transformation. The correct Rabi Hamiltonian in the Coulomb gauge preserving gauge invariance is
\begin{equation}\label{HamCoulomb}
H_C = \omega \crea{a}\ann{a} + \frac{\Omega}{2}(\hat{\sigma}_z\cos[\frac{2g}{\omega_c}(\crea	{a} + \ann{a})] +\hat{\sigma}_y\sin[\frac{2g}{\omega_c}(\crea{a} + \ann{a})]). 
\end{equation}
Interestingly, by expanding the cosine and sine functions in the above expression, one can monitor the ``break-down'' of gauge invariance with increasing coupling strengths (see Fig.~\ref{fig:gauge}).
Note that in a general setting the approach outlined above involve computing the image of some arbitrary operators (functions of $\hat{x}$ and $\hat{p}$) under a given gauge transfomation. It can be performed through the Baker-Campbell-Hausdorff formula.

\subsection{Multi-particle configurations and diamagnetic $A^2$ term}

The debate over the validity of effective models was particularly intense in the context of multi-dipole models of cavity QED. As mentioned above, one of main questions is the role played by the $A^2$ term in the Dicke model and its consequence on the existence of the superradiant phase transition.

More generally, a crucial question in evaluating the validity of effective models is to provide realistic bounds on the possible coupling strength appearing in the model.
 In order to answer to such fundamental questions from first principles, several minimal models for cavity and circuit QED have been recently proposed~\cite{Todorov:2012, Todorov:2014a, Todorov:2014b,Vukics:2015, DeBernardis:2018a}. A microscopic theoretical description based on the Power-Zineau-Woolley transfomation was first developed for cavity QED systems involving 2D~\cite{Todorov:2012} and 3D~\cite{Todorov:2014a} electron gases in solid-state structures. While first designed to describe the regime of high electronic density, this framework was extended to the few-electron regime by considering the interaction of the electron gas with a quantum $LC$ resonator~\cite{Todorov:2014b}. Interestingly, changing the number of electrons in such a model allows to continuously interpolate between effective Rabi and Hopfield Hamiltonians. 
 
 In the context of atomic  QED, starting from the classical Hamitonian Eqs.~(\ref{ClassHam1}), (\ref{ClassHam2}), it was shown that through a proper generalization of the Power-Zineau-Woolley transformation, the Hamiltonian in the multipolar gauge could in be mapped to the Dicke Hamiltonian~\cite{Vukics:2014}. The strategy was to eliminate the $A^2$ term and the inter-atomic dipole interaction exactly, by a proper gauge choice at the classical level, taking into account the specific geometry of cavity QED setups with respect to free space. The obtained general expression for the Hamiltonian in the multipolar gauge reads
\begin{equation}\label{classHamDip}
H' =  \sum_{\alpha}\frac{\mathbf{p}_\alpha^2}{2m_{\alpha}} + H_{\mathrm{field}} + \frac{1}{2\epsilon_0}\int_{\mathcal D}d^3r \mathbf{P^2} -\frac{1}{\epsilon_0}d^3\mathbf{D}\cdot \mathbf{P},
\end{equation}
where $\mathbf{P}$ is the polarization density and $\mathbf{D} = \epsilon_0\mathbf{E} + \mathbf{P}$.
The elimination of dipole-dipole interaction is meaningful only within the long wavelength approximation and the assumptions of well separated atoms. In this case, the kinetic and $\mathbf{P}^2$ terms in Eq.~(\ref{classHamDip}) define the internal structure of the atoms. Within this framework a correspondence with the Dicke model is established through canonical quantization of the resulting Hamiltonian. However finding realistic bounds on the interaction strength requires to find an explicit formula for the atomic polarization field. General estimates for atomic cavity QED showed that the values of the interaction needed to observe critical phenomena in the USC regime come too close to the limit of validity of the independent dipole approximation to provide a definitive answer~\cite{Vukics:2015}.  

Another model for which precise statements could be made was put forward in the context of quantum circuits~\cite{DeBernardis:2018a}. The setup considered is composed of $N$ electric dipoles interacting with the electromagnetic modes of a lumped-element $LC$ resonator. The dipoles are described as effective particles of mass $m$ in a confining potential $V$. For this model the quantization procedure is carried out from the Lagrangian of the circuit~\cite{DeBernardis:2018b}
\begin{align}\label{Lag}
\mathcal{L} = &C\frac{\dot\Phi^2}{2} -\frac{\Phi^2}{2L} + \dot{\Phi}Q_{\mathrm{in}}\nonumber\\ 
&+\sum_{i}[\frac{m}{2}\dot\xi_i^2 - V(\xi_i)]-\frac{m\omega_p^2}{2}\sum_ {i\neq j}\mathcal{D}_{ij}\xi_i\xi_j,
\end{align}
where $\Phi$ is the  is the magnetic flux through the inductor of the $LC$ circuit, $\xi_i$ the displacements between the dipole charges and $Q_{\mathrm{in}}$ the charge induced by the dipole distribution for zero voltage drop accross the capacitor.The dipole-dipole interaction, in particular its geometric aspects are parametrized by the quantity $\mathcal{D}_{ij}$.
The analysis of coupling strength at play in this model reveals that the dipole-dipole direct interaction and the dipole-field coupling cannot be treated independently. Hence the effective model obtained after performing the two-level approximation is an extended-Dicke model, which includes spin-spin interactions. Within this theoretical framework a rich phase diagram is predicted including superradiant and subradiant phases with antiferromagnetic order of the dipoles. 

\section{Conclusion}
\label{sec:conclusion}

We have reviewed in this Progress Report the recent theoretical advances in our understanding of ultrastrong light-matter interactions. The counterintuitive phenomenology offered by this new regime of cavity QED, has led to fruitful developments in many aspects of the theory. Approximation strategies and variational schemes have been developed to compute the effect of counter-rotating terms on spectral properties. In this context, polaron transformations proved to be valuable tools go get physical insight into the eigenstates of systems described by Rabi and spin-boson models. Elegant exact mathematical results on the energy spectrum have also been obtained. In driven-dissipative settings, a consequence of the USC is that the frequency dependence of the noise spectrum cannot be neglected. This may affect drastically the outcome of photodetection signals and correlation measurements. Besides, any output fields can be computed in a meaningful way only with respect to the dressed-basis of the full light-matter system. The counter-rotating terms also impact the treatment of external driving fields, as it is no longer possible to eliminate the time-dependency of the Hamiltonian through a simple change of reference frame. In this context, the Floquet theorem applied to the master equation in the form of the Floquet-Liouville or Floquet-Markov approaches give tools to treat the time-dependency exactly. In the field of waveguide QED, the possibility of reaching the USC regime has led to an extension of scattering theory and to the developments of new numerical tools such as dynamical polaron and multimode coherent states ans\"atze or specifically tailored MPS-based simulations. At a more fundamental level, the prospect of reaching extreme values of the interaction between light and matter has deepened our understanding of the validity effective models for cavity QED.
Inspired by the paradigmatic setting of cavity QED including only a single atom and a single cavity mode, many of the methods presented in this article were primarily designed for systems with only a small number of particles. Although collective effects such as superradiant phase transitions have attracted a great deal of interest in the last decades, the interplay between many-body effects and ultrastrong-coupling phenomenology still offers numerous perspectives~\cite{Schiro:2012, Garbe:2017, Cui:2019}. In this respect, the recent progress in the the field of strongly-correlated photonic phases~\cite{Carusotto:2013, Schmidt:2013, Noh:2016}, will play a significant role.

\bibliography{biblio_revUSCresub}

\begin{thebibliography}{182}%
\makeatletter
\providecommand \@ifxundefined [1]{%
 \@ifx{#1\undefined}
}%
\providecommand \@ifnum [1]{%
 \ifnum #1\expandafter \@firstoftwo
 \else \expandafter \@secondoftwo
 \fi
}%
\providecommand \@ifx [1]{%
 \ifx #1\expandafter \@firstoftwo
 \else \expandafter \@secondoftwo
 \fi
}%
\providecommand \natexlab [1]{#1}%
\providecommand \enquote  [1]{``#1''}%
\providecommand \bibnamefont  [1]{#1}%
\providecommand \bibfnamefont [1]{#1}%
\providecommand \citenamefont [1]{#1}%
\providecommand \href@noop [0]{\@secondoftwo}%
\providecommand \href [0]{\begingroup \@sanitize@url \@href}%
\providecommand \@href[1]{\@@startlink{#1}\@@href}%
\providecommand \@@href[1]{\endgroup#1\@@endlink}%
\providecommand \@sanitize@url [0]{\catcode `\\12\catcode `\$12\catcode
  `\&12\catcode `\#12\catcode `\^12\catcode `\_12\catcode `\%12\relax}%
\providecommand \@@startlink[1]{}%
\providecommand \@@endlink[0]{}%
\providecommand \url  [0]{\begingroup\@sanitize@url \@url }%
\providecommand \@url [1]{\endgroup\@href {#1}{\urlprefix }}%
\providecommand \urlprefix  [0]{URL }%
\providecommand \Eprint [0]{\href }%
\providecommand \doibase [0]{http://dx.doi.org/}%
\providecommand \selectlanguage [0]{\@gobble}%
\providecommand \bibinfo  [0]{\@secondoftwo}%
\providecommand \bibfield  [0]{\@secondoftwo}%
\providecommand \translation [1]{[#1]}%
\providecommand \BibitemOpen [0]{}%
\providecommand \bibitemStop [0]{}%
\providecommand \bibitemNoStop [0]{.\EOS\space}%
\providecommand \EOS [0]{\spacefactor3000\relax}%
\providecommand \BibitemShut  [1]{\csname bibitem#1\endcsname}%
\let\auto@bib@innerbib\@empty
\bibitem [{\citenamefont {Glauber}(1963)}]{Glauber:1963}%
  \BibitemOpen
  \bibfield  {author} {\bibinfo {author} {\bibfnamefont {R.~J.}\ \bibnamefont
  {Glauber}},\ }\href {\doibase 10.1103/PhysRev.130.2529} {\bibfield  {journal}
  {\bibinfo  {journal} {Phys. Rev.}\ }\textbf {\bibinfo {volume} {130}},\
  \bibinfo {pages} {2529} (\bibinfo {year} {1963})}\BibitemShut {NoStop}%
\bibitem [{\citenamefont {Haroche}\ and\ \citenamefont
  {Raimond}(2006)}]{Haroche:2006}%
  \BibitemOpen
  \bibfield  {author} {\bibinfo {author} {\bibfnamefont {S.}~\bibnamefont
  {Haroche}}\ and\ \bibinfo {author} {\bibfnamefont {J.~M.}\ \bibnamefont
  {Raimond}},\ }\href@noop {} {\emph {\bibinfo {title} {Exploring the Quantum:
  Atoms, Cavities and Photons}}}\ (\bibinfo  {publisher} {Oxford University
  Press},\ \bibinfo {address} {Oxford},\ \bibinfo {year} {2006})\BibitemShut
  {NoStop}%
\bibitem [{\citenamefont {Rempe}\ \emph {et~al.}(1987)\citenamefont {Rempe},
  \citenamefont {Walther},\ and\ \citenamefont {Klein}}]{Rempe:1987}%
  \BibitemOpen
  \bibfield  {author} {\bibinfo {author} {\bibfnamefont {G.}~\bibnamefont
  {Rempe}}, \bibinfo {author} {\bibfnamefont {H.}~\bibnamefont {Walther}}, \
  and\ \bibinfo {author} {\bibfnamefont {N.}~\bibnamefont {Klein}},\ }\href
  {\doibase 10.1103/PhysRevLett.58.353} {\bibfield  {journal} {\bibinfo
  {journal} {Phys. Rev. Lett.}\ }\textbf {\bibinfo {volume} {58}},\ \bibinfo
  {pages} {353} (\bibinfo {year} {1987})}\BibitemShut {NoStop}%
\bibitem [{\citenamefont {Reithmaier}\ \emph {et~al.}(2004)\citenamefont
  {Reithmaier}, \citenamefont {Sek}, \citenamefont {L{\"o}ffler}, \citenamefont
  {Hofmann}, \citenamefont {Kuhn}, \citenamefont {Reitzenstein}, \citenamefont
  {Keldysh}, \citenamefont {Kulakovskii}, \citenamefont {Reinecke},\ and\
  \citenamefont {Forchel}}]{Reithmaier:2004}%
  \BibitemOpen
  \bibfield  {author} {\bibinfo {author} {\bibfnamefont {J.~P.}\ \bibnamefont
  {Reithmaier}}, \bibinfo {author} {\bibfnamefont {G.}~\bibnamefont {Sek}},
  \bibinfo {author} {\bibfnamefont {A.}~\bibnamefont {L{\"o}ffler}}, \bibinfo
  {author} {\bibfnamefont {C.}~\bibnamefont {Hofmann}}, \bibinfo {author}
  {\bibfnamefont {S.}~\bibnamefont {Kuhn}}, \bibinfo {author} {\bibfnamefont
  {S.}~\bibnamefont {Reitzenstein}}, \bibinfo {author} {\bibfnamefont {L.~V.}\
  \bibnamefont {Keldysh}}, \bibinfo {author} {\bibfnamefont {V.~D.}\
  \bibnamefont {Kulakovskii}}, \bibinfo {author} {\bibfnamefont {T.~L.}\
  \bibnamefont {Reinecke}}, \ and\ \bibinfo {author} {\bibfnamefont
  {A.}~\bibnamefont {Forchel}},\ }\href {\doibase 10.1038/nature02969}
  {\bibfield  {journal} {\bibinfo  {journal} {Nature}\ }\textbf {\bibinfo
  {volume} {432}},\ \bibinfo {pages} {197} (\bibinfo {year}
  {2004})}\BibitemShut {NoStop}%
\bibitem [{\citenamefont {Peter}\ \emph {et~al.}(2005)\citenamefont {Peter},
  \citenamefont {Senellart}, \citenamefont {Martrou}, \citenamefont
  {Lema\^{\i}tre}, \citenamefont {Hours}, \citenamefont {G\'erard},\ and\
  \citenamefont {Bloch}}]{Peter:2005}%
  \BibitemOpen
  \bibfield  {author} {\bibinfo {author} {\bibfnamefont {E.}~\bibnamefont
  {Peter}}, \bibinfo {author} {\bibfnamefont {P.}~\bibnamefont {Senellart}},
  \bibinfo {author} {\bibfnamefont {D.}~\bibnamefont {Martrou}}, \bibinfo
  {author} {\bibfnamefont {A.}~\bibnamefont {Lema\^{\i}tre}}, \bibinfo {author}
  {\bibfnamefont {J.}~\bibnamefont {Hours}}, \bibinfo {author} {\bibfnamefont
  {J.~M.}\ \bibnamefont {G\'erard}}, \ and\ \bibinfo {author} {\bibfnamefont
  {J.}~\bibnamefont {Bloch}},\ }\href {\doibase 10.1103/PhysRevLett.95.067401}
  {\bibfield  {journal} {\bibinfo  {journal} {Phys. Rev. Lett.}\ }\textbf
  {\bibinfo {volume} {95}},\ \bibinfo {pages} {067401} (\bibinfo {year}
  {2005})}\BibitemShut {NoStop}%
\bibitem [{\citenamefont {Wallraff}\ \emph {et~al.}(2004)\citenamefont
  {Wallraff}, \citenamefont {Schuster}, \citenamefont {Blais}, \citenamefont
  {Frunzio}, \citenamefont {Huang}, \citenamefont {Majer}, \citenamefont
  {Kumar}, \citenamefont {Girvin},\ and\ \citenamefont
  {Schoelkopf}}]{Wallraff:2004}%
  \BibitemOpen
  \bibfield  {author} {\bibinfo {author} {\bibfnamefont {A.}~\bibnamefont
  {Wallraff}}, \bibinfo {author} {\bibfnamefont {D.~I.}\ \bibnamefont
  {Schuster}}, \bibinfo {author} {\bibfnamefont {A.}~\bibnamefont {Blais}},
  \bibinfo {author} {\bibfnamefont {L.}~\bibnamefont {Frunzio}}, \bibinfo
  {author} {\bibfnamefont {R.~S.}\ \bibnamefont {Huang}}, \bibinfo {author}
  {\bibfnamefont {J.}~\bibnamefont {Majer}}, \bibinfo {author} {\bibfnamefont
  {S.}~\bibnamefont {Kumar}}, \bibinfo {author} {\bibfnamefont {S.~M.}\
  \bibnamefont {Girvin}}, \ and\ \bibinfo {author} {\bibfnamefont {R.~J.}\
  \bibnamefont {Schoelkopf}},\ }\href {\doibase 10.1038/nature02851} {\bibfield
   {journal} {\bibinfo  {journal} {Nature}\ }\textbf {\bibinfo {volume}
  {431}},\ \bibinfo {pages} {162} (\bibinfo {year} {2004})}\BibitemShut
  {NoStop}%
\bibitem [{\citenamefont {Imamoglu}\ \emph {et~al.}(1997)\citenamefont
  {Imamoglu}, \citenamefont {Schmidt}, \citenamefont {Woods},\ and\
  \citenamefont {Deutsch}}]{Imamoglu:1997}%
  \BibitemOpen
  \bibfield  {author} {\bibinfo {author} {\bibfnamefont {A.}~\bibnamefont
  {Imamoglu}}, \bibinfo {author} {\bibfnamefont {H.}~\bibnamefont {Schmidt}},
  \bibinfo {author} {\bibfnamefont {G.}~\bibnamefont {Woods}}, \ and\ \bibinfo
  {author} {\bibfnamefont {M.}~\bibnamefont {Deutsch}},\ }\href {\doibase
  10.1103/PhysRevLett.79.1467} {\bibfield  {journal} {\bibinfo  {journal}
  {Phys. Rev. Lett.}\ }\textbf {\bibinfo {volume} {79}},\ \bibinfo {pages}
  {1467} (\bibinfo {year} {1997})}\BibitemShut {NoStop}%
\bibitem [{\citenamefont {Birnbaum}\ \emph {et~al.}(2005)\citenamefont
  {Birnbaum}, \citenamefont {Boca}, \citenamefont {Miller}, \citenamefont
  {Boozer}, \citenamefont {Northup},\ and\ \citenamefont
  {Kimble}}]{Birnbaum:2005}%
  \BibitemOpen
  \bibfield  {author} {\bibinfo {author} {\bibfnamefont {K.~M.}\ \bibnamefont
  {Birnbaum}}, \bibinfo {author} {\bibfnamefont {A.}~\bibnamefont {Boca}},
  \bibinfo {author} {\bibfnamefont {R.}~\bibnamefont {Miller}}, \bibinfo
  {author} {\bibfnamefont {A.~D.}\ \bibnamefont {Boozer}}, \bibinfo {author}
  {\bibfnamefont {T.~E.}\ \bibnamefont {Northup}}, \ and\ \bibinfo {author}
  {\bibfnamefont {H.~J.}\ \bibnamefont {Kimble}},\ }\href {\doibase
  10.1038/nature03804} {\bibfield  {journal} {\bibinfo  {journal} {Nature}\
  }\textbf {\bibinfo {volume} {436}},\ \bibinfo {pages} {87} (\bibinfo {year}
  {2005})}\BibitemShut {NoStop}%
\bibitem [{\citenamefont {Bozyigit}\ \emph {et~al.}(2010)\citenamefont
  {Bozyigit}, \citenamefont {Lang}, \citenamefont {Steffen}, \citenamefont
  {Fink}, \citenamefont {Eichler}, \citenamefont {Baur}, \citenamefont
  {Bianchetti}, \citenamefont {Leek}, \citenamefont {Filipp}, \citenamefont
  {da~Silva}, \citenamefont {Blais},\ and\ \citenamefont
  {Wallraff}}]{Bozyigit:2010}%
  \BibitemOpen
  \bibfield  {author} {\bibinfo {author} {\bibfnamefont {D.}~\bibnamefont
  {Bozyigit}}, \bibinfo {author} {\bibfnamefont {C.}~\bibnamefont {Lang}},
  \bibinfo {author} {\bibfnamefont {L.}~\bibnamefont {Steffen}}, \bibinfo
  {author} {\bibfnamefont {J.~M.}\ \bibnamefont {Fink}}, \bibinfo {author}
  {\bibfnamefont {C.}~\bibnamefont {Eichler}}, \bibinfo {author} {\bibfnamefont
  {M.}~\bibnamefont {Baur}}, \bibinfo {author} {\bibfnamefont {R.}~\bibnamefont
  {Bianchetti}}, \bibinfo {author} {\bibfnamefont {P.~J.}\ \bibnamefont
  {Leek}}, \bibinfo {author} {\bibfnamefont {S.}~\bibnamefont {Filipp}},
  \bibinfo {author} {\bibfnamefont {M.~P.}\ \bibnamefont {da~Silva}}, \bibinfo
  {author} {\bibfnamefont {A.}~\bibnamefont {Blais}}, \ and\ \bibinfo {author}
  {\bibfnamefont {A.}~\bibnamefont {Wallraff}},\ }\href
  {https://doi.org/10.1038/nphys1845} {\bibfield  {journal} {\bibinfo
  {journal} {Nature Physics}\ }\textbf {\bibinfo {volume} {7}},\ \bibinfo
  {pages} {154 EP } (\bibinfo {year} {2010})}\BibitemShut {NoStop}%
\bibitem [{\citenamefont {Lang}\ \emph {et~al.}(2011)\citenamefont {Lang},
  \citenamefont {Bozyigit}, \citenamefont {Eichler}, \citenamefont {Steffen},
  \citenamefont {Fink}, \citenamefont {Abdumalikov}, \citenamefont {Baur},
  \citenamefont {Filipp}, \citenamefont {da~Silva}, \citenamefont {Blais},\
  and\ \citenamefont {Wallraff}}]{Lang:2011}%
  \BibitemOpen
  \bibfield  {author} {\bibinfo {author} {\bibfnamefont {C.}~\bibnamefont
  {Lang}}, \bibinfo {author} {\bibfnamefont {D.}~\bibnamefont {Bozyigit}},
  \bibinfo {author} {\bibfnamefont {C.}~\bibnamefont {Eichler}}, \bibinfo
  {author} {\bibfnamefont {L.}~\bibnamefont {Steffen}}, \bibinfo {author}
  {\bibfnamefont {J.~M.}\ \bibnamefont {Fink}}, \bibinfo {author}
  {\bibfnamefont {A.~A.}\ \bibnamefont {Abdumalikov}}, \bibinfo {author}
  {\bibfnamefont {M.}~\bibnamefont {Baur}}, \bibinfo {author} {\bibfnamefont
  {S.}~\bibnamefont {Filipp}}, \bibinfo {author} {\bibfnamefont {M.~P.}\
  \bibnamefont {da~Silva}}, \bibinfo {author} {\bibfnamefont {A.}~\bibnamefont
  {Blais}}, \ and\ \bibinfo {author} {\bibfnamefont {A.}~\bibnamefont
  {Wallraff}},\ }\href {\doibase 10.1103/PhysRevLett.106.243601} {\bibfield
  {journal} {\bibinfo  {journal} {Phys. Rev. Lett.}\ }\textbf {\bibinfo
  {volume} {106}},\ \bibinfo {pages} {243601} (\bibinfo {year}
  {2011})}\BibitemShut {NoStop}%
\bibitem [{\citenamefont {Hoffman}\ \emph {et~al.}(2011)\citenamefont
  {Hoffman}, \citenamefont {Srinivasan}, \citenamefont {Schmidt}, \citenamefont
  {Spietz}, \citenamefont {Aumentado}, \citenamefont {T\"ureci},\ and\
  \citenamefont {Houck}}]{Hoffman:2011}%
  \BibitemOpen
  \bibfield  {author} {\bibinfo {author} {\bibfnamefont {A.~J.}\ \bibnamefont
  {Hoffman}}, \bibinfo {author} {\bibfnamefont {S.~J.}\ \bibnamefont
  {Srinivasan}}, \bibinfo {author} {\bibfnamefont {S.}~\bibnamefont {Schmidt}},
  \bibinfo {author} {\bibfnamefont {L.}~\bibnamefont {Spietz}}, \bibinfo
  {author} {\bibfnamefont {J.}~\bibnamefont {Aumentado}}, \bibinfo {author}
  {\bibfnamefont {H.~E.}\ \bibnamefont {T\"ureci}}, \ and\ \bibinfo {author}
  {\bibfnamefont {A.~A.}\ \bibnamefont {Houck}},\ }\href {\doibase
  10.1103/PhysRevLett.107.053602} {\bibfield  {journal} {\bibinfo  {journal}
  {Phys. Rev. Lett.}\ }\textbf {\bibinfo {volume} {107}},\ \bibinfo {pages}
  {053602} (\bibinfo {year} {2011})}\BibitemShut {NoStop}%
\bibitem [{\citenamefont {Jaynes}\ and\ \citenamefont
  {Cummings}(1963)}]{Jaynes:1963}%
  \BibitemOpen
  \bibfield  {author} {\bibinfo {author} {\bibfnamefont {E.~T.}\ \bibnamefont
  {Jaynes}}\ and\ \bibinfo {author} {\bibfnamefont {F.~W.}\ \bibnamefont
  {Cummings}},\ }\href@noop {} {\bibfield  {journal} {\bibinfo  {journal}
  {Proc. IEEE}\ }\textbf {\bibinfo {volume} {51}},\ \bibinfo {pages} {89}
  (\bibinfo {year} {1963})}\BibitemShut {NoStop}%
\bibitem [{\citenamefont {Anappara}\ \emph {et~al.}(2009)\citenamefont
  {Anappara}, \citenamefont {De~Liberato}, \citenamefont {Tredicucci},
  \citenamefont {Ciuti}, \citenamefont {Biasiol}, \citenamefont {Sorba},\ and\
  \citenamefont {Beltram}}]{Anappara:2009}%
  \BibitemOpen
  \bibfield  {author} {\bibinfo {author} {\bibfnamefont {A.~A.}\ \bibnamefont
  {Anappara}}, \bibinfo {author} {\bibfnamefont {S.}~\bibnamefont
  {De~Liberato}}, \bibinfo {author} {\bibfnamefont {A.}~\bibnamefont
  {Tredicucci}}, \bibinfo {author} {\bibfnamefont {C.}~\bibnamefont {Ciuti}},
  \bibinfo {author} {\bibfnamefont {G.}~\bibnamefont {Biasiol}}, \bibinfo
  {author} {\bibfnamefont {L.}~\bibnamefont {Sorba}}, \ and\ \bibinfo {author}
  {\bibfnamefont {F.}~\bibnamefont {Beltram}},\ }\href {\doibase
  10.1103/PhysRevB.79.201303} {\bibfield  {journal} {\bibinfo  {journal} {Phys.
  Rev. B}\ }\textbf {\bibinfo {volume} {79}},\ \bibinfo {pages} {201303}
  (\bibinfo {year} {2009})}\BibitemShut {NoStop}%
\bibitem [{\citenamefont {Todorov}\ \emph {et~al.}(2010)\citenamefont
  {Todorov}, \citenamefont {Andrews}, \citenamefont {Colombelli}, \citenamefont
  {De~Liberato}, \citenamefont {Ciuti}, \citenamefont {Klang}, \citenamefont
  {Strasser},\ and\ \citenamefont {Sirtori}}]{Todorov:2010}%
  \BibitemOpen
  \bibfield  {author} {\bibinfo {author} {\bibfnamefont {Y.}~\bibnamefont
  {Todorov}}, \bibinfo {author} {\bibfnamefont {A.~M.}\ \bibnamefont
  {Andrews}}, \bibinfo {author} {\bibfnamefont {R.}~\bibnamefont {Colombelli}},
  \bibinfo {author} {\bibfnamefont {S.}~\bibnamefont {De~Liberato}}, \bibinfo
  {author} {\bibfnamefont {C.}~\bibnamefont {Ciuti}}, \bibinfo {author}
  {\bibfnamefont {P.}~\bibnamefont {Klang}}, \bibinfo {author} {\bibfnamefont
  {G.}~\bibnamefont {Strasser}}, \ and\ \bibinfo {author} {\bibfnamefont
  {C.}~\bibnamefont {Sirtori}},\ }\href {\doibase
  10.1103/PhysRevLett.105.196402} {\bibfield  {journal} {\bibinfo  {journal}
  {Phys. Rev. Lett.}\ }\textbf {\bibinfo {volume} {105}},\ \bibinfo {pages}
  {196402} (\bibinfo {year} {2010})}\BibitemShut {NoStop}%
\bibitem [{\citenamefont {Niemczyk}\ \emph {et~al.}(2010)\citenamefont
  {Niemczyk}, \citenamefont {Deppe}, \citenamefont {Huebl}, \citenamefont
  {Menzel}, \citenamefont {Hocke}, \citenamefont {Schwarz}, \citenamefont
  {Garcia-Ripoll}, \citenamefont {Zueco}, \citenamefont {H{\"u}mmer},
  \citenamefont {Solano}, \citenamefont {Marx},\ and\ \citenamefont
  {Gross}}]{Niemczyk:2010}%
  \BibitemOpen
  \bibfield  {author} {\bibinfo {author} {\bibfnamefont {T.}~\bibnamefont
  {Niemczyk}}, \bibinfo {author} {\bibfnamefont {F.}~\bibnamefont {Deppe}},
  \bibinfo {author} {\bibfnamefont {H.}~\bibnamefont {Huebl}}, \bibinfo
  {author} {\bibfnamefont {E.~P.}\ \bibnamefont {Menzel}}, \bibinfo {author}
  {\bibfnamefont {F.}~\bibnamefont {Hocke}}, \bibinfo {author} {\bibfnamefont
  {M.~J.}\ \bibnamefont {Schwarz}}, \bibinfo {author} {\bibfnamefont {J.~J.}\
  \bibnamefont {Garcia-Ripoll}}, \bibinfo {author} {\bibfnamefont
  {D.}~\bibnamefont {Zueco}}, \bibinfo {author} {\bibfnamefont
  {T.}~\bibnamefont {H{\"u}mmer}}, \bibinfo {author} {\bibfnamefont
  {E.}~\bibnamefont {Solano}}, \bibinfo {author} {\bibfnamefont
  {A.}~\bibnamefont {Marx}}, \ and\ \bibinfo {author} {\bibfnamefont
  {R.}~\bibnamefont {Gross}},\ }\href {https://doi.org/10.1038/nphys1730}
  {\bibfield  {journal} {\bibinfo  {journal} {Nature Physics}\ }\textbf
  {\bibinfo {volume} {6}},\ \bibinfo {pages} {772 EP } (\bibinfo {year}
  {2010})}\BibitemShut {NoStop}%
\bibitem [{\citenamefont {Forn-D\'{\i}az}\ \emph {et~al.}(2010)\citenamefont
  {Forn-D\'{\i}az}, \citenamefont {Lisenfeld}, \citenamefont {Marcos},
  \citenamefont {Garc\'{\i}a-Ripoll}, \citenamefont {Solano}, \citenamefont
  {Harmans},\ and\ \citenamefont {Mooij}}]{Forn-Diaz:2010}%
  \BibitemOpen
  \bibfield  {author} {\bibinfo {author} {\bibfnamefont {P.}~\bibnamefont
  {Forn-D\'{\i}az}}, \bibinfo {author} {\bibfnamefont {J.}~\bibnamefont
  {Lisenfeld}}, \bibinfo {author} {\bibfnamefont {D.}~\bibnamefont {Marcos}},
  \bibinfo {author} {\bibfnamefont {J.~J.}\ \bibnamefont {Garc\'{\i}a-Ripoll}},
  \bibinfo {author} {\bibfnamefont {E.}~\bibnamefont {Solano}}, \bibinfo
  {author} {\bibfnamefont {C.~J. P.~M.}\ \bibnamefont {Harmans}}, \ and\
  \bibinfo {author} {\bibfnamefont {J.~E.}\ \bibnamefont {Mooij}},\ }\href
  {\doibase 10.1103/PhysRevLett.105.237001} {\bibfield  {journal} {\bibinfo
  {journal} {Phys. Rev. Lett.}\ }\textbf {\bibinfo {volume} {105}},\ \bibinfo
  {pages} {237001} (\bibinfo {year} {2010})}\BibitemShut {NoStop}%
\bibitem [{\citenamefont {Forn-D{\'\i}az}\ \emph {et~al.}(2017)\citenamefont
  {Forn-D{\'\i}az}, \citenamefont {Garc{\'\i}a-Ripoll}, \citenamefont
  {Peropadre}, \citenamefont {Orgiazzi}, \citenamefont {Yurtalan},
  \citenamefont {Belyansky}, \citenamefont {Wilson},\ and\ \citenamefont
  {Lupascu}}]{Forn-Diaz:2017}%
  \BibitemOpen
  \bibfield  {author} {\bibinfo {author} {\bibfnamefont {P.}~\bibnamefont
  {Forn-D{\'\i}az}}, \bibinfo {author} {\bibfnamefont {J.~J.}\ \bibnamefont
  {Garc{\'\i}a-Ripoll}}, \bibinfo {author} {\bibfnamefont {B.}~\bibnamefont
  {Peropadre}}, \bibinfo {author} {\bibfnamefont {J.~L.}\ \bibnamefont
  {Orgiazzi}}, \bibinfo {author} {\bibfnamefont {M.~A.}\ \bibnamefont
  {Yurtalan}}, \bibinfo {author} {\bibfnamefont {R.}~\bibnamefont {Belyansky}},
  \bibinfo {author} {\bibfnamefont {C.~M.}\ \bibnamefont {Wilson}}, \ and\
  \bibinfo {author} {\bibfnamefont {A.}~\bibnamefont {Lupascu}},\ }\href
  {https://doi.org/10.1038/nphys3905} {\bibfield  {journal} {\bibinfo
  {journal} {Nature Physics}\ }\textbf {\bibinfo {volume} {13}},\ \bibinfo
  {pages} {39 EP } (\bibinfo {year} {2017})}\BibitemShut {NoStop}%
\bibitem [{\citenamefont {Yoshihara}\ \emph {et~al.}(2017)\citenamefont
  {Yoshihara}, \citenamefont {Fuse}, \citenamefont {Ashhab}, \citenamefont
  {Kakuyanagi}, \citenamefont {Saito},\ and\ \citenamefont
  {Semba}}]{Yoshihara:2017}%
  \BibitemOpen
  \bibfield  {author} {\bibinfo {author} {\bibfnamefont {F.}~\bibnamefont
  {Yoshihara}}, \bibinfo {author} {\bibfnamefont {T.}~\bibnamefont {Fuse}},
  \bibinfo {author} {\bibfnamefont {S.}~\bibnamefont {Ashhab}}, \bibinfo
  {author} {\bibfnamefont {K.}~\bibnamefont {Kakuyanagi}}, \bibinfo {author}
  {\bibfnamefont {S.}~\bibnamefont {Saito}}, \ and\ \bibinfo {author}
  {\bibfnamefont {K.}~\bibnamefont {Semba}},\ }\href
  {https://doi.org/10.1038/nphys3906} {\bibfield  {journal} {\bibinfo
  {journal} {Nature Physics}\ }\textbf {\bibinfo {volume} {13}},\ \bibinfo
  {pages} {44 EP } (\bibinfo {year} {2017})}\BibitemShut {NoStop}%
\bibitem [{\citenamefont {Ciuti}\ \emph {et~al.}(2005)\citenamefont {Ciuti},
  \citenamefont {Bastard},\ and\ \citenamefont {Carusotto}}]{Ciuti:2005}%
  \BibitemOpen
  \bibfield  {author} {\bibinfo {author} {\bibfnamefont {C.}~\bibnamefont
  {Ciuti}}, \bibinfo {author} {\bibfnamefont {G.}~\bibnamefont {Bastard}}, \
  and\ \bibinfo {author} {\bibfnamefont {I.}~\bibnamefont {Carusotto}},\ }\href
  {\doibase 10.1103/PhysRevB.72.115303} {\bibfield  {journal} {\bibinfo
  {journal} {Phys. Rev. B}\ }\textbf {\bibinfo {volume} {72}},\ \bibinfo
  {pages} {115303} (\bibinfo {year} {2005})}\BibitemShut {NoStop}%
\bibitem [{\citenamefont {Devoret}\ \emph {et~al.}(2007)\citenamefont
  {Devoret}, \citenamefont {Girvin},\ and\ \citenamefont
  {Schoelkopf}}]{Devoret:2007}%
  \BibitemOpen
  \bibfield  {author} {\bibinfo {author} {\bibfnamefont {M.~H.}\ \bibnamefont
  {Devoret}}, \bibinfo {author} {\bibfnamefont {S.}~\bibnamefont {Girvin}}, \
  and\ \bibinfo {author} {\bibfnamefont {R.}~\bibnamefont {Schoelkopf}},\
  }\href {\doibase 10.1002/andp.200710261} {\bibfield  {journal} {\bibinfo
  {journal} {Ann. Phys. (Leipzig)}\ }\textbf {\bibinfo {volume} {16}},\
  \bibinfo {pages} {7067} (\bibinfo {year} {2007})}\BibitemShut {NoStop}%
\bibitem [{\citenamefont {Bourassa}\ \emph {et~al.}(2009)\citenamefont
  {Bourassa}, \citenamefont {Gambetta}, \citenamefont {Abdumalikov},
  \citenamefont {Astafiev}, \citenamefont {Nakamura},\ and\ \citenamefont
  {Blais}}]{Bourassa:2009}%
  \BibitemOpen
  \bibfield  {author} {\bibinfo {author} {\bibfnamefont {J.}~\bibnamefont
  {Bourassa}}, \bibinfo {author} {\bibfnamefont {J.~M.}\ \bibnamefont
  {Gambetta}}, \bibinfo {author} {\bibfnamefont {A.~A.}\ \bibnamefont
  {Abdumalikov}}, \bibinfo {author} {\bibfnamefont {O.}~\bibnamefont
  {Astafiev}}, \bibinfo {author} {\bibfnamefont {Y.}~\bibnamefont {Nakamura}},
  \ and\ \bibinfo {author} {\bibfnamefont {A.}~\bibnamefont {Blais}},\ }\href
  {\doibase 10.1103/PhysRevA.80.032109} {\bibfield  {journal} {\bibinfo
  {journal} {Phys. Rev. A}\ }\textbf {\bibinfo {volume} {80}},\ \bibinfo
  {pages} {032109} (\bibinfo {year} {2009})}\BibitemShut {NoStop}%
\bibitem [{\citenamefont {De~Liberato}\ \emph {et~al.}(2009)\citenamefont
  {De~Liberato}, \citenamefont {Gerace}, \citenamefont {Carusotto},\ and\
  \citenamefont {Ciuti}}]{DeLiberato:2009}%
  \BibitemOpen
  \bibfield  {author} {\bibinfo {author} {\bibfnamefont {S.}~\bibnamefont
  {De~Liberato}}, \bibinfo {author} {\bibfnamefont {D.}~\bibnamefont {Gerace}},
  \bibinfo {author} {\bibfnamefont {I.}~\bibnamefont {Carusotto}}, \ and\
  \bibinfo {author} {\bibfnamefont {C.}~\bibnamefont {Ciuti}},\ }\href
  {\doibase 10.1103/PhysRevA.80.053810} {\bibfield  {journal} {\bibinfo
  {journal} {Phys. Rev. A}\ }\textbf {\bibinfo {volume} {80}},\ \bibinfo
  {pages} {053810} (\bibinfo {year} {2009})}\BibitemShut {NoStop}%
\bibitem [{\citenamefont {S\'anchez-Burillo}\ \emph {et~al.}(2019)\citenamefont
  {S\'anchez-Burillo}, \citenamefont {Mart\'{\i}n-Moreno}, \citenamefont
  {Garc\'{\i}a-Ripoll},\ and\ \citenamefont {Zueco}}]{Sanchez-Burillo:2019}%
  \BibitemOpen
  \bibfield  {author} {\bibinfo {author} {\bibfnamefont {E.}~\bibnamefont
  {S\'anchez-Burillo}}, \bibinfo {author} {\bibfnamefont {L.}~\bibnamefont
  {Mart\'{\i}n-Moreno}}, \bibinfo {author} {\bibfnamefont {J.~J.}\ \bibnamefont
  {Garc\'{\i}a-Ripoll}}, \ and\ \bibinfo {author} {\bibfnamefont
  {D.}~\bibnamefont {Zueco}},\ }\href {\doibase 10.1103/PhysRevLett.123.013601}
  {\bibfield  {journal} {\bibinfo  {journal} {Phys. Rev. Lett.}\ }\textbf
  {\bibinfo {volume} {123}},\ \bibinfo {pages} {013601} (\bibinfo {year}
  {2019})}\BibitemShut {NoStop}%
\bibitem [{\citenamefont {Ridolfo}\ \emph {et~al.}(2012)\citenamefont
  {Ridolfo}, \citenamefont {Leib}, \citenamefont {Savasta},\ and\ \citenamefont
  {Hartmann}}]{Ridolfo:2012}%
  \BibitemOpen
  \bibfield  {author} {\bibinfo {author} {\bibfnamefont {A.}~\bibnamefont
  {Ridolfo}}, \bibinfo {author} {\bibfnamefont {M.}~\bibnamefont {Leib}},
  \bibinfo {author} {\bibfnamefont {S.}~\bibnamefont {Savasta}}, \ and\
  \bibinfo {author} {\bibfnamefont {M.~J.}\ \bibnamefont {Hartmann}},\ }\href
  {\doibase 10.1103/PhysRevLett.109.193602} {\bibfield  {journal} {\bibinfo
  {journal} {Phys. Rev. Lett.}\ }\textbf {\bibinfo {volume} {109}},\ \bibinfo
  {pages} {193602} (\bibinfo {year} {2012})}\BibitemShut {NoStop}%
\bibitem [{\citenamefont {Le~Boit\'e}\ \emph {et~al.}(2016)\citenamefont
  {Le~Boit\'e}, \citenamefont {Hwang}, \citenamefont {Nha},\ and\ \citenamefont
  {Plenio}}]{LeBoite:2016}%
  \BibitemOpen
  \bibfield  {author} {\bibinfo {author} {\bibfnamefont {A.}~\bibnamefont
  {Le~Boit\'e}}, \bibinfo {author} {\bibfnamefont {M.-J.}\ \bibnamefont
  {Hwang}}, \bibinfo {author} {\bibfnamefont {H.}~\bibnamefont {Nha}}, \ and\
  \bibinfo {author} {\bibfnamefont {M.~B.}\ \bibnamefont {Plenio}},\ }\href
  {\doibase 10.1103/PhysRevA.94.033827} {\bibfield  {journal} {\bibinfo
  {journal} {Phys. Rev. A}\ }\textbf {\bibinfo {volume} {94}},\ \bibinfo
  {pages} {033827} (\bibinfo {year} {2016})}\BibitemShut {NoStop}%
\bibitem [{\citenamefont {Shi}\ \emph {et~al.}(2018)\citenamefont {Shi},
  \citenamefont {Chang},\ and\ \citenamefont {Garc\'{\i}a-Ripoll}}]{Shi:2018}%
  \BibitemOpen
  \bibfield  {author} {\bibinfo {author} {\bibfnamefont {T.}~\bibnamefont
  {Shi}}, \bibinfo {author} {\bibfnamefont {Y.}~\bibnamefont {Chang}}, \ and\
  \bibinfo {author} {\bibfnamefont {J.~J.}\ \bibnamefont
  {Garc\'{\i}a-Ripoll}},\ }\href {\doibase 10.1103/PhysRevLett.120.153602}
  {\bibfield  {journal} {\bibinfo  {journal} {Phys. Rev. Lett.}\ }\textbf
  {\bibinfo {volume} {120}},\ \bibinfo {pages} {153602} (\bibinfo {year}
  {2018})}\BibitemShut {NoStop}%
\bibitem [{\citenamefont {Felicetti}\ \emph {et~al.}(2014)\citenamefont
  {Felicetti}, \citenamefont {Romero}, \citenamefont {Rossini}, \citenamefont
  {Fazio},\ and\ \citenamefont {Solano}}]{Felicetti:2014}%
  \BibitemOpen
  \bibfield  {author} {\bibinfo {author} {\bibfnamefont {S.}~\bibnamefont
  {Felicetti}}, \bibinfo {author} {\bibfnamefont {G.}~\bibnamefont {Romero}},
  \bibinfo {author} {\bibfnamefont {D.}~\bibnamefont {Rossini}}, \bibinfo
  {author} {\bibfnamefont {R.}~\bibnamefont {Fazio}}, \ and\ \bibinfo {author}
  {\bibfnamefont {E.}~\bibnamefont {Solano}},\ }\href {\doibase
  10.1103/PhysRevA.89.013853} {\bibfield  {journal} {\bibinfo  {journal} {Phys.
  Rev. A}\ }\textbf {\bibinfo {volume} {89}},\ \bibinfo {pages} {013853}
  (\bibinfo {year} {2014})}\BibitemShut {NoStop}%
\bibitem [{\citenamefont {Bartolo}\ and\ \citenamefont
  {Ciuti}(2018)}]{Bartolo:2018}%
  \BibitemOpen
  \bibfield  {author} {\bibinfo {author} {\bibfnamefont {N.}~\bibnamefont
  {Bartolo}}\ and\ \bibinfo {author} {\bibfnamefont {C.}~\bibnamefont
  {Ciuti}},\ }\href {\doibase 10.1103/PhysRevB.98.205301} {\bibfield  {journal}
  {\bibinfo  {journal} {Phys. Rev. B}\ }\textbf {\bibinfo {volume} {98}},\
  \bibinfo {pages} {205301} (\bibinfo {year} {2018})}\BibitemShut {NoStop}%
\bibitem [{\citenamefont {Frisk~Kockum}\ \emph {et~al.}(2019)\citenamefont
  {Frisk~Kockum}, \citenamefont {Miranowicz}, \citenamefont {De~Liberato},
  \citenamefont {Savasta},\ and\ \citenamefont {Nori}}]{Frisk-Kockum:2019}%
  \BibitemOpen
  \bibfield  {author} {\bibinfo {author} {\bibfnamefont {A.}~\bibnamefont
  {Frisk~Kockum}}, \bibinfo {author} {\bibfnamefont {A.}~\bibnamefont
  {Miranowicz}}, \bibinfo {author} {\bibfnamefont {S.}~\bibnamefont
  {De~Liberato}}, \bibinfo {author} {\bibfnamefont {S.}~\bibnamefont
  {Savasta}}, \ and\ \bibinfo {author} {\bibfnamefont {F.}~\bibnamefont
  {Nori}},\ }\href {\doibase 10.1038/s42254-018-0006-2} {\bibfield  {journal}
  {\bibinfo  {journal} {Nature Reviews Physics}\ }\textbf {\bibinfo {volume}
  {1}},\ \bibinfo {pages} {19} (\bibinfo {year} {2019})}\BibitemShut {NoStop}%
\bibitem [{\citenamefont {Forn-D\'{\i}az}\ \emph {et~al.}(2019)\citenamefont
  {Forn-D\'{\i}az}, \citenamefont {Lamata}, \citenamefont {Rico}, \citenamefont
  {Kono},\ and\ \citenamefont {Solano}}]{Forn-Diaz:2019}%
  \BibitemOpen
  \bibfield  {author} {\bibinfo {author} {\bibfnamefont {P.}~\bibnamefont
  {Forn-D\'{\i}az}}, \bibinfo {author} {\bibfnamefont {L.}~\bibnamefont
  {Lamata}}, \bibinfo {author} {\bibfnamefont {E.}~\bibnamefont {Rico}},
  \bibinfo {author} {\bibfnamefont {J.}~\bibnamefont {Kono}}, \ and\ \bibinfo
  {author} {\bibfnamefont {E.}~\bibnamefont {Solano}},\ }\href {\doibase
  10.1103/RevModPhys.91.025005} {\bibfield  {journal} {\bibinfo  {journal}
  {Rev. Mod. Phys.}\ }\textbf {\bibinfo {volume} {91}},\ \bibinfo {pages}
  {025005} (\bibinfo {year} {2019})}\BibitemShut {NoStop}%
\bibitem [{\citenamefont {Langford}\ \emph {et~al.}(2017)\citenamefont
  {Langford}, \citenamefont {Sagastizabal}, \citenamefont {Kounalakis},
  \citenamefont {Dickel}, \citenamefont {Bruno}, \citenamefont {Luthi},
  \citenamefont {Thoen}, \citenamefont {Endo},\ and\ \citenamefont
  {DiCarlo}}]{Langford:2017}%
  \BibitemOpen
  \bibfield  {author} {\bibinfo {author} {\bibfnamefont {N.~K.}\ \bibnamefont
  {Langford}}, \bibinfo {author} {\bibfnamefont {R.}~\bibnamefont
  {Sagastizabal}}, \bibinfo {author} {\bibfnamefont {M.}~\bibnamefont
  {Kounalakis}}, \bibinfo {author} {\bibfnamefont {C.}~\bibnamefont {Dickel}},
  \bibinfo {author} {\bibfnamefont {A.}~\bibnamefont {Bruno}}, \bibinfo
  {author} {\bibfnamefont {F.}~\bibnamefont {Luthi}}, \bibinfo {author}
  {\bibfnamefont {D.~J.}\ \bibnamefont {Thoen}}, \bibinfo {author}
  {\bibfnamefont {A.}~\bibnamefont {Endo}}, \ and\ \bibinfo {author}
  {\bibfnamefont {L.}~\bibnamefont {DiCarlo}},\ }\href {\doibase
  10.1038/s41467-017-01061-x} {\bibfield  {journal} {\bibinfo  {journal}
  {Nature Communications}\ }\textbf {\bibinfo {volume} {8}},\ \bibinfo {pages}
  {1715} (\bibinfo {year} {2017})}\BibitemShut {NoStop}%
\bibitem [{\citenamefont {Braum\"uller}\ \emph {et~al.}(2017)\citenamefont
  {Braum\"uller}, \citenamefont {Marthaler}, \citenamefont {Schneider},
  \citenamefont {Stehli}, \citenamefont {Rotzinger}, \citenamefont {Weides},\
  and\ \citenamefont {Ustinov}}]{Braumuller:2017}%
  \BibitemOpen
  \bibfield  {author} {\bibinfo {author} {\bibfnamefont {J.}~\bibnamefont
  {Braum\"uller}}, \bibinfo {author} {\bibfnamefont {M.}~\bibnamefont
  {Marthaler}}, \bibinfo {author} {\bibfnamefont {A.}~\bibnamefont
  {Schneider}}, \bibinfo {author} {\bibfnamefont {A.}~\bibnamefont {Stehli}},
  \bibinfo {author} {\bibfnamefont {H.}~\bibnamefont {Rotzinger}}, \bibinfo
  {author} {\bibfnamefont {M.}~\bibnamefont {Weides}}, \ and\ \bibinfo {author}
  {\bibfnamefont {A.~V.}\ \bibnamefont {Ustinov}},\ }\href {\doibase
  10.1038/s41467-017-00894-w} {\bibfield  {journal} {\bibinfo  {journal} {Nat.
  Commun.}\ }\textbf {\bibinfo {volume} {8}},\ \bibinfo {pages} {779} (\bibinfo
  {year} {2017})}\BibitemShut {NoStop}%
\bibitem [{\citenamefont {Markovi\'{c}}\ \emph {et~al.}(2018)\citenamefont
  {Markovi\'{c}}, \citenamefont {Jezouin}, \citenamefont {Ficheux},
  \citenamefont {Fedortchenko}, \citenamefont {Felicetti}, \citenamefont
  {Coudreau}, \citenamefont {Milman}, \citenamefont {Leghtas},\ and\
  \citenamefont {Huard}}]{Markovic:2018}%
  \BibitemOpen
  \bibfield  {author} {\bibinfo {author} {\bibfnamefont {D.}~\bibnamefont
  {Markovi\'{c}}}, \bibinfo {author} {\bibfnamefont {S.}~\bibnamefont
  {Jezouin}}, \bibinfo {author} {\bibfnamefont {Q.}~\bibnamefont {Ficheux}},
  \bibinfo {author} {\bibfnamefont {S.}~\bibnamefont {Fedortchenko}}, \bibinfo
  {author} {\bibfnamefont {S.}~\bibnamefont {Felicetti}}, \bibinfo {author}
  {\bibfnamefont {T.}~\bibnamefont {Coudreau}}, \bibinfo {author}
  {\bibfnamefont {P.}~\bibnamefont {Milman}}, \bibinfo {author} {\bibfnamefont
  {Z.}~\bibnamefont {Leghtas}}, \ and\ \bibinfo {author} {\bibfnamefont
  {B.}~\bibnamefont {Huard}},\ }\href {\doibase 10.1103/PhysRevLett.121.040505}
  {\bibfield  {journal} {\bibinfo  {journal} {Phys. Rev. Lett.}\ }\textbf
  {\bibinfo {volume} {121}},\ \bibinfo {pages} {040505} (\bibinfo {year}
  {2018})}\BibitemShut {NoStop}%
\bibitem [{\citenamefont {Lv}\ \emph {et~al.}(2018)\citenamefont {Lv},
  \citenamefont {An}, \citenamefont {Liu}, \citenamefont {Zhang}, \citenamefont
  {Pedernales}, \citenamefont {Lamata}, \citenamefont {Solano},\ and\
  \citenamefont {Kim}}]{Lv:2018}%
  \BibitemOpen
  \bibfield  {author} {\bibinfo {author} {\bibfnamefont {D.}~\bibnamefont
  {Lv}}, \bibinfo {author} {\bibfnamefont {S.}~\bibnamefont {An}}, \bibinfo
  {author} {\bibfnamefont {Z.}~\bibnamefont {Liu}}, \bibinfo {author}
  {\bibfnamefont {J.-N.}\ \bibnamefont {Zhang}}, \bibinfo {author}
  {\bibfnamefont {J.~S.}\ \bibnamefont {Pedernales}}, \bibinfo {author}
  {\bibfnamefont {L.}~\bibnamefont {Lamata}}, \bibinfo {author} {\bibfnamefont
  {E.}~\bibnamefont {Solano}}, \ and\ \bibinfo {author} {\bibfnamefont
  {K.}~\bibnamefont {Kim}},\ }\href {\doibase 10.1103/PhysRevX.8.021027}
  {\bibfield  {journal} {\bibinfo  {journal} {Phys. Rev. X}\ }\textbf {\bibinfo
  {volume} {8}},\ \bibinfo {pages} {021027} (\bibinfo {year}
  {2018})}\BibitemShut {NoStop}%
\bibitem [{\citenamefont {Peterson}\ \emph {et~al.}(2019)\citenamefont
  {Peterson}, \citenamefont {Kotler}, \citenamefont {Lecocq}, \citenamefont
  {Cicak}, \citenamefont {Jin}, \citenamefont {Simmonds}, \citenamefont
  {Aumentado},\ and\ \citenamefont {Teufel}}]{Peterson:2019}%
  \BibitemOpen
  \bibfield  {author} {\bibinfo {author} {\bibfnamefont {G.~A.}\ \bibnamefont
  {Peterson}}, \bibinfo {author} {\bibfnamefont {S.}~\bibnamefont {Kotler}},
  \bibinfo {author} {\bibfnamefont {F.}~\bibnamefont {Lecocq}}, \bibinfo
  {author} {\bibfnamefont {K.}~\bibnamefont {Cicak}}, \bibinfo {author}
  {\bibfnamefont {X.~Y.}\ \bibnamefont {Jin}}, \bibinfo {author} {\bibfnamefont
  {R.~W.}\ \bibnamefont {Simmonds}}, \bibinfo {author} {\bibfnamefont
  {J.}~\bibnamefont {Aumentado}}, \ and\ \bibinfo {author} {\bibfnamefont
  {J.~D.}\ \bibnamefont {Teufel}},\ }\href {\doibase
  10.1103/PhysRevLett.123.247701} {\bibfield  {journal} {\bibinfo  {journal}
  {Phys. Rev. Lett.}\ }\textbf {\bibinfo {volume} {123}},\ \bibinfo {pages}
  {247701} (\bibinfo {year} {2019})}\BibitemShut {NoStop}%
\bibitem [{\citenamefont {Casanova}\ \emph {et~al.}(2010)\citenamefont
  {Casanova}, \citenamefont {Romero}, \citenamefont {Lizuain}, \citenamefont
  {Garc\'{\i}a-Ripoll},\ and\ \citenamefont {Solano}}]{Casanova:2010}%
  \BibitemOpen
  \bibfield  {author} {\bibinfo {author} {\bibfnamefont {J.}~\bibnamefont
  {Casanova}}, \bibinfo {author} {\bibfnamefont {G.}~\bibnamefont {Romero}},
  \bibinfo {author} {\bibfnamefont {I.}~\bibnamefont {Lizuain}}, \bibinfo
  {author} {\bibfnamefont {J.~J.}\ \bibnamefont {Garc\'{\i}a-Ripoll}}, \ and\
  \bibinfo {author} {\bibfnamefont {E.}~\bibnamefont {Solano}},\ }\href
  {\doibase 10.1103/PhysRevLett.105.263603} {\bibfield  {journal} {\bibinfo
  {journal} {Phys. Rev. Lett.}\ }\textbf {\bibinfo {volume} {105}},\ \bibinfo
  {pages} {263603} (\bibinfo {year} {2010})}\BibitemShut {NoStop}%
\bibitem [{\citenamefont {Cohen‐Tannoudji}\ \emph {et~al.}(1997)\citenamefont
  {Cohen‐Tannoudji}, \citenamefont {Dupont‐Roc},\ and\ \citenamefont
  {Grynberg}}]{Cohen-Tannoudji:1997}%
  \BibitemOpen
  \bibfield  {author} {\bibinfo {author} {\bibfnamefont {C.}~\bibnamefont
  {Cohen‐Tannoudji}}, \bibinfo {author} {\bibfnamefont {J.}~\bibnamefont
  {Dupont‐Roc}}, \ and\ \bibinfo {author} {\bibfnamefont {G.}~\bibnamefont
  {Grynberg}},\ }\href {\doibase 10.1002/9783527618422} {\emph {\bibinfo
  {title} {Photons and Atoms: Introduction to Quantum Electrodynamics}}}\
  (\bibinfo  {publisher} {Wiley-VCH},\ \bibinfo {address} {Weinheim},\ \bibinfo
  {year} {1997})\BibitemShut {NoStop}%
\bibitem [{\citenamefont {Rabi}(1937)}]{Rabi:1937}%
  \BibitemOpen
  \bibfield  {author} {\bibinfo {author} {\bibfnamefont {I.~I.}\ \bibnamefont
  {Rabi}},\ }\href {\doibase 10.1103/PhysRev.51.652} {\bibfield  {journal}
  {\bibinfo  {journal} {Phys. Rev.}\ }\textbf {\bibinfo {volume} {51}},\
  \bibinfo {pages} {652} (\bibinfo {year} {1937})}\BibitemShut {NoStop}%
\bibitem [{\citenamefont {Trav\v{e}nec}(2012)}]{Travenec:2012}%
  \BibitemOpen
  \bibfield  {author} {\bibinfo {author} {\bibfnamefont {I.}~\bibnamefont
  {Trav\v{e}nec}},\ }\href {\doibase 10.1103/PhysRevA.85.043805} {\bibfield
  {journal} {\bibinfo  {journal} {Phys. Rev. A}\ }\textbf {\bibinfo {volume}
  {85}},\ \bibinfo {pages} {043805} (\bibinfo {year} {2012})}\BibitemShut
  {NoStop}%
\bibitem [{\citenamefont {Albert}(2012)}]{Albert:2012}%
  \BibitemOpen
  \bibfield  {author} {\bibinfo {author} {\bibfnamefont {V.~V.}\ \bibnamefont
  {Albert}},\ }\href {\doibase 10.1103/PhysRevLett.108.180401} {\bibfield
  {journal} {\bibinfo  {journal} {Phys. Rev. Lett.}\ }\textbf {\bibinfo
  {volume} {108}},\ \bibinfo {pages} {180401} (\bibinfo {year}
  {2012})}\BibitemShut {NoStop}%
\bibitem [{\citenamefont {Mahmoodian}(2019)}]{Mahmoodian:2019}%
  \BibitemOpen
  \bibfield  {author} {\bibinfo {author} {\bibfnamefont {S.}~\bibnamefont
  {Mahmoodian}},\ }\href {\doibase 10.1103/PhysRevLett.123.133603} {\bibfield
  {journal} {\bibinfo  {journal} {Phys. Rev. Lett.}\ }\textbf {\bibinfo
  {volume} {123}},\ \bibinfo {pages} {133603} (\bibinfo {year}
  {2019})}\BibitemShut {NoStop}%
\bibitem [{\citenamefont {Leggett}\ \emph {et~al.}(1987)\citenamefont
  {Leggett}, \citenamefont {Chakravarty}, \citenamefont {Dorsey}, \citenamefont
  {Fisher}, \citenamefont {Garg},\ and\ \citenamefont
  {Zwerger}}]{Leggett:1987}%
  \BibitemOpen
  \bibfield  {author} {\bibinfo {author} {\bibfnamefont {A.~J.}\ \bibnamefont
  {Leggett}}, \bibinfo {author} {\bibfnamefont {S.}~\bibnamefont
  {Chakravarty}}, \bibinfo {author} {\bibfnamefont {A.~T.}\ \bibnamefont
  {Dorsey}}, \bibinfo {author} {\bibfnamefont {M.~P.~A.}\ \bibnamefont
  {Fisher}}, \bibinfo {author} {\bibfnamefont {A.}~\bibnamefont {Garg}}, \ and\
  \bibinfo {author} {\bibfnamefont {W.}~\bibnamefont {Zwerger}},\ }\href
  {\doibase 10.1103/RevModPhys.59.1} {\bibfield  {journal} {\bibinfo  {journal}
  {Rev. Mod. Phys.}\ }\textbf {\bibinfo {volume} {59}},\ \bibinfo {pages} {1}
  (\bibinfo {year} {1987})}\BibitemShut {NoStop}%
\bibitem [{\citenamefont {Hopfield}(1958)}]{Hopfield:1958}%
  \BibitemOpen
  \bibfield  {author} {\bibinfo {author} {\bibfnamefont {J.~J.}\ \bibnamefont
  {Hopfield}},\ }\href {\doibase 10.1103/PhysRev.112.1555} {\bibfield
  {journal} {\bibinfo  {journal} {Phys. Rev.}\ }\textbf {\bibinfo {volume}
  {112}},\ \bibinfo {pages} {1555} (\bibinfo {year} {1958})}\BibitemShut
  {NoStop}%
\bibitem [{\citenamefont {Ciuti}\ and\ \citenamefont
  {Carusotto}(2006)}]{Ciuti:2006}%
  \BibitemOpen
  \bibfield  {author} {\bibinfo {author} {\bibfnamefont {C.}~\bibnamefont
  {Ciuti}}\ and\ \bibinfo {author} {\bibfnamefont {I.}~\bibnamefont
  {Carusotto}},\ }\href {\doibase 10.1103/PhysRevA.74.033811} {\bibfield
  {journal} {\bibinfo  {journal} {Phys. Rev. A}\ }\textbf {\bibinfo {volume}
  {74}},\ \bibinfo {pages} {033811} (\bibinfo {year} {2006})}\BibitemShut
  {NoStop}%
\bibitem [{\citenamefont {Bamba}\ and\ \citenamefont
  {Ogawa}(2012)}]{Bamba:2012}%
  \BibitemOpen
  \bibfield  {author} {\bibinfo {author} {\bibfnamefont {M.}~\bibnamefont
  {Bamba}}\ and\ \bibinfo {author} {\bibfnamefont {T.}~\bibnamefont {Ogawa}},\
  }\href {\doibase 10.1103/PhysRevA.86.063831} {\bibfield  {journal} {\bibinfo
  {journal} {Phys. Rev. A}\ }\textbf {\bibinfo {volume} {86}},\ \bibinfo
  {pages} {063831} (\bibinfo {year} {2012})}\BibitemShut {NoStop}%
\bibitem [{\citenamefont {Benivegna}\ and\ \citenamefont
  {Messina}(1987)}]{Benivegna:1987}%
  \BibitemOpen
  \bibfield  {author} {\bibinfo {author} {\bibfnamefont {G.}~\bibnamefont
  {Benivegna}}\ and\ \bibinfo {author} {\bibfnamefont {A.}~\bibnamefont
  {Messina}},\ }\href {\doibase 10.1103/PhysRevA.35.3313} {\bibfield  {journal}
  {\bibinfo  {journal} {Phys. Rev. A}\ }\textbf {\bibinfo {volume} {35}},\
  \bibinfo {pages} {3313} (\bibinfo {year} {1987})}\BibitemShut {NoStop}%
\bibitem [{\citenamefont {Braak}(2011)}]{Braak:2011}%
  \BibitemOpen
  \bibfield  {author} {\bibinfo {author} {\bibfnamefont {D.}~\bibnamefont
  {Braak}},\ }\href {\doibase 10.1103/PhysRevLett.107.100401} {\bibfield
  {journal} {\bibinfo  {journal} {Phys. Rev. Lett.}\ }\textbf {\bibinfo
  {volume} {107}},\ \bibinfo {pages} {100401} (\bibinfo {year}
  {2011})}\BibitemShut {NoStop}%
\bibitem [{\citenamefont {Cohen‐Tannoudji}\ \emph {et~al.}(1998)\citenamefont
  {Cohen‐Tannoudji}, \citenamefont {Dupont‐Roc},\ and\ \citenamefont
  {Grynberg}}]{Cohen-Tannoudji:1998}%
  \BibitemOpen
  \bibfield  {author} {\bibinfo {author} {\bibfnamefont {C.}~\bibnamefont
  {Cohen‐Tannoudji}}, \bibinfo {author} {\bibfnamefont {J.}~\bibnamefont
  {Dupont‐Roc}}, \ and\ \bibinfo {author} {\bibfnamefont {G.}~\bibnamefont
  {Grynberg}},\ }\href@noop {} {\emph {\bibinfo {title} {Atom-Photon
  Interactions: Basic Process and Appilcations}}}\ (\bibinfo  {publisher}
  {Wiley-VCH},\ \bibinfo {address} {Weinheim},\ \bibinfo {year}
  {1998})\BibitemShut {NoStop}%
\bibitem [{\citenamefont {Schrieffer}\ and\ \citenamefont
  {Wolff}(1966)}]{Schrieffer:1966}%
  \BibitemOpen
  \bibfield  {author} {\bibinfo {author} {\bibfnamefont {J.~R.}\ \bibnamefont
  {Schrieffer}}\ and\ \bibinfo {author} {\bibfnamefont {P.~A.}\ \bibnamefont
  {Wolff}},\ }\href {\doibase 10.1103/PhysRev.149.491} {\bibfield  {journal}
  {\bibinfo  {journal} {Phys. Rev.}\ }\textbf {\bibinfo {volume} {149}},\
  \bibinfo {pages} {491} (\bibinfo {year} {1966})}\BibitemShut {NoStop}%
\bibitem [{\citenamefont {Bravyi}\ \emph {et~al.}(2011)\citenamefont {Bravyi},
  \citenamefont {DiVincenzo},\ and\ \citenamefont {Loss}}]{Bravyi:2011}%
  \BibitemOpen
  \bibfield  {author} {\bibinfo {author} {\bibfnamefont {S.}~\bibnamefont
  {Bravyi}}, \bibinfo {author} {\bibfnamefont {D.~P.}\ \bibnamefont
  {DiVincenzo}}, \ and\ \bibinfo {author} {\bibfnamefont {D.}~\bibnamefont
  {Loss}},\ }\href {\doibase 10.1016/j.aop.2011.06.004} {\bibfield  {journal}
  {\bibinfo  {journal} {Ann. of Phys.}\ }\textbf {\bibinfo {volume} {326}},\
  \bibinfo {pages} {2793} (\bibinfo {year} {2011})}\BibitemShut {NoStop}%
\bibitem [{\citenamefont {Jauslin}\ \emph {et~al.}(2000)\citenamefont
  {Jauslin}, \citenamefont {Gu\'erin},\ and\ \citenamefont
  {Thomas}}]{Jauslin:2000}%
  \BibitemOpen
  \bibfield  {author} {\bibinfo {author} {\bibfnamefont {H.~R.}\ \bibnamefont
  {Jauslin}}, \bibinfo {author} {\bibfnamefont {S.}~\bibnamefont {Gu\'erin}}, \
  and\ \bibinfo {author} {\bibfnamefont {S.}~\bibnamefont {Thomas}},\ }\href
  {\doibase 10.1016/S0378-4371(99)00540-3} {\bibfield  {journal} {\bibinfo
  {journal} {Physica A}\ }\textbf {\bibinfo {volume} {279}},\ \bibinfo {pages}
  {432} (\bibinfo {year} {2000})}\BibitemShut {NoStop}%
\bibitem [{\citenamefont {Klimov}\ and\ \citenamefont
  {Sanchez-Soto}(2000)}]{Klimov:2000}%
  \BibitemOpen
  \bibfield  {author} {\bibinfo {author} {\bibfnamefont {A.~B.}\ \bibnamefont
  {Klimov}}\ and\ \bibinfo {author} {\bibfnamefont {L.~L.}\ \bibnamefont
  {Sanchez-Soto}},\ }\href {\doibase 10.1103/PhysRevA.61.063802} {\bibfield
  {journal} {\bibinfo  {journal} {Phys. Rev. A}\ }\textbf {\bibinfo {volume}
  {61}},\ \bibinfo {pages} {063802} (\bibinfo {year} {2000})}\BibitemShut
  {NoStop}%
\bibitem [{\citenamefont {Klimov}\ \emph {et~al.}(2002)\citenamefont {Klimov},
  \citenamefont {Sanchez-Soto}, \citenamefont {Navarro},\ and\ \citenamefont
  {Yustas}}]{Klimov:2002a}%
  \BibitemOpen
  \bibfield  {author} {\bibinfo {author} {\bibfnamefont {A.~B.}\ \bibnamefont
  {Klimov}}, \bibinfo {author} {\bibfnamefont {L.~L.}\ \bibnamefont
  {Sanchez-Soto}}, \bibinfo {author} {\bibfnamefont {A.}~\bibnamefont
  {Navarro}}, \ and\ \bibinfo {author} {\bibfnamefont {E.~C.}\ \bibnamefont
  {Yustas}},\ }\href {\doibase 10.1080/09500340210134675} {\bibfield  {journal}
  {\bibinfo  {journal} {J. Mod. Opt.}\ }\textbf {\bibinfo {volume} {49}},\
  \bibinfo {pages} {2211} (\bibinfo {year} {2002})}\BibitemShut {NoStop}%
\bibitem [{\citenamefont {Klimov}\ and\ \citenamefont
  {Chumakov}(2009)}]{Klimov:2009}%
  \BibitemOpen
  \bibfield  {author} {\bibinfo {author} {\bibfnamefont {A.~B.}\ \bibnamefont
  {Klimov}}\ and\ \bibinfo {author} {\bibfnamefont {S.~M.}\ \bibnamefont
  {Chumakov}},\ }\href@noop {} {\emph {\bibinfo {title} {A Group-Theoretical
  Approach to Quantum Optics}}}\ (\bibinfo  {publisher} {Wiley-VCH},\ \bibinfo
  {address} {Weinheim},\ \bibinfo {year} {2009})\BibitemShut {NoStop}%
\bibitem [{\citenamefont {Hwang}\ \emph {et~al.}(2015)\citenamefont {Hwang},
  \citenamefont {Puebla},\ and\ \citenamefont {Plenio}}]{Hwang:2015}%
  \BibitemOpen
  \bibfield  {author} {\bibinfo {author} {\bibfnamefont {M.-J.}\ \bibnamefont
  {Hwang}}, \bibinfo {author} {\bibfnamefont {R.}~\bibnamefont {Puebla}}, \
  and\ \bibinfo {author} {\bibfnamefont {M.~B.}\ \bibnamefont {Plenio}},\
  }\href@noop {} {\bibfield  {journal} {\bibinfo  {journal} {Phys. Rev. Lett.}\
  }\textbf {\bibinfo {volume} {115}},\ \bibinfo {pages} {180404} (\bibinfo
  {year} {2015})}\BibitemShut {NoStop}%
\bibitem [{\citenamefont {Zhang}\ \emph {et~al.}(2019)\citenamefont {Zhang},
  \citenamefont {Mao}, \citenamefont {Xu}, \citenamefont {Zhang}, \citenamefont
  {You}, \citenamefont {Liu},\ and\ \citenamefont {Luo}}]{Zhang:2019}%
  \BibitemOpen
  \bibfield  {author} {\bibinfo {author} {\bibfnamefont {Y.}~\bibnamefont
  {Zhang}}, \bibinfo {author} {\bibfnamefont {B.-B.}\ \bibnamefont {Mao}},
  \bibinfo {author} {\bibfnamefont {D.}~\bibnamefont {Xu}}, \bibinfo {author}
  {\bibfnamefont {Y.-Y.}\ \bibnamefont {Zhang}}, \bibinfo {author}
  {\bibfnamefont {W.-L.}\ \bibnamefont {You}}, \bibinfo {author} {\bibfnamefont
  {M.}~\bibnamefont {Liu}}, \ and\ \bibinfo {author} {\bibfnamefont {H.-G.}\
  \bibnamefont {Luo}},\ }\href {https://arxiv.org/abs/1910.13043} {\enquote
  {\bibinfo {title} {Quantum phase transitions and critical behaviors in the
  two-mode three-level quantum rabi model},}\ } (\bibinfo {year} {2019}),\
  \Eprint {http://arxiv.org/abs/1910.13043} {arXiv:1910.13043 [quant-ph]}
  \BibitemShut {NoStop}%
\bibitem [{\citenamefont {Amniat-Talab}\ \emph {et~al.}(2005)\citenamefont
  {Amniat-Talab}, \citenamefont {Gu\'erin},\ and\ \citenamefont
  {Jauslin}}]{Amniat-Talab:2005}%
  \BibitemOpen
  \bibfield  {author} {\bibinfo {author} {\bibfnamefont {M.}~\bibnamefont
  {Amniat-Talab}}, \bibinfo {author} {\bibfnamefont {S.}~\bibnamefont
  {Gu\'erin}}, \ and\ \bibinfo {author} {\bibfnamefont {H.~R.}\ \bibnamefont
  {Jauslin}},\ }\href {\doibase 10.1063/1.1864252} {\bibfield  {journal}
  {\bibinfo  {journal} {J. Math. Phys.}\ }\textbf {\bibinfo {volume} {46}},\
  \bibinfo {pages} {042311} (\bibinfo {year} {2005})}\BibitemShut {NoStop}%
\bibitem [{\citenamefont {Bloch}\ and\ \citenamefont
  {Siegert}(1940)}]{Bloch:1940}%
  \BibitemOpen
  \bibfield  {author} {\bibinfo {author} {\bibfnamefont {F.}~\bibnamefont
  {Bloch}}\ and\ \bibinfo {author} {\bibfnamefont {A.}~\bibnamefont
  {Siegert}},\ }\href {\doibase 10.1103/PhysRev.57.522} {\bibfield  {journal}
  {\bibinfo  {journal} {Phys. Rev.}\ }\textbf {\bibinfo {volume} {57}},\
  \bibinfo {pages} {522} (\bibinfo {year} {1940})}\BibitemShut {NoStop}%
\bibitem [{\citenamefont {Irish}\ \emph {et~al.}(2005)\citenamefont {Irish},
  \citenamefont {Gea-Banacloche}, \citenamefont {Martin},\ and\ \citenamefont
  {Schwab}}]{Irish:2005}%
  \BibitemOpen
  \bibfield  {author} {\bibinfo {author} {\bibfnamefont {E.~K.}\ \bibnamefont
  {Irish}}, \bibinfo {author} {\bibfnamefont {J.}~\bibnamefont
  {Gea-Banacloche}}, \bibinfo {author} {\bibfnamefont {I.}~\bibnamefont
  {Martin}}, \ and\ \bibinfo {author} {\bibfnamefont {K.~C.}\ \bibnamefont
  {Schwab}},\ }\href {\doibase 10.1103/PhysRevB.72.195410} {\bibfield
  {journal} {\bibinfo  {journal} {Phys. Rev. B}\ }\textbf {\bibinfo {volume}
  {72}},\ \bibinfo {pages} {195410} (\bibinfo {year} {2005})}\BibitemShut
  {NoStop}%
\bibitem [{\citenamefont {Irish}(2007)}]{Irish:2007}%
  \BibitemOpen
  \bibfield  {author} {\bibinfo {author} {\bibfnamefont {E.}~\bibnamefont
  {Irish}},\ }\href@noop {} {\bibfield  {journal} {\bibinfo  {journal} {Phys.
  Rev. Lett.}\ }\textbf {\bibinfo {volume} {99}},\ \bibinfo {pages} {173601}
  (\bibinfo {year} {2007})}\BibitemShut {NoStop}%
\bibitem [{\citenamefont {Rossatto}\ \emph {et~al.}(2017)\citenamefont
  {Rossatto}, \citenamefont {Villas-B\^oas}, \citenamefont {Sanz},\ and\
  \citenamefont {Solano}}]{Rossatto:2017}%
  \BibitemOpen
  \bibfield  {author} {\bibinfo {author} {\bibfnamefont {D.~Z.}\ \bibnamefont
  {Rossatto}}, \bibinfo {author} {\bibfnamefont {C.~J.}\ \bibnamefont
  {Villas-B\^oas}}, \bibinfo {author} {\bibfnamefont {M.}~\bibnamefont {Sanz}},
  \ and\ \bibinfo {author} {\bibfnamefont {E.}~\bibnamefont {Solano}},\ }\href
  {\doibase 10.1103/PhysRevA.96.013849} {\bibfield  {journal} {\bibinfo
  {journal} {Phys. Rev. A}\ }\textbf {\bibinfo {volume} {96}},\ \bibinfo
  {pages} {013849} (\bibinfo {year} {2017})}\BibitemShut {NoStop}%
\bibitem [{\citenamefont {Judd}(1979)}]{Judd:1979}%
  \BibitemOpen
  \bibfield  {author} {\bibinfo {author} {\bibfnamefont {B.~R.}\ \bibnamefont
  {Judd}},\ }\href {\doibase 10.1088/0022-3719/12/9/010} {\bibfield  {journal}
  {\bibinfo  {journal} {J. Phys. C : Solid State Phys.}\ }\textbf {\bibinfo
  {volume} {12}},\ \bibinfo {pages} {1685} (\bibinfo {year}
  {1979})}\BibitemShut {NoStop}%
\bibitem [{\citenamefont {Klimov}\ \emph {et~al.}(2003)\citenamefont {Klimov},
  \citenamefont {Sainz},\ and\ \citenamefont {Chumakov}}]{Klimov:2003}%
  \BibitemOpen
  \bibfield  {author} {\bibinfo {author} {\bibfnamefont {A.~B.}\ \bibnamefont
  {Klimov}}, \bibinfo {author} {\bibfnamefont {I.}~\bibnamefont {Sainz}}, \
  and\ \bibinfo {author} {\bibfnamefont {S.~M.}\ \bibnamefont {Chumakov}},\
  }\href {\doibase 10.1103/PhysRevA.68.063811} {\bibfield  {journal} {\bibinfo
  {journal} {Phys. Rev. A}\ }\textbf {\bibinfo {volume} {68}},\ \bibinfo
  {pages} {063811} (\bibinfo {year} {2003})}\BibitemShut {NoStop}%
\bibitem [{\citenamefont {Ma}\ and\ \citenamefont {Law}(2015)}]{Ma:2015}%
  \BibitemOpen
  \bibfield  {author} {\bibinfo {author} {\bibfnamefont {K.~K.~W.}\
  \bibnamefont {Ma}}\ and\ \bibinfo {author} {\bibfnamefont {C.~K.}\
  \bibnamefont {Law}},\ }\href {\doibase 10.1103/PhysRevA.92.023842} {\bibfield
   {journal} {\bibinfo  {journal} {Phys. Rev. A}\ }\textbf {\bibinfo {volume}
  {92}},\ \bibinfo {pages} {023842} (\bibinfo {year} {2015})}\BibitemShut
  {NoStop}%
\bibitem [{\citenamefont {Garziano}\ \emph {et~al.}(2015)\citenamefont
  {Garziano}, \citenamefont {Stassi}, \citenamefont {Macr\`{\i}}, \citenamefont
  {Kockum}, \citenamefont {Savasta},\ and\ \citenamefont
  {Nori}}]{Garziano:2015}%
  \BibitemOpen
  \bibfield  {author} {\bibinfo {author} {\bibfnamefont {L.}~\bibnamefont
  {Garziano}}, \bibinfo {author} {\bibfnamefont {R.}~\bibnamefont {Stassi}},
  \bibinfo {author} {\bibfnamefont {V.}~\bibnamefont {Macr\`{\i}}}, \bibinfo
  {author} {\bibfnamefont {A.~F.}\ \bibnamefont {Kockum}}, \bibinfo {author}
  {\bibfnamefont {S.}~\bibnamefont {Savasta}}, \ and\ \bibinfo {author}
  {\bibfnamefont {F.}~\bibnamefont {Nori}},\ }\href {\doibase
  10.1103/PhysRevA.92.063830} {\bibfield  {journal} {\bibinfo  {journal} {Phys.
  Rev. A}\ }\textbf {\bibinfo {volume} {92}},\ \bibinfo {pages} {063830}
  (\bibinfo {year} {2015})}\BibitemShut {NoStop}%
\bibitem [{\citenamefont {Garziano}\ \emph {et~al.}(2016)\citenamefont
  {Garziano}, \citenamefont {Macr\`{\i}}, \citenamefont {Stassi}, \citenamefont
  {Di~Stefano}, \citenamefont {Nori},\ and\ \citenamefont
  {Savasta}}]{Garziano:2016}%
  \BibitemOpen
  \bibfield  {author} {\bibinfo {author} {\bibfnamefont {L.}~\bibnamefont
  {Garziano}}, \bibinfo {author} {\bibfnamefont {V.}~\bibnamefont
  {Macr\`{\i}}}, \bibinfo {author} {\bibfnamefont {R.}~\bibnamefont {Stassi}},
  \bibinfo {author} {\bibfnamefont {O.}~\bibnamefont {Di~Stefano}}, \bibinfo
  {author} {\bibfnamefont {F.}~\bibnamefont {Nori}}, \ and\ \bibinfo {author}
  {\bibfnamefont {S.}~\bibnamefont {Savasta}},\ }\href {\doibase
  10.1103/PhysRevLett.117.043601} {\bibfield  {journal} {\bibinfo  {journal}
  {Phys. Rev. Lett.}\ }\textbf {\bibinfo {volume} {117}},\ \bibinfo {pages}
  {043601} (\bibinfo {year} {2016})}\BibitemShut {NoStop}%
\bibitem [{\citenamefont {Frisk~Kockum}\ \emph {et~al.}(2017)\citenamefont
  {Frisk~Kockum}, \citenamefont {Miranowicz}, \citenamefont {Macr\`{\i}},
  \citenamefont {Savasta},\ and\ \citenamefont {Nori}}]{Frisk-Kockum:2017}%
  \BibitemOpen
  \bibfield  {author} {\bibinfo {author} {\bibfnamefont {A.}~\bibnamefont
  {Frisk~Kockum}}, \bibinfo {author} {\bibfnamefont {A.}~\bibnamefont
  {Miranowicz}}, \bibinfo {author} {\bibfnamefont {V.}~\bibnamefont
  {Macr\`{\i}}}, \bibinfo {author} {\bibfnamefont {S.}~\bibnamefont {Savasta}},
  \ and\ \bibinfo {author} {\bibfnamefont {F.}~\bibnamefont {Nori}},\ }\href
  {\doibase 10.1103/PhysRevA.95.063849} {\bibfield  {journal} {\bibinfo
  {journal} {Phys. Rev. A}\ }\textbf {\bibinfo {volume} {95}},\ \bibinfo
  {pages} {063849} (\bibinfo {year} {2017})}\BibitemShut {NoStop}%
\bibitem [{\citenamefont {Stassi}\ \emph {et~al.}(2017)\citenamefont {Stassi},
  \citenamefont {Macr\`{\i}}, \citenamefont {Kockum}, \citenamefont
  {Di~Stefano}, \citenamefont {Miranowicz}, \citenamefont {Savasta},\ and\
  \citenamefont {Nori}}]{Stassi:2017}%
  \BibitemOpen
  \bibfield  {author} {\bibinfo {author} {\bibfnamefont {R.}~\bibnamefont
  {Stassi}}, \bibinfo {author} {\bibfnamefont {V.}~\bibnamefont {Macr\`{\i}}},
  \bibinfo {author} {\bibfnamefont {A.~F.}\ \bibnamefont {Kockum}}, \bibinfo
  {author} {\bibfnamefont {O.}~\bibnamefont {Di~Stefano}}, \bibinfo {author}
  {\bibfnamefont {A.}~\bibnamefont {Miranowicz}}, \bibinfo {author}
  {\bibfnamefont {S.}~\bibnamefont {Savasta}}, \ and\ \bibinfo {author}
  {\bibfnamefont {F.}~\bibnamefont {Nori}},\ }\href {\doibase
  10.1103/PhysRevA.96.023818} {\bibfield  {journal} {\bibinfo  {journal} {Phys.
  Rev. A}\ }\textbf {\bibinfo {volume} {96}},\ \bibinfo {pages} {023818}
  (\bibinfo {year} {2017})}\BibitemShut {NoStop}%
\bibitem [{\citenamefont {Mu{\~n}oz}\ \emph {et~al.}(2019)\citenamefont
  {Mu{\~n}oz}, \citenamefont {Kockum}, \citenamefont {Miranowicz},\ and\
  \citenamefont {Nori}}]{Munoz:2019}%
  \BibitemOpen
  \bibfield  {author} {\bibinfo {author} {\bibfnamefont {C.~S.}\ \bibnamefont
  {Mu{\~n}oz}}, \bibinfo {author} {\bibfnamefont {A.~F.}\ \bibnamefont
  {Kockum}}, \bibinfo {author} {\bibfnamefont {A.}~\bibnamefont {Miranowicz}},
  \ and\ \bibinfo {author} {\bibfnamefont {F.}~\bibnamefont {Nori}},\
  }\href@noop {} {\enquote {\bibinfo {title} {Ultrastrong-coupling effects
  induced by a single classical drive in jaynes-cummings-type systems},}\ }
  (\bibinfo {year} {2019}),\ \Eprint {http://arxiv.org/abs/1910.12875}
  {arXiv:1910.12875 [quant-ph]} \BibitemShut {NoStop}%
\bibitem [{\citenamefont {Feranchuk}\ \emph {et~al.}(1996)\citenamefont
  {Feranchuk}, \citenamefont {Komarov},\ and\ \citenamefont
  {Ulyanenkov}}]{Feranchuk:1996}%
  \BibitemOpen
  \bibfield  {author} {\bibinfo {author} {\bibfnamefont {I.~D.}\ \bibnamefont
  {Feranchuk}}, \bibinfo {author} {\bibfnamefont {L.~I.}\ \bibnamefont
  {Komarov}}, \ and\ \bibinfo {author} {\bibfnamefont {A.~P.}\ \bibnamefont
  {Ulyanenkov}},\ }\href {\doibase 10.1088/0305-4470/29/14/026} {\bibfield
  {journal} {\bibinfo  {journal} {J. Phys. A: Math. Gen.}\ }\textbf {\bibinfo
  {volume} {29}},\ \bibinfo {pages} {4035} (\bibinfo {year}
  {1996})}\BibitemShut {NoStop}%
\bibitem [{\citenamefont {Zhang}\ and\ \citenamefont
  {Chen}(2015)}]{Zhang:2015}%
  \BibitemOpen
  \bibfield  {author} {\bibinfo {author} {\bibfnamefont {Y.-Y.}\ \bibnamefont
  {Zhang}}\ and\ \bibinfo {author} {\bibfnamefont {Q.-H.}\ \bibnamefont
  {Chen}},\ }\href {\doibase 10.1103/PhysRevA.91.013814} {\bibfield  {journal}
  {\bibinfo  {journal} {Phys. Rev. A}\ }\textbf {\bibinfo {volume} {91}},\
  \bibinfo {pages} {013814} (\bibinfo {year} {2015})}\BibitemShut {NoStop}%
\bibitem [{\citenamefont {Zhang}\ \emph {et~al.}(2013)\citenamefont {Zhang},
  \citenamefont {Chen},\ and\ \citenamefont {Zhao}}]{Zhang:2013}%
  \BibitemOpen
  \bibfield  {author} {\bibinfo {author} {\bibfnamefont {Y.-Y.}\ \bibnamefont
  {Zhang}}, \bibinfo {author} {\bibfnamefont {Q.-H.}\ \bibnamefont {Chen}}, \
  and\ \bibinfo {author} {\bibfnamefont {Y.}~\bibnamefont {Zhao}},\ }\href
  {\doibase 10.1103/PhysRevA.87.033827} {\bibfield  {journal} {\bibinfo
  {journal} {Phys. Rev. A}\ }\textbf {\bibinfo {volume} {87}},\ \bibinfo
  {pages} {033827} (\bibinfo {year} {2013})}\BibitemShut {NoStop}%
\bibitem [{\citenamefont {Albert}\ \emph {et~al.}(2011)\citenamefont {Albert},
  \citenamefont {Scholes},\ and\ \citenamefont {Brumer}}]{Albert:2011}%
  \BibitemOpen
  \bibfield  {author} {\bibinfo {author} {\bibfnamefont {V.~V.}\ \bibnamefont
  {Albert}}, \bibinfo {author} {\bibfnamefont {G.~D.}\ \bibnamefont {Scholes}},
  \ and\ \bibinfo {author} {\bibfnamefont {P.}~\bibnamefont {Brumer}},\ }\href
  {\doibase 10.1103/PhysRevA.84.042110} {\bibfield  {journal} {\bibinfo
  {journal} {Phys. Rev. A}\ }\textbf {\bibinfo {volume} {84}},\ \bibinfo
  {pages} {042110} (\bibinfo {year} {2011})}\BibitemShut {NoStop}%
\bibitem [{\citenamefont {Zhang}(2016)}]{Zhang:2016}%
  \BibitemOpen
  \bibfield  {author} {\bibinfo {author} {\bibfnamefont {Y.-Y.}\ \bibnamefont
  {Zhang}},\ }\href {\doibase 10.1103/PhysRevA.94.063824} {\bibfield  {journal}
  {\bibinfo  {journal} {Phys. Rev. A}\ }\textbf {\bibinfo {volume} {94}},\
  \bibinfo {pages} {063824} (\bibinfo {year} {2016})}\BibitemShut {NoStop}%
\bibitem [{\citenamefont {Rivera}\ \emph {et~al.}(2019)\citenamefont {Rivera},
  \citenamefont {Flick},\ and\ \citenamefont {Narang}}]{Rivera:2019}%
  \BibitemOpen
  \bibfield  {author} {\bibinfo {author} {\bibfnamefont {N.}~\bibnamefont
  {Rivera}}, \bibinfo {author} {\bibfnamefont {J.}~\bibnamefont {Flick}}, \
  and\ \bibinfo {author} {\bibfnamefont {P.}~\bibnamefont {Narang}},\ }\href
  {\doibase 10.1103/PhysRevLett.122.193603} {\bibfield  {journal} {\bibinfo
  {journal} {Phys. Rev. Lett.}\ }\textbf {\bibinfo {volume} {122}},\ \bibinfo
  {pages} {193603} (\bibinfo {year} {2019})}\BibitemShut {NoStop}%
\bibitem [{\citenamefont {Bera}\ \emph {et~al.}(2014)\citenamefont {Bera},
  \citenamefont {Nazir}, \citenamefont {Chin}, \citenamefont {Baranger},\ and\
  \citenamefont {Florens}}]{Bera:2014}%
  \BibitemOpen
  \bibfield  {author} {\bibinfo {author} {\bibfnamefont {S.}~\bibnamefont
  {Bera}}, \bibinfo {author} {\bibfnamefont {A.}~\bibnamefont {Nazir}},
  \bibinfo {author} {\bibfnamefont {A.~W.}\ \bibnamefont {Chin}}, \bibinfo
  {author} {\bibfnamefont {H.~U.}\ \bibnamefont {Baranger}}, \ and\ \bibinfo
  {author} {\bibfnamefont {S.}~\bibnamefont {Florens}},\ }\href {\doibase
  10.1103/PhysRevB.90.075110} {\bibfield  {journal} {\bibinfo  {journal} {Phys.
  Rev. B}\ }\textbf {\bibinfo {volume} {90}},\ \bibinfo {pages} {075110}
  (\bibinfo {year} {2014})}\BibitemShut {NoStop}%
\bibitem [{\citenamefont {Hwang}\ and\ \citenamefont
  {Choi}(2010)}]{Hwang:2010}%
  \BibitemOpen
  \bibfield  {author} {\bibinfo {author} {\bibfnamefont {M.-J.}\ \bibnamefont
  {Hwang}}\ and\ \bibinfo {author} {\bibfnamefont {M.-S.}\ \bibnamefont
  {Choi}},\ }\href {\doibase 10.1103/PhysRevA.82.025802} {\bibfield  {journal}
  {\bibinfo  {journal} {Phys. Rev. A}\ }\textbf {\bibinfo {volume} {82}},\
  \bibinfo {pages} {025802} (\bibinfo {year} {2010})}\BibitemShut {NoStop}%
\bibitem [{\citenamefont {Ying}\ \emph {et~al.}(2015)\citenamefont {Ying},
  \citenamefont {Liu}, \citenamefont {Luo}, \citenamefont {Lin},\ and\
  \citenamefont {You}}]{Ying:2015}%
  \BibitemOpen
  \bibfield  {author} {\bibinfo {author} {\bibfnamefont {Z.-J.}\ \bibnamefont
  {Ying}}, \bibinfo {author} {\bibfnamefont {M.}~\bibnamefont {Liu}}, \bibinfo
  {author} {\bibfnamefont {H.-G.}\ \bibnamefont {Luo}}, \bibinfo {author}
  {\bibfnamefont {H.-Q.}\ \bibnamefont {Lin}}, \ and\ \bibinfo {author}
  {\bibfnamefont {J.~Q.}\ \bibnamefont {You}},\ }\href {\doibase
  10.1103/PhysRevA.92.053823} {\bibfield  {journal} {\bibinfo  {journal} {Phys.
  Rev. A}\ }\textbf {\bibinfo {volume} {92}},\ \bibinfo {pages} {053823}
  (\bibinfo {year} {2015})}\BibitemShut {NoStop}%
\bibitem [{\citenamefont {Cong}\ \emph {et~al.}(2017)\citenamefont {Cong},
  \citenamefont {Sun}, \citenamefont {Liu}, \citenamefont {Ying},\ and\
  \citenamefont {Luo}}]{Cong:2017}%
  \BibitemOpen
  \bibfield  {author} {\bibinfo {author} {\bibfnamefont {L.}~\bibnamefont
  {Cong}}, \bibinfo {author} {\bibfnamefont {X.-M.}\ \bibnamefont {Sun}},
  \bibinfo {author} {\bibfnamefont {M.}~\bibnamefont {Liu}}, \bibinfo {author}
  {\bibfnamefont {Z.-J.}\ \bibnamefont {Ying}}, \ and\ \bibinfo {author}
  {\bibfnamefont {H.-G.}\ \bibnamefont {Luo}},\ }\href {\doibase
  10.1103/PhysRevA.95.063803} {\bibfield  {journal} {\bibinfo  {journal} {Phys.
  Rev. A}\ }\textbf {\bibinfo {volume} {95}},\ \bibinfo {pages} {063803}
  (\bibinfo {year} {2017})}\BibitemShut {NoStop}%
\bibitem [{\citenamefont {Mao}\ \emph {et~al.}(2019)\citenamefont {Mao},
  \citenamefont {Li}, \citenamefont {Wang}, \citenamefont {You}, \citenamefont
  {Wu}, \citenamefont {Liu},\ and\ \citenamefont {Luo}}]{Mao:2019}%
  \BibitemOpen
  \bibfield  {author} {\bibinfo {author} {\bibfnamefont {B.-B.}\ \bibnamefont
  {Mao}}, \bibinfo {author} {\bibfnamefont {L.}~\bibnamefont {Li}}, \bibinfo
  {author} {\bibfnamefont {Y.}~\bibnamefont {Wang}}, \bibinfo {author}
  {\bibfnamefont {W.-L.}\ \bibnamefont {You}}, \bibinfo {author} {\bibfnamefont
  {W.}~\bibnamefont {Wu}}, \bibinfo {author} {\bibfnamefont {M.}~\bibnamefont
  {Liu}}, \ and\ \bibinfo {author} {\bibfnamefont {H.-G.}\ \bibnamefont
  {Luo}},\ }\href {\doibase 10.1103/PhysRevA.99.033834} {\bibfield  {journal}
  {\bibinfo  {journal} {Phys. Rev. A}\ }\textbf {\bibinfo {volume} {99}},\
  \bibinfo {pages} {033834} (\bibinfo {year} {2019})}\BibitemShut {NoStop}%
\bibitem [{\citenamefont {Cong}\ \emph {et~al.}(2019)\citenamefont {Cong},
  \citenamefont {Sun}, \citenamefont {Liu}, \citenamefont {Ying},\ and\
  \citenamefont {Luo}}]{Cong:2019}%
  \BibitemOpen
  \bibfield  {author} {\bibinfo {author} {\bibfnamefont {L.}~\bibnamefont
  {Cong}}, \bibinfo {author} {\bibfnamefont {X.-M.}\ \bibnamefont {Sun}},
  \bibinfo {author} {\bibfnamefont {M.}~\bibnamefont {Liu}}, \bibinfo {author}
  {\bibfnamefont {Z.-J.}\ \bibnamefont {Ying}}, \ and\ \bibinfo {author}
  {\bibfnamefont {H.-G.}\ \bibnamefont {Luo}},\ }\href {\doibase
  10.1103/PhysRevA.99.013815} {\bibfield  {journal} {\bibinfo  {journal} {Phys.
  Rev. A}\ }\textbf {\bibinfo {volume} {99}},\ \bibinfo {pages} {013815}
  (\bibinfo {year} {2019})}\BibitemShut {NoStop}%
\bibitem [{\citenamefont {Tokatly}(2013)}]{Tokatly:2013}%
  \BibitemOpen
  \bibfield  {author} {\bibinfo {author} {\bibfnamefont {I.~V.}\ \bibnamefont
  {Tokatly}},\ }\href {\doibase 10.1103/PhysRevLett.110.233001} {\bibfield
  {journal} {\bibinfo  {journal} {Phys. Rev. Lett.}\ }\textbf {\bibinfo
  {volume} {110}},\ \bibinfo {pages} {233001} (\bibinfo {year}
  {2013})}\BibitemShut {NoStop}%
\bibitem [{\citenamefont {Flick}\ and\ \citenamefont
  {Narang}(2018)}]{Flick:2018}%
  \BibitemOpen
  \bibfield  {author} {\bibinfo {author} {\bibfnamefont {J.}~\bibnamefont
  {Flick}}\ and\ \bibinfo {author} {\bibfnamefont {P.}~\bibnamefont {Narang}},\
  }\href {\doibase 10.1103/PhysRevLett.121.113002} {\bibfield  {journal}
  {\bibinfo  {journal} {Phys. Rev. Lett.}\ }\textbf {\bibinfo {volume} {121}},\
  \bibinfo {pages} {113002} (\bibinfo {year} {2018})}\BibitemShut {NoStop}%
\bibitem [{\citenamefont {Di~Paolo}\ \emph {et~al.}(2019)\citenamefont
  {Di~Paolo}, \citenamefont {Barkoutsos}, \citenamefont {Tavernelli},\ and\
  \citenamefont {Blais}}]{DiPaolo:2019}%
  \BibitemOpen
  \bibfield  {author} {\bibinfo {author} {\bibfnamefont {A.}~\bibnamefont
  {Di~Paolo}}, \bibinfo {author} {\bibfnamefont {P.~K.}\ \bibnamefont
  {Barkoutsos}}, \bibinfo {author} {\bibfnamefont {I.}~\bibnamefont
  {Tavernelli}}, \ and\ \bibinfo {author} {\bibfnamefont {A.}~\bibnamefont
  {Blais}},\ }\href {https://arxiv.org/abs/1909.08640} {\bibfield  {journal}
  {\bibinfo  {journal} {arXiv preprint arXiv:1909.08640}\ } (\bibinfo {year}
  {2019})}\BibitemShut {NoStop}%
\bibitem [{\citenamefont {Braak}\ \emph {et~al.}(2016)\citenamefont {Braak},
  \citenamefont {Chen}, \citenamefont {Batchelor},\ and\ \citenamefont
  {Solano}}]{Braak:2016}%
  \BibitemOpen
  \bibfield  {author} {\bibinfo {author} {\bibfnamefont {D.}~\bibnamefont
  {Braak}}, \bibinfo {author} {\bibfnamefont {Q.-H.}\ \bibnamefont {Chen}},
  \bibinfo {author} {\bibfnamefont {M.~T.}\ \bibnamefont {Batchelor}}, \ and\
  \bibinfo {author} {\bibfnamefont {E.}~\bibnamefont {Solano}},\ }\href
  {\doibase 10.1088/1751-8113/49/30/300301} {\bibfield  {journal} {\bibinfo
  {journal} {J. Phys. A: Math. Theor.}\ }\textbf {\bibinfo {volume} {49}},\
  \bibinfo {pages} {300301} (\bibinfo {year} {2016})}\BibitemShut {NoStop}%
\bibitem [{\citenamefont {Bargmann}(1961)}]{Bargmann:1961}%
  \BibitemOpen
  \bibfield  {author} {\bibinfo {author} {\bibfnamefont {V.}~\bibnamefont
  {Bargmann}},\ }\href {\doibase 10.1002/cpa.3160140303} {\bibfield  {journal}
  {\bibinfo  {journal} {Commun. Pure Appl. Math.}\ }\textbf {\bibinfo {volume}
  {14}},\ \bibinfo {pages} {187} (\bibinfo {year} {1961})}\BibitemShut
  {NoStop}%
\bibitem [{\citenamefont {Schweber}(1967)}]{Schweber:1967}%
  \BibitemOpen
  \bibfield  {author} {\bibinfo {author} {\bibfnamefont {S.}~\bibnamefont
  {Schweber}},\ }\href {\doibase 10.1016/0003-4916(67)90234-5} {\bibfield
  {journal} {\bibinfo  {journal} {Ann. of Phys.}\ }\textbf {\bibinfo {volume}
  {41}},\ \bibinfo {pages} {205} (\bibinfo {year} {1967})}\BibitemShut
  {NoStop}%
\bibitem [{\citenamefont {Reik}\ and\ \citenamefont
  {Doucha}(1986)}]{Reik:1986}%
  \BibitemOpen
  \bibfield  {author} {\bibinfo {author} {\bibfnamefont {H.~G.}\ \bibnamefont
  {Reik}}\ and\ \bibinfo {author} {\bibfnamefont {M.}~\bibnamefont {Doucha}},\
  }\href {\doibase 10.1103/PhysRevLett.57.787} {\bibfield  {journal} {\bibinfo
  {journal} {Phys. Rev. Lett.}\ }\textbf {\bibinfo {volume} {57}},\ \bibinfo
  {pages} {787} (\bibinfo {year} {1986})}\BibitemShut {NoStop}%
\bibitem [{\citenamefont {Ku\'s}\ and\ \citenamefont
  {Lewenstein}(1986)}]{Kus:1986}%
  \BibitemOpen
  \bibfield  {author} {\bibinfo {author} {\bibfnamefont {M.}~\bibnamefont
  {Ku\'s}}\ and\ \bibinfo {author} {\bibfnamefont {M.}~\bibnamefont
  {Lewenstein}},\ }\href {\doibase 10.1088/0305-4470/19/2/023} {\bibfield
  {journal} {\bibinfo  {journal} {J. Phys. A: Math. Gen.}\ }\textbf {\bibinfo
  {volume} {19}},\ \bibinfo {pages} {305} (\bibinfo {year} {1986})}\BibitemShut
  {NoStop}%
\bibitem [{\citenamefont {Maciejewski}\ \emph {et~al.}(2014)\citenamefont
  {Maciejewski}, \citenamefont {Przybylska},\ and\ \citenamefont
  {Stachowiak}}]{Maciejewski:2014}%
  \BibitemOpen
  \bibfield  {author} {\bibinfo {author} {\bibfnamefont {A.~J.}\ \bibnamefont
  {Maciejewski}}, \bibinfo {author} {\bibfnamefont {M.}~\bibnamefont
  {Przybylska}}, \ and\ \bibinfo {author} {\bibfnamefont {T.}~\bibnamefont
  {Stachowiak}},\ }\href {\doibase 10.1016/j.physleta.2013.10.032} {\bibfield
  {journal} {\bibinfo  {journal} {Phys. Lett. A}\ }\textbf {\bibinfo {volume}
  {378}},\ \bibinfo {pages} {16} (\bibinfo {year} {2014})}\BibitemShut
  {NoStop}%
\bibitem [{\citenamefont {Ronveaux}(1995)}]{Ronveaux:1995}%
  \BibitemOpen
  \bibfield  {author} {\bibinfo {author} {\bibfnamefont {A.}~\bibnamefont
  {Ronveaux}},\ }\href@noop {} {\emph {\bibinfo {title} {Heun's Differential
  Equations}}}\ (\bibinfo  {publisher} {Oxford University Press},\ \bibinfo
  {address} {Oxford},\ \bibinfo {year} {1995})\BibitemShut {NoStop}%
\bibitem [{\citenamefont {Slavyanov}\ and\ \citenamefont
  {Lay}(2000)}]{Slavyanov:2000}%
  \BibitemOpen
  \bibfield  {author} {\bibinfo {author} {\bibfnamefont {S.~Y.}\ \bibnamefont
  {Slavyanov}}\ and\ \bibinfo {author} {\bibfnamefont {W.}~\bibnamefont
  {Lay}},\ }\href@noop {} {\emph {\bibinfo {title} {Special Functions}}}\
  (\bibinfo  {publisher} {Oxford University Press},\ \bibinfo {address}
  {Oxford},\ \bibinfo {year} {2000})\BibitemShut {NoStop}%
\bibitem [{\citenamefont {Xie}\ \emph {et~al.}(2014)\citenamefont {Xie},
  \citenamefont {Cui}, \citenamefont {Cao}, \citenamefont {Amico},\ and\
  \citenamefont {Fan}}]{Xie:2014}%
  \BibitemOpen
  \bibfield  {author} {\bibinfo {author} {\bibfnamefont {Q.-T.}\ \bibnamefont
  {Xie}}, \bibinfo {author} {\bibfnamefont {S.}~\bibnamefont {Cui}}, \bibinfo
  {author} {\bibfnamefont {J.-P.}\ \bibnamefont {Cao}}, \bibinfo {author}
  {\bibfnamefont {L.}~\bibnamefont {Amico}}, \ and\ \bibinfo {author}
  {\bibfnamefont {H.}~\bibnamefont {Fan}},\ }\href {\doibase
  10.1103/PhysRevX.4.021046} {\bibfield  {journal} {\bibinfo  {journal} {Phys.
  Rev. X}\ }\textbf {\bibinfo {volume} {4}},\ \bibinfo {pages} {021046}
  (\bibinfo {year} {2014})}\BibitemShut {NoStop}%
\bibitem [{\citenamefont {Tomka}\ \emph {et~al.}(2014)\citenamefont {Tomka},
  \citenamefont {El~Araby}, \citenamefont {Pletyukhov},\ and\ \citenamefont
  {Gritsev}}]{Tomka:2014}%
  \BibitemOpen
  \bibfield  {author} {\bibinfo {author} {\bibfnamefont {M.}~\bibnamefont
  {Tomka}}, \bibinfo {author} {\bibfnamefont {O.}~\bibnamefont {El~Araby}},
  \bibinfo {author} {\bibfnamefont {M.}~\bibnamefont {Pletyukhov}}, \ and\
  \bibinfo {author} {\bibfnamefont {V.}~\bibnamefont {Gritsev}},\ }\href
  {\doibase 10.1103/PhysRevA.90.063839} {\bibfield  {journal} {\bibinfo
  {journal} {Phys. Rev. A}\ }\textbf {\bibinfo {volume} {90}},\ \bibinfo
  {pages} {063839} (\bibinfo {year} {2014})}\BibitemShut {NoStop}%
\bibitem [{\citenamefont {Xie}\ \emph {et~al.}(2017)\citenamefont {Xie},
  \citenamefont {Zhong}, \citenamefont {Batchelor},\ and\ \citenamefont
  {Lee}}]{Xie:2017}%
  \BibitemOpen
  \bibfield  {author} {\bibinfo {author} {\bibfnamefont {Q.}~\bibnamefont
  {Xie}}, \bibinfo {author} {\bibfnamefont {H.}~\bibnamefont {Zhong}}, \bibinfo
  {author} {\bibfnamefont {M.~T.}\ \bibnamefont {Batchelor}}, \ and\ \bibinfo
  {author} {\bibfnamefont {C.}~\bibnamefont {Lee}},\ }\href@noop {} {\bibfield
  {journal} {\bibinfo  {journal} {J. Phys. A: Math. Theor.}\ }\textbf {\bibinfo
  {volume} {50}},\ \bibinfo {pages} {113001} (\bibinfo {year}
  {2017})}\BibitemShut {NoStop}%
\bibitem [{\citenamefont {Chen}\ \emph {et~al.}(2008)\citenamefont {Chen},
  \citenamefont {Zhang}, \citenamefont {Liu},\ and\ \citenamefont
  {Wang}}]{Chen:2008}%
  \BibitemOpen
  \bibfield  {author} {\bibinfo {author} {\bibfnamefont {Q.-H.}\ \bibnamefont
  {Chen}}, \bibinfo {author} {\bibfnamefont {Y.-Y.}\ \bibnamefont {Zhang}},
  \bibinfo {author} {\bibfnamefont {T.}~\bibnamefont {Liu}}, \ and\ \bibinfo
  {author} {\bibfnamefont {K.-L.}\ \bibnamefont {Wang}},\ }\href {\doibase
  10.1103/PhysRevA.78.051801} {\bibfield  {journal} {\bibinfo  {journal} {Phys.
  Rev. A}\ }\textbf {\bibinfo {volume} {78}},\ \bibinfo {pages} {051801}
  (\bibinfo {year} {2008})}\BibitemShut {NoStop}%
\bibitem [{\citenamefont {Chen}\ \emph {et~al.}(2012)\citenamefont {Chen},
  \citenamefont {Wang}, \citenamefont {He}, \citenamefont {Liu},\ and\
  \citenamefont {Wang}}]{Chen:2012}%
  \BibitemOpen
  \bibfield  {author} {\bibinfo {author} {\bibfnamefont {Q.-H.}\ \bibnamefont
  {Chen}}, \bibinfo {author} {\bibfnamefont {C.}~\bibnamefont {Wang}}, \bibinfo
  {author} {\bibfnamefont {S.}~\bibnamefont {He}}, \bibinfo {author}
  {\bibfnamefont {T.}~\bibnamefont {Liu}}, \ and\ \bibinfo {author}
  {\bibfnamefont {K.-L.}\ \bibnamefont {Wang}},\ }\href {\doibase
  10.1103/PhysRevA.86.023822} {\bibfield  {journal} {\bibinfo  {journal} {Phys.
  Rev. A}\ }\textbf {\bibinfo {volume} {86}},\ \bibinfo {pages} {023822}
  (\bibinfo {year} {2012})}\BibitemShut {NoStop}%
\bibitem [{\citenamefont {Duan}\ \emph {et~al.}(2015)\citenamefont {Duan},
  \citenamefont {He}, \citenamefont {Braak},\ and\ \citenamefont
  {Chen}}]{Duan:2015}%
  \BibitemOpen
  \bibfield  {author} {\bibinfo {author} {\bibfnamefont {L.}~\bibnamefont
  {Duan}}, \bibinfo {author} {\bibfnamefont {S.}~\bibnamefont {He}}, \bibinfo
  {author} {\bibfnamefont {D.}~\bibnamefont {Braak}}, \ and\ \bibinfo {author}
  {\bibfnamefont {Q.-H.}\ \bibnamefont {Chen}},\ }\href {\doibase
  10.1209/0295-5075/112/34003} {\bibfield  {journal} {\bibinfo  {journal}
  {Europhys. Lett.}\ }\textbf {\bibinfo {volume} {112}},\ \bibinfo {pages}
  {34003} (\bibinfo {year} {2015})}\BibitemShut {NoStop}%
\bibitem [{\citenamefont {Duan}\ \emph {et~al.}(2016)\citenamefont {Duan},
  \citenamefont {Xie}, \citenamefont {Braak},\ and\ \citenamefont
  {Chen}}]{Duan:2016}%
  \BibitemOpen
  \bibfield  {author} {\bibinfo {author} {\bibfnamefont {L.}~\bibnamefont
  {Duan}}, \bibinfo {author} {\bibfnamefont {Y.-F.}\ \bibnamefont {Xie}},
  \bibinfo {author} {\bibfnamefont {D.}~\bibnamefont {Braak}}, \ and\ \bibinfo
  {author} {\bibfnamefont {Q.-H.}\ \bibnamefont {Chen}},\ }\href {\doibase
  10.1088/1751-8113/49/46/464002} {\bibfield  {journal} {\bibinfo  {journal}
  {J. Phys. A: Math. Theor.}\ }\textbf {\bibinfo {volume} {49}},\ \bibinfo
  {pages} {464002} (\bibinfo {year} {2016})}\BibitemShut {NoStop}%
\bibitem [{\citenamefont {Cui}\ \emph {et~al.}(2017)\citenamefont {Cui},
  \citenamefont {Cao}, \citenamefont {Fan},\ and\ \citenamefont
  {Amico}}]{Cui:2017}%
  \BibitemOpen
  \bibfield  {author} {\bibinfo {author} {\bibfnamefont {S.}~\bibnamefont
  {Cui}}, \bibinfo {author} {\bibfnamefont {J.-P.}\ \bibnamefont {Cao}},
  \bibinfo {author} {\bibfnamefont {H.}~\bibnamefont {Fan}}, \ and\ \bibinfo
  {author} {\bibfnamefont {L.}~\bibnamefont {Amico}},\ }\href {\doibase
  10.1088/1751-8121/aa6a6f} {\bibfield  {journal} {\bibinfo  {journal} {J.
  Phys. A: Math. Theor.}\ }\textbf {\bibinfo {volume} {50}},\ \bibinfo {pages}
  {204001} (\bibinfo {year} {2017})}\BibitemShut {NoStop}%
\bibitem [{\citenamefont {Xie}\ and\ \citenamefont {Chen}(2019)}]{Xie:2019}%
  \BibitemOpen
  \bibfield  {author} {\bibinfo {author} {\bibfnamefont {Y.-F.}\ \bibnamefont
  {Xie}}\ and\ \bibinfo {author} {\bibfnamefont {Q.-H.}\ \bibnamefont {Chen}},\
  }\href {\doibase 10.1088/0253-6102/71/5/623} {\bibfield  {journal} {\bibinfo
  {journal} {Comm. Theor. Phys.}\ }\textbf {\bibinfo {volume} {71}},\ \bibinfo
  {pages} {623} (\bibinfo {year} {2019})}\BibitemShut {NoStop}%
\bibitem [{\citenamefont {Beaudoin}\ \emph {et~al.}(2011)\citenamefont
  {Beaudoin}, \citenamefont {Gambetta},\ and\ \citenamefont
  {Blais}}]{Beaudoin:2011}%
  \BibitemOpen
  \bibfield  {author} {\bibinfo {author} {\bibfnamefont {F.}~\bibnamefont
  {Beaudoin}}, \bibinfo {author} {\bibfnamefont {J.~M.}\ \bibnamefont
  {Gambetta}}, \ and\ \bibinfo {author} {\bibfnamefont {A.}~\bibnamefont
  {Blais}},\ }\href {\doibase 10.1103/PhysRevA.84.043832} {\bibfield  {journal}
  {\bibinfo  {journal} {Phys. Rev. A}\ }\textbf {\bibinfo {volume} {84}},\
  \bibinfo {pages} {043832} (\bibinfo {year} {2011})}\BibitemShut {NoStop}%
\bibitem [{\citenamefont {De~Liberato}\ \emph {et~al.}(2007)\citenamefont
  {De~Liberato}, \citenamefont {Ciuti},\ and\ \citenamefont
  {Carusotto}}]{DeLiberato:2007}%
  \BibitemOpen
  \bibfield  {author} {\bibinfo {author} {\bibfnamefont {S.}~\bibnamefont
  {De~Liberato}}, \bibinfo {author} {\bibfnamefont {C.}~\bibnamefont {Ciuti}},
  \ and\ \bibinfo {author} {\bibfnamefont {I.}~\bibnamefont {Carusotto}},\
  }\href {\doibase 10.1103/PhysRevLett.98.103602} {\bibfield  {journal}
  {\bibinfo  {journal} {Phys. Rev. Lett.}\ }\textbf {\bibinfo {volume} {98}},\
  \bibinfo {pages} {103602} (\bibinfo {year} {2007})}\BibitemShut {NoStop}%
\bibitem [{\citenamefont {Lolli}\ \emph {et~al.}(2015)\citenamefont {Lolli},
  \citenamefont {Baksic}, \citenamefont {Nagy}, \citenamefont {Manucharyan},\
  and\ \citenamefont {Ciuti}}]{Lolli:2015}%
  \BibitemOpen
  \bibfield  {author} {\bibinfo {author} {\bibfnamefont {J.}~\bibnamefont
  {Lolli}}, \bibinfo {author} {\bibfnamefont {A.}~\bibnamefont {Baksic}},
  \bibinfo {author} {\bibfnamefont {D.}~\bibnamefont {Nagy}}, \bibinfo {author}
  {\bibfnamefont {V.~E.}\ \bibnamefont {Manucharyan}}, \ and\ \bibinfo {author}
  {\bibfnamefont {C.}~\bibnamefont {Ciuti}},\ }\href {\doibase
  10.1103/PhysRevLett.114.183601} {\bibfield  {journal} {\bibinfo  {journal}
  {Phys. Rev. Lett.}\ }\textbf {\bibinfo {volume} {114}},\ \bibinfo {pages}
  {183601} (\bibinfo {year} {2015})}\BibitemShut {NoStop}%
\bibitem [{\citenamefont {Felicetti}\ \emph {et~al.}(2015)\citenamefont
  {Felicetti}, \citenamefont {Douce}, \citenamefont {Romero}, \citenamefont
  {Milman},\ and\ \citenamefont {Solano}}]{Felicetti:2015}%
  \BibitemOpen
  \bibfield  {author} {\bibinfo {author} {\bibfnamefont {S.}~\bibnamefont
  {Felicetti}}, \bibinfo {author} {\bibfnamefont {T.}~\bibnamefont {Douce}},
  \bibinfo {author} {\bibfnamefont {G.}~\bibnamefont {Romero}}, \bibinfo
  {author} {\bibfnamefont {P.}~\bibnamefont {Milman}}, \ and\ \bibinfo {author}
  {\bibfnamefont {E.}~\bibnamefont {Solano}},\ }\href {\doibase
  10.1038/srep11818} {\bibfield  {journal} {\bibinfo  {journal} {Sci. Rep.}\
  }\textbf {\bibinfo {volume} {5}},\ \bibinfo {pages} {11818} (\bibinfo {year}
  {2015})}\BibitemShut {NoStop}%
\bibitem [{\citenamefont {Werlang}\ \emph {et~al.}(2008)\citenamefont
  {Werlang}, \citenamefont {Dodonov}, \citenamefont {Duzzioni},\ and\
  \citenamefont {Villas-B\^oas}}]{Werlang:2008}%
  \BibitemOpen
  \bibfield  {author} {\bibinfo {author} {\bibfnamefont {T.}~\bibnamefont
  {Werlang}}, \bibinfo {author} {\bibfnamefont {A.~V.}\ \bibnamefont
  {Dodonov}}, \bibinfo {author} {\bibfnamefont {E.~I.}\ \bibnamefont
  {Duzzioni}}, \ and\ \bibinfo {author} {\bibfnamefont {C.~J.}\ \bibnamefont
  {Villas-B\^oas}},\ }\href {\doibase 10.1103/PhysRevA.78.053805} {\bibfield
  {journal} {\bibinfo  {journal} {Phys. Rev. A}\ }\textbf {\bibinfo {volume}
  {78}},\ \bibinfo {pages} {053805} (\bibinfo {year} {2008})}\BibitemShut
  {NoStop}%
\bibitem [{\citenamefont {Breuer}\ and\ \citenamefont
  {Petruccione}(2002)}]{Breuer:2002}%
  \BibitemOpen
  \bibfield  {author} {\bibinfo {author} {\bibfnamefont {H.~P.}\ \bibnamefont
  {Breuer}}\ and\ \bibinfo {author} {\bibfnamefont {F.}~\bibnamefont
  {Petruccione}},\ }\href@noop {} {\emph {\bibinfo {title} {The Theory of Open
  Quantum Systems}}}\ (\bibinfo  {publisher} {Oxford University Press},\
  \bibinfo {year} {2002})\BibitemShut {NoStop}%
\bibitem [{\citenamefont {Rau}\ \emph {et~al.}(2004)\citenamefont {Rau},
  \citenamefont {Johansson},\ and\ \citenamefont {Shnirman}}]{Rau:2004}%
  \BibitemOpen
  \bibfield  {author} {\bibinfo {author} {\bibfnamefont {I.}~\bibnamefont
  {Rau}}, \bibinfo {author} {\bibfnamefont {G.}~\bibnamefont {Johansson}}, \
  and\ \bibinfo {author} {\bibfnamefont {A.}~\bibnamefont {Shnirman}},\ }\href
  {\doibase 10.1103/PhysRevB.70.054521} {\bibfield  {journal} {\bibinfo
  {journal} {Phys. Rev. B}\ }\textbf {\bibinfo {volume} {70}},\ \bibinfo
  {pages} {054521} (\bibinfo {year} {2004})}\BibitemShut {NoStop}%
\bibitem [{\citenamefont {Scala}\ \emph
  {et~al.}(2007{\natexlab{a}})\citenamefont {Scala}, \citenamefont {Militello},
  \citenamefont {Messina}, \citenamefont {Piilo},\ and\ \citenamefont
  {Maniscalco}}]{Scala:2007a}%
  \BibitemOpen
  \bibfield  {author} {\bibinfo {author} {\bibfnamefont {M.}~\bibnamefont
  {Scala}}, \bibinfo {author} {\bibfnamefont {B.}~\bibnamefont {Militello}},
  \bibinfo {author} {\bibfnamefont {A.}~\bibnamefont {Messina}}, \bibinfo
  {author} {\bibfnamefont {J.}~\bibnamefont {Piilo}}, \ and\ \bibinfo {author}
  {\bibfnamefont {S.}~\bibnamefont {Maniscalco}},\ }\href {\doibase
  10.1103/PhysRevA.75.013811} {\bibfield  {journal} {\bibinfo  {journal} {Phys.
  Rev. A}\ }\textbf {\bibinfo {volume} {75}},\ \bibinfo {pages} {013811}
  (\bibinfo {year} {2007}{\natexlab{a}})}\BibitemShut {NoStop}%
\bibitem [{\citenamefont {Scala}\ \emph
  {et~al.}(2007{\natexlab{b}})\citenamefont {Scala}, \citenamefont {Militello},
  \citenamefont {Messina}, \citenamefont {Maniscalco}, \citenamefont {Piilo},\
  and\ \citenamefont {Suominen}}]{Scala:2007b}%
  \BibitemOpen
  \bibfield  {author} {\bibinfo {author} {\bibfnamefont {M.}~\bibnamefont
  {Scala}}, \bibinfo {author} {\bibfnamefont {B.}~\bibnamefont {Militello}},
  \bibinfo {author} {\bibfnamefont {A.}~\bibnamefont {Messina}}, \bibinfo
  {author} {\bibfnamefont {S.}~\bibnamefont {Maniscalco}}, \bibinfo {author}
  {\bibfnamefont {J.}~\bibnamefont {Piilo}}, \ and\ \bibinfo {author}
  {\bibfnamefont {K.-A.}\ \bibnamefont {Suominen}},\ }\href {\doibase
  10.1088/1751-8113/40/48/015} {\bibfield  {journal} {\bibinfo  {journal} {J.
  Phys. A: Math. Theor.}\ }\textbf {\bibinfo {volume} {40}},\ \bibinfo {pages}
  {14527} (\bibinfo {year} {2007}{\natexlab{b}})}\BibitemShut {NoStop}%
\bibitem [{\citenamefont {Felicetti}\ and\ \citenamefont
  {Le~Boit\'e}(2020)}]{Felicetti:2020}%
  \BibitemOpen
  \bibfield  {author} {\bibinfo {author} {\bibfnamefont {S.}~\bibnamefont
  {Felicetti}}\ and\ \bibinfo {author} {\bibfnamefont {A.}~\bibnamefont
  {Le~Boit\'e}},\ }\href {\doibase 10.1103/PhysRevLett.124.040404} {\bibfield
  {journal} {\bibinfo  {journal} {Phys. Rev. Lett.}\ }\textbf {\bibinfo
  {volume} {124}},\ \bibinfo {pages} {040404} (\bibinfo {year}
  {2020})}\BibitemShut {NoStop}%
\bibitem [{\citenamefont {Le~Boit\'e}\ \emph {et~al.}(2017)\citenamefont
  {Le~Boit\'e}, \citenamefont {Hwang},\ and\ \citenamefont
  {Plenio}}]{LeBoite:2017}%
  \BibitemOpen
  \bibfield  {author} {\bibinfo {author} {\bibfnamefont {A.}~\bibnamefont
  {Le~Boit\'e}}, \bibinfo {author} {\bibfnamefont {M.-J.}\ \bibnamefont
  {Hwang}}, \ and\ \bibinfo {author} {\bibfnamefont {M.~B.}\ \bibnamefont
  {Plenio}},\ }\href {\doibase 10.1103/PhysRevA.95.023829} {\bibfield
  {journal} {\bibinfo  {journal} {Phys. Rev. A}\ }\textbf {\bibinfo {volume}
  {95}},\ \bibinfo {pages} {023829} (\bibinfo {year} {2017})}\BibitemShut
  {NoStop}%
\bibitem [{\citenamefont {Settineri}\ \emph {et~al.}(2018)\citenamefont
  {Settineri}, \citenamefont {Macr\'{\i}}, \citenamefont {Ridolfo},
  \citenamefont {Di~Stefano}, \citenamefont {Kockum}, \citenamefont {Nori},\
  and\ \citenamefont {Savasta}}]{Settineri:2018}%
  \BibitemOpen
  \bibfield  {author} {\bibinfo {author} {\bibfnamefont {A.}~\bibnamefont
  {Settineri}}, \bibinfo {author} {\bibfnamefont {V.}~\bibnamefont
  {Macr\'{\i}}}, \bibinfo {author} {\bibfnamefont {A.}~\bibnamefont {Ridolfo}},
  \bibinfo {author} {\bibfnamefont {O.}~\bibnamefont {Di~Stefano}}, \bibinfo
  {author} {\bibfnamefont {A.~F.}\ \bibnamefont {Kockum}}, \bibinfo {author}
  {\bibfnamefont {F.}~\bibnamefont {Nori}}, \ and\ \bibinfo {author}
  {\bibfnamefont {S.}~\bibnamefont {Savasta}},\ }\href {\doibase
  10.1103/PhysRevA.98.053834} {\bibfield  {journal} {\bibinfo  {journal} {Phys.
  Rev. A}\ }\textbf {\bibinfo {volume} {98}},\ \bibinfo {pages} {053834}
  (\bibinfo {year} {2018})}\BibitemShut {NoStop}%
\bibitem [{\citenamefont {Nataf}\ and\ \citenamefont
  {Ciuti}(2011)}]{Nataf:2011}%
  \BibitemOpen
  \bibfield  {author} {\bibinfo {author} {\bibfnamefont {P.}~\bibnamefont
  {Nataf}}\ and\ \bibinfo {author} {\bibfnamefont {C.}~\bibnamefont {Ciuti}},\
  }\href@noop {} {\bibfield  {journal} {\bibinfo  {journal} {Phys. Rev. Lett.}\
  }\textbf {\bibinfo {volume} {107}},\ \bibinfo {pages} {190402} (\bibinfo
  {year} {2011})}\BibitemShut {NoStop}%
\bibitem [{\citenamefont {Zueco}\ and\ \citenamefont
  {Garc\'{\i}a-Ripoll}(2019)}]{Zueco:2019}%
  \BibitemOpen
  \bibfield  {author} {\bibinfo {author} {\bibfnamefont {D.}~\bibnamefont
  {Zueco}}\ and\ \bibinfo {author} {\bibfnamefont {J.}~\bibnamefont
  {Garc\'{\i}a-Ripoll}},\ }\href {\doibase 10.1103/PhysRevA.99.013807}
  {\bibfield  {journal} {\bibinfo  {journal} {Phys. Rev. A}\ }\textbf {\bibinfo
  {volume} {99}},\ \bibinfo {pages} {013807} (\bibinfo {year}
  {2019})}\BibitemShut {NoStop}%
\bibitem [{\citenamefont {Bamba}\ and\ \citenamefont
  {Ogawa}(2014)}]{Bamba:2014}%
  \BibitemOpen
  \bibfield  {author} {\bibinfo {author} {\bibfnamefont {M.}~\bibnamefont
  {Bamba}}\ and\ \bibinfo {author} {\bibfnamefont {T.}~\bibnamefont {Ogawa}},\
  }\href {\doibase 10.1103/PhysRevA.89.023817} {\bibfield  {journal} {\bibinfo
  {journal} {Phys. Rev. A}\ }\textbf {\bibinfo {volume} {89}},\ \bibinfo
  {pages} {023817} (\bibinfo {year} {2014})}\BibitemShut {NoStop}%
\bibitem [{\citenamefont {Felicetti}\ \emph {et~al.}(2018)\citenamefont
  {Felicetti}, \citenamefont {Hwang},\ and\ \citenamefont
  {Le~Boit\'e}}]{Felicetti:2018}%
  \BibitemOpen
  \bibfield  {author} {\bibinfo {author} {\bibfnamefont {S.}~\bibnamefont
  {Felicetti}}, \bibinfo {author} {\bibfnamefont {M.-J.}\ \bibnamefont
  {Hwang}}, \ and\ \bibinfo {author} {\bibfnamefont {A.}~\bibnamefont
  {Le~Boit\'e}},\ }\href {\doibase 10.1103/PhysRevA.98.053859} {\bibfield
  {journal} {\bibinfo  {journal} {Phys. Rev. A}\ }\textbf {\bibinfo {volume}
  {98}},\ \bibinfo {pages} {053859} (\bibinfo {year} {2018})}\BibitemShut
  {NoStop}%
\bibitem [{\citenamefont {Garziano}\ \emph {et~al.}(2013)\citenamefont
  {Garziano}, \citenamefont {Ridolfo}, \citenamefont {Stassi}, \citenamefont
  {Di~Stefano},\ and\ \citenamefont {Savasta}}]{Garziano:2013}%
  \BibitemOpen
  \bibfield  {author} {\bibinfo {author} {\bibfnamefont {L.}~\bibnamefont
  {Garziano}}, \bibinfo {author} {\bibfnamefont {A.}~\bibnamefont {Ridolfo}},
  \bibinfo {author} {\bibfnamefont {R.}~\bibnamefont {Stassi}}, \bibinfo
  {author} {\bibfnamefont {O.}~\bibnamefont {Di~Stefano}}, \ and\ \bibinfo
  {author} {\bibfnamefont {S.}~\bibnamefont {Savasta}},\ }\href {\doibase
  10.1103/PhysRevA.88.063829} {\bibfield  {journal} {\bibinfo  {journal} {Phys.
  Rev. A}\ }\textbf {\bibinfo {volume} {88}},\ \bibinfo {pages} {063829}
  (\bibinfo {year} {2013})}\BibitemShut {NoStop}%
\bibitem [{\citenamefont {Di~Stefano}\ \emph {et~al.}(2017)\citenamefont
  {Di~Stefano}, \citenamefont {Stassi}, \citenamefont {Garziano}, \citenamefont
  {Frisk~Kockum}, \citenamefont {Savasta},\ and\ \citenamefont
  {Nori}}]{DiStefano:2017}%
  \BibitemOpen
  \bibfield  {author} {\bibinfo {author} {\bibfnamefont {O.}~\bibnamefont
  {Di~Stefano}}, \bibinfo {author} {\bibfnamefont {R.}~\bibnamefont {Stassi}},
  \bibinfo {author} {\bibfnamefont {L.}~\bibnamefont {Garziano}}, \bibinfo
  {author} {\bibfnamefont {A.}~\bibnamefont {Frisk~Kockum}}, \bibinfo {author}
  {\bibfnamefont {S.}~\bibnamefont {Savasta}}, \ and\ \bibinfo {author}
  {\bibfnamefont {F.}~\bibnamefont {Nori}},\ }\href {\doibase
  10.1088/1367-2630/aa6cd7} {\bibfield  {journal} {\bibinfo  {journal} {New
  Journal of Physics}\ }\textbf {\bibinfo {volume} {19}},\ \bibinfo {pages}
  {053010} (\bibinfo {year} {2017})}\BibitemShut {NoStop}%
\bibitem [{\citenamefont {Lalumi\`ere}\ \emph {et~al.}(2013)\citenamefont
  {Lalumi\`ere}, \citenamefont {Sanders}, \citenamefont {van Loo},
  \citenamefont {Fedorov}, \citenamefont {Wallraff},\ and\ \citenamefont
  {Blais}}]{Lalumiere:2013}%
  \BibitemOpen
  \bibfield  {author} {\bibinfo {author} {\bibfnamefont {K.}~\bibnamefont
  {Lalumi\`ere}}, \bibinfo {author} {\bibfnamefont {B.~C.}\ \bibnamefont
  {Sanders}}, \bibinfo {author} {\bibfnamefont {A.~F.}\ \bibnamefont {van
  Loo}}, \bibinfo {author} {\bibfnamefont {A.}~\bibnamefont {Fedorov}},
  \bibinfo {author} {\bibfnamefont {A.}~\bibnamefont {Wallraff}}, \ and\
  \bibinfo {author} {\bibfnamefont {A.}~\bibnamefont {Blais}},\ }\href
  {\doibase 10.1103/PhysRevA.88.043806} {\bibfield  {journal} {\bibinfo
  {journal} {Phys. Rev. A}\ }\textbf {\bibinfo {volume} {88}},\ \bibinfo
  {pages} {043806} (\bibinfo {year} {2013})}\BibitemShut {NoStop}%
\bibitem [{\citenamefont {Huppert}\ \emph {et~al.}(2016)\citenamefont
  {Huppert}, \citenamefont {Vasanelli}, \citenamefont {Pegolotti},
  \citenamefont {Todorov},\ and\ \citenamefont {Sirtori}}]{Huppert:2016}%
  \BibitemOpen
  \bibfield  {author} {\bibinfo {author} {\bibfnamefont {S.}~\bibnamefont
  {Huppert}}, \bibinfo {author} {\bibfnamefont {A.}~\bibnamefont {Vasanelli}},
  \bibinfo {author} {\bibfnamefont {G.}~\bibnamefont {Pegolotti}}, \bibinfo
  {author} {\bibfnamefont {Y.}~\bibnamefont {Todorov}}, \ and\ \bibinfo
  {author} {\bibfnamefont {C.}~\bibnamefont {Sirtori}},\ }\href {\doibase
  10.1103/PhysRevB.94.155418} {\bibfield  {journal} {\bibinfo  {journal} {Phys.
  Rev. B}\ }\textbf {\bibinfo {volume} {94}},\ \bibinfo {pages} {155418}
  (\bibinfo {year} {2016})}\BibitemShut {NoStop}%
\bibitem [{\citenamefont {Di~Stefano}\ \emph {et~al.}(2018)\citenamefont
  {Di~Stefano}, \citenamefont {Frisk~Kockum}, \citenamefont {Ridolfo},
  \citenamefont {Savasta},\ and\ \citenamefont {Nori}}]{DiStefano:2018b}%
  \BibitemOpen
  \bibfield  {author} {\bibinfo {author} {\bibfnamefont {O.}~\bibnamefont
  {Di~Stefano}}, \bibinfo {author} {\bibfnamefont {A.}~\bibnamefont
  {Frisk~Kockum}}, \bibinfo {author} {\bibfnamefont {A.}~\bibnamefont
  {Ridolfo}}, \bibinfo {author} {\bibfnamefont {S.}~\bibnamefont {Savasta}}, \
  and\ \bibinfo {author} {\bibfnamefont {F.}~\bibnamefont {Nori}},\ }\href@noop
  {} {\bibfield  {journal} {\bibinfo  {journal} {Sci. Rep.}\ }\textbf {\bibinfo
  {volume} {8}},\ \bibinfo {pages} {17825} (\bibinfo {year}
  {2018})}\BibitemShut {NoStop}%
\bibitem [{\citenamefont {Shirley}(1965)}]{Shirley:1965}%
  \BibitemOpen
  \bibfield  {author} {\bibinfo {author} {\bibfnamefont {J.~H.}\ \bibnamefont
  {Shirley}},\ }\href {\doibase 10.1103/PhysRev.138.B979} {\bibfield  {journal}
  {\bibinfo  {journal} {Phys. Rev.}\ }\textbf {\bibinfo {volume} {138}},\
  \bibinfo {pages} {B979} (\bibinfo {year} {1965})}\BibitemShut {NoStop}%
\bibitem [{\citenamefont {Grifoni}\ and\ \citenamefont
  {H{\"a}nggi}(1998)}]{Grifoni:1998}%
  \BibitemOpen
  \bibfield  {author} {\bibinfo {author} {\bibfnamefont {M.}~\bibnamefont
  {Grifoni}}\ and\ \bibinfo {author} {\bibfnamefont {P.}~\bibnamefont
  {H{\"a}nggi}},\ }\href {\doibase
  https://doi.org/10.1016/S0370-1573(98)00022-2} {\bibfield  {journal}
  {\bibinfo  {journal} {Physics Reports}\ }\textbf {\bibinfo {volume} {304}},\
  \bibinfo {pages} {229 } (\bibinfo {year} {1998})}\BibitemShut {NoStop}%
\bibitem [{\citenamefont {Hausinger}\ and\ \citenamefont
  {Grifoni}(2011)}]{Hausinger:2011}%
  \BibitemOpen
  \bibfield  {author} {\bibinfo {author} {\bibfnamefont {J.}~\bibnamefont
  {Hausinger}}\ and\ \bibinfo {author} {\bibfnamefont {M.}~\bibnamefont
  {Grifoni}},\ }\href {\doibase 10.1103/PhysRevA.83.030301} {\bibfield
  {journal} {\bibinfo  {journal} {Phys. Rev. A}\ }\textbf {\bibinfo {volume}
  {83}},\ \bibinfo {pages} {030301} (\bibinfo {year} {2011})}\BibitemShut
  {NoStop}%
\bibitem [{\citenamefont {Hausinger}\ and\ \citenamefont
  {Grifoni}(2008)}]{Hausinger:2008}%
  \BibitemOpen
  \bibfield  {author} {\bibinfo {author} {\bibfnamefont {J.}~\bibnamefont
  {Hausinger}}\ and\ \bibinfo {author} {\bibfnamefont {M.}~\bibnamefont
  {Grifoni}},\ }\href {\doibase 10.1088/1367-2630/10/11/115015} {\bibfield
  {journal} {\bibinfo  {journal} {New. J. Phys.}\ }\textbf {\bibinfo {volume}
  {10}},\ \bibinfo {pages} {115015} (\bibinfo {year} {2008})}\BibitemShut
  {NoStop}%
\bibitem [{\citenamefont {Restrepo}\ \emph {et~al.}(2016)\citenamefont
  {Restrepo}, \citenamefont {Cerrillo}, \citenamefont {Bastidas}, \citenamefont
  {Angelakis},\ and\ \citenamefont {Brandes}}]{Restrepo:2016}%
  \BibitemOpen
  \bibfield  {author} {\bibinfo {author} {\bibfnamefont {S.}~\bibnamefont
  {Restrepo}}, \bibinfo {author} {\bibfnamefont {J.}~\bibnamefont {Cerrillo}},
  \bibinfo {author} {\bibfnamefont {V.~M.}\ \bibnamefont {Bastidas}}, \bibinfo
  {author} {\bibfnamefont {D.~G.}\ \bibnamefont {Angelakis}}, \ and\ \bibinfo
  {author} {\bibfnamefont {T.}~\bibnamefont {Brandes}},\ }\href {\doibase
  10.1103/PhysRevLett.117.250401} {\bibfield  {journal} {\bibinfo  {journal}
  {Phys. Rev. Lett.}\ }\textbf {\bibinfo {volume} {117}},\ \bibinfo {pages}
  {250401} (\bibinfo {year} {2016})}\BibitemShut {NoStop}%
\bibitem [{\citenamefont {Rivas}\ \emph {et~al.}(2010)\citenamefont {Rivas},
  \citenamefont {K~Plato}, \citenamefont {Huelga},\ and\ \citenamefont
  {B~Plenio}}]{Rivas:2010}%
  \BibitemOpen
  \bibfield  {author} {\bibinfo {author} {\bibfnamefont {{\'A}.}~\bibnamefont
  {Rivas}}, \bibinfo {author} {\bibfnamefont {A.~D.}\ \bibnamefont {K~Plato}},
  \bibinfo {author} {\bibfnamefont {S.~F.}\ \bibnamefont {Huelga}}, \ and\
  \bibinfo {author} {\bibfnamefont {M.}~\bibnamefont {B~Plenio}},\ }\href
  {\doibase 10.1088/1367-2630/12/11/113032} {\bibfield  {journal} {\bibinfo
  {journal} {New. J. Phys.}\ }\textbf {\bibinfo {volume} {12}},\ \bibinfo
  {pages} {113032} (\bibinfo {year} {2010})}\BibitemShut {NoStop}%
\bibitem [{\citenamefont {Ho}\ \emph {et~al.}(1986)\citenamefont {Ho},
  \citenamefont {Wang},\ and\ \citenamefont {Chu}}]{Ho:1986}%
  \BibitemOpen
  \bibfield  {author} {\bibinfo {author} {\bibfnamefont {T.-S.}\ \bibnamefont
  {Ho}}, \bibinfo {author} {\bibfnamefont {K.}~\bibnamefont {Wang}}, \ and\
  \bibinfo {author} {\bibfnamefont {S.-I.}\ \bibnamefont {Chu}},\ }\href
  {\doibase 10.1103/PhysRevA.33.1798} {\bibfield  {journal} {\bibinfo
  {journal} {Phys. Rev. A}\ }\textbf {\bibinfo {volume} {33}},\ \bibinfo
  {pages} {1798} (\bibinfo {year} {1986})}\BibitemShut {NoStop}%
\bibitem [{\citenamefont {Chu}\ and\ \citenamefont {Telnov}(2004)}]{Chu:2004}%
  \BibitemOpen
  \bibfield  {author} {\bibinfo {author} {\bibfnamefont {S.-I.}\ \bibnamefont
  {Chu}}\ and\ \bibinfo {author} {\bibfnamefont {D.~A.}\ \bibnamefont
  {Telnov}},\ }\href {\doibase 10.1016/j.physrep.2003.10.001} {\bibfield
  {journal} {\bibinfo  {journal} {Phys. Rep.}\ }\textbf {\bibinfo {volume}
  {390}},\ \bibinfo {pages} {1} (\bibinfo {year} {2004})}\BibitemShut {NoStop}%
\bibitem [{\citenamefont {Floquet}(1883)}]{Floquet:1883}%
  \BibitemOpen
  \bibfield  {author} {\bibinfo {author} {\bibfnamefont {G.}~\bibnamefont
  {Floquet}},\ }\href {http://www.numdam.org/article/ASENS_1883_2_12__47_0.pdf}
  {\bibfield  {journal} {\bibinfo  {journal} {Ann. Sci. de L'\'Ecole Norm.
  Sup.}\ }\textbf {\bibinfo {volume} {12}},\ \bibinfo {pages} {47} (\bibinfo
  {year} {1883})}\BibitemShut {NoStop}%
\bibitem [{\citenamefont {Hausinger}\ and\ \citenamefont
  {Grifoni}(2010)}]{Hausinger:2010}%
  \BibitemOpen
  \bibfield  {author} {\bibinfo {author} {\bibfnamefont {J.}~\bibnamefont
  {Hausinger}}\ and\ \bibinfo {author} {\bibfnamefont {M.}~\bibnamefont
  {Grifoni}},\ }\href {\doibase 10.1103/PhysRevA.81.022117} {\bibfield
  {journal} {\bibinfo  {journal} {Phys. Rev. A}\ }\textbf {\bibinfo {volume}
  {81}},\ \bibinfo {pages} {022117} (\bibinfo {year} {2010})}\BibitemShut
  {NoStop}%
\bibitem [{\citenamefont {Bl\"umel}\ \emph {et~al.}(1991)\citenamefont
  {Bl\"umel}, \citenamefont {Buchleitner}, \citenamefont {Graham},
  \citenamefont {Sirko}, \citenamefont {Smilansky},\ and\ \citenamefont
  {Walther}}]{Blumel:1991}%
  \BibitemOpen
  \bibfield  {author} {\bibinfo {author} {\bibfnamefont {R.}~\bibnamefont
  {Bl\"umel}}, \bibinfo {author} {\bibfnamefont {A.}~\bibnamefont
  {Buchleitner}}, \bibinfo {author} {\bibfnamefont {R.}~\bibnamefont {Graham}},
  \bibinfo {author} {\bibfnamefont {L.}~\bibnamefont {Sirko}}, \bibinfo
  {author} {\bibfnamefont {U.}~\bibnamefont {Smilansky}}, \ and\ \bibinfo
  {author} {\bibfnamefont {H.}~\bibnamefont {Walther}},\ }\href {\doibase
  10.1103/PhysRevA.44.4521} {\bibfield  {journal} {\bibinfo  {journal} {Phys.
  Rev. A}\ }\textbf {\bibinfo {volume} {44}},\ \bibinfo {pages} {4521}
  (\bibinfo {year} {1991})}\BibitemShut {NoStop}%
\bibitem [{\citenamefont {Breuer}\ and\ \citenamefont
  {Petruccione}(1997)}]{Breuer:1997}%
  \BibitemOpen
  \bibfield  {author} {\bibinfo {author} {\bibfnamefont {H.-P.}\ \bibnamefont
  {Breuer}}\ and\ \bibinfo {author} {\bibfnamefont {F.}~\bibnamefont
  {Petruccione}},\ }\href {\doibase 10.1103/PhysRevA.55.3101} {\bibfield
  {journal} {\bibinfo  {journal} {Phys. Rev. A}\ }\textbf {\bibinfo {volume}
  {55}},\ \bibinfo {pages} {3101} (\bibinfo {year} {1997})}\BibitemShut
  {NoStop}%
\bibitem [{\citenamefont {Breuer}\ \emph {et~al.}(2000)\citenamefont {Breuer},
  \citenamefont {Huber},\ and\ \citenamefont {Petruccione}}]{Breuer:2000}%
  \BibitemOpen
  \bibfield  {author} {\bibinfo {author} {\bibfnamefont {H.-P.}\ \bibnamefont
  {Breuer}}, \bibinfo {author} {\bibfnamefont {W.}~\bibnamefont {Huber}}, \
  and\ \bibinfo {author} {\bibfnamefont {F.}~\bibnamefont {Petruccione}},\
  }\href {\doibase 10.1103/PhysRevE.61.4883} {\bibfield  {journal} {\bibinfo
  {journal} {Phys. Rev. E}\ }\textbf {\bibinfo {volume} {61}},\ \bibinfo
  {pages} {4883} (\bibinfo {year} {2000})}\BibitemShut {NoStop}%
\bibitem [{\citenamefont {Astafiev}\ \emph {et~al.}(2010)\citenamefont
  {Astafiev}, \citenamefont {Zagoskin}, \citenamefont {Jr.}, \citenamefont
  {Pashkin}, \citenamefont {Yamamoto}, \citenamefont {Inomata}, \citenamefont
  {Nakamura},\ and\ \citenamefont {Tsai}}]{Astafiev:2010}%
  \BibitemOpen
  \bibfield  {author} {\bibinfo {author} {\bibfnamefont {O.}~\bibnamefont
  {Astafiev}}, \bibinfo {author} {\bibfnamefont {A.~M.}\ \bibnamefont
  {Zagoskin}}, \bibinfo {author} {\bibfnamefont {A.~A.~A.}\ \bibnamefont
  {Jr.}}, \bibinfo {author} {\bibfnamefont {Y.~A.}\ \bibnamefont {Pashkin}},
  \bibinfo {author} {\bibfnamefont {T.}~\bibnamefont {Yamamoto}}, \bibinfo
  {author} {\bibfnamefont {K.}~\bibnamefont {Inomata}}, \bibinfo {author}
  {\bibfnamefont {Y.}~\bibnamefont {Nakamura}}, \ and\ \bibinfo {author}
  {\bibfnamefont {J.~S.}\ \bibnamefont {Tsai}},\ }\href {\doibase
  10.1126/science.1181918} {\bibfield  {journal} {\bibinfo  {journal}
  {Science}\ }\textbf {\bibinfo {volume} {327}},\ \bibinfo {pages} {840}
  (\bibinfo {year} {2010})}\BibitemShut {NoStop}%
\bibitem [{\citenamefont {Hoi}\ \emph {et~al.}(2012)\citenamefont {Hoi},
  \citenamefont {Palomaki}, \citenamefont {Lindkvist}, \citenamefont
  {Johansson}, \citenamefont {Delsing},\ and\ \citenamefont
  {Wilson}}]{Hoi:2012}%
  \BibitemOpen
  \bibfield  {author} {\bibinfo {author} {\bibfnamefont {I.-C.}\ \bibnamefont
  {Hoi}}, \bibinfo {author} {\bibfnamefont {T.}~\bibnamefont {Palomaki}},
  \bibinfo {author} {\bibfnamefont {J.}~\bibnamefont {Lindkvist}}, \bibinfo
  {author} {\bibfnamefont {G.}~\bibnamefont {Johansson}}, \bibinfo {author}
  {\bibfnamefont {P.}~\bibnamefont {Delsing}}, \ and\ \bibinfo {author}
  {\bibfnamefont {C.~M.}\ \bibnamefont {Wilson}},\ }\href {\doibase
  10.1103/PhysRevLett.108.263601} {\bibfield  {journal} {\bibinfo  {journal}
  {Phys. Rev. Lett.}\ }\textbf {\bibinfo {volume} {108}},\ \bibinfo {pages}
  {263601} (\bibinfo {year} {2012})}\BibitemShut {NoStop}%
\bibitem [{\citenamefont {Hoi}\ \emph {et~al.}(2013)\citenamefont {Hoi},
  \citenamefont {Wilson}, \citenamefont {Johansson}, \citenamefont {Lindkvis},
  \citenamefont {Peropadre}, \citenamefont {Palomaki},\ and\ \citenamefont
  {Delsing}}]{Hoi:2013}%
  \BibitemOpen
  \bibfield  {author} {\bibinfo {author} {\bibfnamefont {I.-C.}\ \bibnamefont
  {Hoi}}, \bibinfo {author} {\bibfnamefont {C.~M.}\ \bibnamefont {Wilson}},
  \bibinfo {author} {\bibfnamefont {G.}~\bibnamefont {Johansson}}, \bibinfo
  {author} {\bibfnamefont {J.}~\bibnamefont {Lindkvis}}, \bibinfo {author}
  {\bibfnamefont {B.}~\bibnamefont {Peropadre}}, \bibinfo {author}
  {\bibfnamefont {T.}~\bibnamefont {Palomaki}}, \ and\ \bibinfo {author}
  {\bibfnamefont {P.}~\bibnamefont {Delsing}},\ }\href {\doibase
  10.1088/1367-2630/15/2/025011} {\bibfield  {journal} {\bibinfo  {journal}
  {New. J. Phys.}\ }\textbf {\bibinfo {volume} {15}},\ \bibinfo {pages}
  {025011} (\bibinfo {year} {2013})}\BibitemShut {NoStop}%
\bibitem [{\citenamefont {Pletyukhov}\ and\ \citenamefont
  {Gritsev}(2015)}]{Pletyukhov:2015}%
  \BibitemOpen
  \bibfield  {author} {\bibinfo {author} {\bibfnamefont {M.}~\bibnamefont
  {Pletyukhov}}\ and\ \bibinfo {author} {\bibfnamefont {V.}~\bibnamefont
  {Gritsev}},\ }\href {\doibase 10.1103/PhysRevA.91.063841} {\bibfield
  {journal} {\bibinfo  {journal} {Phys. Rev. A}\ }\textbf {\bibinfo {volume}
  {91}},\ \bibinfo {pages} {063841} (\bibinfo {year} {2015})}\BibitemShut
  {NoStop}%
\bibitem [{\citenamefont {Shi}\ and\ \citenamefont {Sun}(2009)}]{Shi:2009}%
  \BibitemOpen
  \bibfield  {author} {\bibinfo {author} {\bibfnamefont {T.}~\bibnamefont
  {Shi}}\ and\ \bibinfo {author} {\bibfnamefont {C.~P.}\ \bibnamefont {Sun}},\
  }\href {\doibase 10.1103/PhysRevB.79.205111} {\bibfield  {journal} {\bibinfo
  {journal} {Phys. Rev. B}\ }\textbf {\bibinfo {volume} {79}},\ \bibinfo
  {pages} {205111} (\bibinfo {year} {2009})}\BibitemShut {NoStop}%
\bibitem [{\citenamefont {Le~Hur}(2012)}]{LeHur:2012}%
  \BibitemOpen
  \bibfield  {author} {\bibinfo {author} {\bibfnamefont {K.}~\bibnamefont
  {Le~Hur}},\ }\href {\doibase 10.1103/PhysRevB.85.140506} {\bibfield
  {journal} {\bibinfo  {journal} {Phys. Rev. B}\ }\textbf {\bibinfo {volume}
  {85}},\ \bibinfo {pages} {140506} (\bibinfo {year} {2012})}\BibitemShut
  {NoStop}%
\bibitem [{\citenamefont {Peropadre}\ \emph
  {et~al.}(2013{\natexlab{a}})\citenamefont {Peropadre}, \citenamefont
  {Lindkvis}, \citenamefont {Hoi}, \citenamefont {Wilson}, \citenamefont
  {Garc\'{\i}a-Ripoll}, \citenamefont {Delsing},\ and\ \citenamefont
  {Johansson}}]{Peropadre:2013a}%
  \BibitemOpen
  \bibfield  {author} {\bibinfo {author} {\bibfnamefont {B.}~\bibnamefont
  {Peropadre}}, \bibinfo {author} {\bibfnamefont {J.}~\bibnamefont {Lindkvis}},
  \bibinfo {author} {\bibfnamefont {I.-C.}\ \bibnamefont {Hoi}}, \bibinfo
  {author} {\bibfnamefont {C.~M.}\ \bibnamefont {Wilson}}, \bibinfo {author}
  {\bibfnamefont {J.~J.}\ \bibnamefont {Garc\'{\i}a-Ripoll}}, \bibinfo {author}
  {\bibfnamefont {P.}~\bibnamefont {Delsing}}, \ and\ \bibinfo {author}
  {\bibfnamefont {G.}~\bibnamefont {Johansson}},\ }\href {\doibase
  10.1088/1367-2630/15/3/035009} {\bibfield  {journal} {\bibinfo  {journal}
  {New. J. Phys.}\ }\textbf {\bibinfo {volume} {15}},\ \bibinfo {pages}
  {035009} (\bibinfo {year} {2013}{\natexlab{a}})}\BibitemShut {NoStop}%
\bibitem [{\citenamefont {Or\'us}(2014)}]{Orus:2014}%
  \BibitemOpen
  \bibfield  {author} {\bibinfo {author} {\bibfnamefont {R.}~\bibnamefont
  {Or\'us}},\ }\href {\doibase 10.1016/j.aop.2014.06.013} {\bibfield  {journal}
  {\bibinfo  {journal} {Ann. of Phys.}\ }\textbf {\bibinfo {volume} {349}},\
  \bibinfo {pages} {117} (\bibinfo {year} {2014})}\BibitemShut {NoStop}%
\bibitem [{\citenamefont {Peropadre}\ \emph
  {et~al.}(2013{\natexlab{b}})\citenamefont {Peropadre}, \citenamefont {Zueco},
  \citenamefont {Porras},\ and\ \citenamefont
  {Garc\'{\i}a-Ripoll}}]{Peropadre:2013b}%
  \BibitemOpen
  \bibfield  {author} {\bibinfo {author} {\bibfnamefont {B.}~\bibnamefont
  {Peropadre}}, \bibinfo {author} {\bibfnamefont {D.}~\bibnamefont {Zueco}},
  \bibinfo {author} {\bibfnamefont {D.}~\bibnamefont {Porras}}, \ and\ \bibinfo
  {author} {\bibfnamefont {J.~J.}\ \bibnamefont {Garc\'{\i}a-Ripoll}},\ }\href
  {\doibase 10.1103/PhysRevLett.111.243602} {\bibfield  {journal} {\bibinfo
  {journal} {Phys. Rev. Lett.}\ }\textbf {\bibinfo {volume} {111}},\ \bibinfo
  {pages} {243602} (\bibinfo {year} {2013}{\natexlab{b}})}\BibitemShut
  {NoStop}%
\bibitem [{\citenamefont {Sanchez-Burillo}\ \emph {et~al.}(2014)\citenamefont
  {Sanchez-Burillo}, \citenamefont {Zueco}, \citenamefont {Garcia-Ripoll},\
  and\ \citenamefont {Martin-Moreno}}]{Sanchez-Burillo:2014}%
  \BibitemOpen
  \bibfield  {author} {\bibinfo {author} {\bibfnamefont {E.}~\bibnamefont
  {Sanchez-Burillo}}, \bibinfo {author} {\bibfnamefont {D.}~\bibnamefont
  {Zueco}}, \bibinfo {author} {\bibfnamefont {J.~J.}\ \bibnamefont
  {Garcia-Ripoll}}, \ and\ \bibinfo {author} {\bibfnamefont {L.}~\bibnamefont
  {Martin-Moreno}},\ }\href {\doibase 10.1103/PhysRevLett.113.263604}
  {\bibfield  {journal} {\bibinfo  {journal} {Phys. Rev. Lett.}\ }\textbf
  {\bibinfo {volume} {113}},\ \bibinfo {pages} {263604} (\bibinfo {year}
  {2014})}\BibitemShut {NoStop}%
\bibitem [{\citenamefont {D\'{\i}az-Camacho}\ \emph {et~al.}(2016)\citenamefont
  {D\'{\i}az-Camacho}, \citenamefont {Bermudez},\ and\ \citenamefont
  {Garc\'{\i}a-Ripoll}}]{Diaz-Camacho:2016}%
  \BibitemOpen
  \bibfield  {author} {\bibinfo {author} {\bibfnamefont {G.}~\bibnamefont
  {D\'{\i}az-Camacho}}, \bibinfo {author} {\bibfnamefont {A.}~\bibnamefont
  {Bermudez}}, \ and\ \bibinfo {author} {\bibfnamefont {J.~J.}\ \bibnamefont
  {Garc\'{\i}a-Ripoll}},\ }\href {\doibase 10.1103/PhysRevA.93.043843}
  {\bibfield  {journal} {\bibinfo  {journal} {Phys. Rev. A}\ }\textbf {\bibinfo
  {volume} {93}},\ \bibinfo {pages} {043843} (\bibinfo {year}
  {2016})}\BibitemShut {NoStop}%
\bibitem [{\citenamefont {Gheeraert}\ \emph {et~al.}(2017)\citenamefont
  {Gheeraert}, \citenamefont {Bera},\ and\ \citenamefont
  {Florens}}]{Gheeraert:2017}%
  \BibitemOpen
  \bibfield  {author} {\bibinfo {author} {\bibfnamefont {N.}~\bibnamefont
  {Gheeraert}}, \bibinfo {author} {\bibfnamefont {S.}~\bibnamefont {Bera}}, \
  and\ \bibinfo {author} {\bibfnamefont {S.}~\bibnamefont {Florens}},\ }\href
  {\doibase 10.1088/1367-2630/aa5dea} {\bibfield  {journal} {\bibinfo
  {journal} {New. J. Phys.}\ }\textbf {\bibinfo {volume} {19}},\ \bibinfo
  {pages} {023036} (\bibinfo {year} {2017})}\BibitemShut {NoStop}%
\bibitem [{\citenamefont {Gheeraert}\ \emph {et~al.}(2018)\citenamefont
  {Gheeraert}, \citenamefont {Zhang}, \citenamefont {S\'epulcre}, \citenamefont
  {Bera}, \citenamefont {Roch}, \citenamefont {Baranger},\ and\ \citenamefont
  {Florens}}]{Gheeraert:2018}%
  \BibitemOpen
  \bibfield  {author} {\bibinfo {author} {\bibfnamefont {N.}~\bibnamefont
  {Gheeraert}}, \bibinfo {author} {\bibfnamefont {X.~H.~H.}\ \bibnamefont
  {Zhang}}, \bibinfo {author} {\bibfnamefont {T.}~\bibnamefont {S\'epulcre}},
  \bibinfo {author} {\bibfnamefont {S.}~\bibnamefont {Bera}}, \bibinfo {author}
  {\bibfnamefont {N.}~\bibnamefont {Roch}}, \bibinfo {author} {\bibfnamefont
  {H.~U.}\ \bibnamefont {Baranger}}, \ and\ \bibinfo {author} {\bibfnamefont
  {S.}~\bibnamefont {Florens}},\ }\href {\doibase 10.1103/PhysRevA.98.043816}
  {\bibfield  {journal} {\bibinfo  {journal} {Phys. Rev. A}\ }\textbf {\bibinfo
  {volume} {98}},\ \bibinfo {pages} {043816} (\bibinfo {year}
  {2018})}\BibitemShut {NoStop}%
\bibitem [{\citenamefont {Pletyukhov}\ and\ \citenamefont
  {Gritsev}(2012)}]{Pletyukhov:2012}%
  \BibitemOpen
  \bibfield  {author} {\bibinfo {author} {\bibfnamefont {M.}~\bibnamefont
  {Pletyukhov}}\ and\ \bibinfo {author} {\bibfnamefont {V.}~\bibnamefont
  {Gritsev}},\ }\href {\doibase 10.1088/1367-2630/14/9/095028} {\bibfield
  {journal} {\bibinfo  {journal} {New. J. Phys.}\ }\textbf {\bibinfo {volume}
  {14}},\ \bibinfo {pages} {095028} (\bibinfo {year} {2012})}\BibitemShut
  {NoStop}%
\bibitem [{\citenamefont {Yudson}(1985)}]{Yudson:1985}%
  \BibitemOpen
  \bibfield  {author} {\bibinfo {author} {\bibfnamefont {V.~I.}\ \bibnamefont
  {Yudson}},\ }\href@noop {} {\bibfield  {journal} {\bibinfo  {journal} {Zh.
  Eksp. Teor. Fiz}\ }\textbf {\bibinfo {volume} {88}},\ \bibinfo {pages} {1757}
  (\bibinfo {year} {1985})}\BibitemShut {NoStop}%
\bibitem [{\citenamefont {Yudson}\ and\ \citenamefont
  {Reineker}(2008)}]{Yudson:2008}%
  \BibitemOpen
  \bibfield  {author} {\bibinfo {author} {\bibfnamefont {V.~I.}\ \bibnamefont
  {Yudson}}\ and\ \bibinfo {author} {\bibfnamefont {P.}~\bibnamefont
  {Reineker}},\ }\href {\doibase 10.1103/PhysRevA.78.052713} {\bibfield
  {journal} {\bibinfo  {journal} {Phys. Rev. A}\ }\textbf {\bibinfo {volume}
  {78}},\ \bibinfo {pages} {052713} (\bibinfo {year} {2008})}\BibitemShut
  {NoStop}%
\bibitem [{\citenamefont {Shen}\ and\ \citenamefont
  {Fan}(2007{\natexlab{a}})}]{Shen:2007a}%
  \BibitemOpen
  \bibfield  {author} {\bibinfo {author} {\bibfnamefont {J.-T.}\ \bibnamefont
  {Shen}}\ and\ \bibinfo {author} {\bibfnamefont {S.}~\bibnamefont {Fan}},\
  }\href {\doibase 10.1103/PhysRevLett.98.153003} {\bibfield  {journal}
  {\bibinfo  {journal} {Phys. Rev. Lett.}\ }\textbf {\bibinfo {volume} {98}},\
  \bibinfo {pages} {153003} (\bibinfo {year} {2007}{\natexlab{a}})}\BibitemShut
  {NoStop}%
\bibitem [{\citenamefont {Shen}\ and\ \citenamefont
  {Fan}(2007{\natexlab{b}})}]{Shen:2007b}%
  \BibitemOpen
  \bibfield  {author} {\bibinfo {author} {\bibfnamefont {J.-T.}\ \bibnamefont
  {Shen}}\ and\ \bibinfo {author} {\bibfnamefont {S.}~\bibnamefont {Fan}},\
  }\href {\doibase 10.1103/PhysRevA.76.062709} {\bibfield  {journal} {\bibinfo
  {journal} {Phys. Rev. A}\ }\textbf {\bibinfo {volume} {76}},\ \bibinfo
  {pages} {062709} (\bibinfo {year} {2007}{\natexlab{b}})}\BibitemShut
  {NoStop}%
\bibitem [{\citenamefont {Zheng}\ \emph {et~al.}(2010)\citenamefont {Zheng},
  \citenamefont {Gauthier},\ and\ \citenamefont {Baranger}}]{Zheng:2010}%
  \BibitemOpen
  \bibfield  {author} {\bibinfo {author} {\bibfnamefont {H.}~\bibnamefont
  {Zheng}}, \bibinfo {author} {\bibfnamefont {D.~J.}\ \bibnamefont {Gauthier}},
  \ and\ \bibinfo {author} {\bibfnamefont {H.~U.}\ \bibnamefont {Baranger}},\
  }\href {\doibase 10.1103/PhysRevA.82.063816} {\bibfield  {journal} {\bibinfo
  {journal} {Phys. Rev. A}\ }\textbf {\bibinfo {volume} {82}},\ \bibinfo
  {pages} {063816} (\bibinfo {year} {2010})}\BibitemShut {NoStop}%
\bibitem [{\citenamefont {Di~Stefano}\ \emph {et~al.}(2019)\citenamefont
  {Di~Stefano}, \citenamefont {Settineri}, \citenamefont {Macr{\`\i}},
  \citenamefont {Garziano}, \citenamefont {Stassi}, \citenamefont {Savasta},\
  and\ \citenamefont {Nori}}]{DiStefano:2019}%
  \BibitemOpen
  \bibfield  {author} {\bibinfo {author} {\bibfnamefont {O.}~\bibnamefont
  {Di~Stefano}}, \bibinfo {author} {\bibfnamefont {A.}~\bibnamefont
  {Settineri}}, \bibinfo {author} {\bibfnamefont {V.}~\bibnamefont
  {Macr{\`\i}}}, \bibinfo {author} {\bibfnamefont {L.}~\bibnamefont
  {Garziano}}, \bibinfo {author} {\bibfnamefont {R.}~\bibnamefont {Stassi}},
  \bibinfo {author} {\bibfnamefont {S.}~\bibnamefont {Savasta}}, \ and\
  \bibinfo {author} {\bibfnamefont {F.}~\bibnamefont {Nori}},\ }\href {\doibase
  10.1038/s41567-019-0534-4} {\bibfield  {journal} {\bibinfo  {journal} {Nat.
  Phys.}\ }\textbf {\bibinfo {volume} {15}},\ \bibinfo {pages} {803} (\bibinfo
  {year} {2019})}\BibitemShut {NoStop}%
\bibitem [{\citenamefont {Manucharyan}\ \emph {et~al.}(2017)\citenamefont
  {Manucharyan}, \citenamefont {Baksic},\ and\ \citenamefont
  {Ciuti}}]{Manucharyan:2017}%
  \BibitemOpen
  \bibfield  {author} {\bibinfo {author} {\bibfnamefont {V.~E.}\ \bibnamefont
  {Manucharyan}}, \bibinfo {author} {\bibfnamefont {A.}~\bibnamefont {Baksic}},
  \ and\ \bibinfo {author} {\bibfnamefont {C.}~\bibnamefont {Ciuti}},\ }\href
  {\doibase 10.1088/1751-8121/aa6fbc} {\bibfield  {journal} {\bibinfo
  {journal} {J. Phys. A: Math. Theor.}\ }\textbf {\bibinfo {volume} {50}},\
  \bibinfo {pages} {294001} (\bibinfo {year} {2017})}\BibitemShut {NoStop}%
\bibitem [{\citenamefont {De~Liberato}(2014)}]{DeLiberato:2014}%
  \BibitemOpen
  \bibfield  {author} {\bibinfo {author} {\bibfnamefont {S.}~\bibnamefont
  {De~Liberato}},\ }\href {\doibase 10.1103/PhysRevLett.112.016401} {\bibfield
  {journal} {\bibinfo  {journal} {Phys. Rev. Lett.}\ }\textbf {\bibinfo
  {volume} {112}},\ \bibinfo {pages} {016401} (\bibinfo {year}
  {2014})}\BibitemShut {NoStop}%
\bibitem [{\citenamefont {Garc{\'\i}a-Ripoll}\ \emph
  {et~al.}(2015)\citenamefont {Garc{\'\i}a-Ripoll}, \citenamefont {Peropadre},\
  and\ \citenamefont {De~Liberato}}]{Garcia-Ripoll:2015}%
  \BibitemOpen
  \bibfield  {author} {\bibinfo {author} {\bibfnamefont {J.~J.}\ \bibnamefont
  {Garc{\'\i}a-Ripoll}}, \bibinfo {author} {\bibfnamefont {B.}~\bibnamefont
  {Peropadre}}, \ and\ \bibinfo {author} {\bibfnamefont {S.}~\bibnamefont
  {De~Liberato}},\ }\href {\doibase 10.1038/srep16055} {\bibfield  {journal}
  {\bibinfo  {journal} {Scientific Reports}\ }\textbf {\bibinfo {volume} {5}},\
  \bibinfo {pages} {16055} (\bibinfo {year} {2015})}\BibitemShut {NoStop}%
\bibitem [{\citenamefont {Nataf}\ and\ \citenamefont
  {Ciuti}(2010)}]{Nataf:2010b}%
  \BibitemOpen
  \bibfield  {author} {\bibinfo {author} {\bibfnamefont {P.}~\bibnamefont
  {Nataf}}\ and\ \bibinfo {author} {\bibfnamefont {C.}~\bibnamefont {Ciuti}},\
  }\href@noop {} {\bibfield  {journal} {\bibinfo  {journal} {Nat. Commun.}\
  }\textbf {\bibinfo {volume} {1}},\ \bibinfo {pages} {72} (\bibinfo {year}
  {2010})}\BibitemShut {NoStop}%
\bibitem [{\citenamefont {Viehmann}\ \emph {et~al.}(2011)\citenamefont
  {Viehmann}, \citenamefont {von Delft},\ and\ \citenamefont
  {Marquardt}}]{Viehmann:2011}%
  \BibitemOpen
  \bibfield  {author} {\bibinfo {author} {\bibfnamefont {O.}~\bibnamefont
  {Viehmann}}, \bibinfo {author} {\bibfnamefont {J.}~\bibnamefont {von Delft}},
  \ and\ \bibinfo {author} {\bibfnamefont {F.}~\bibnamefont {Marquardt}},\
  }\href {\doibase 10.1103/PhysRevLett.107.113602} {\bibfield  {journal}
  {\bibinfo  {journal} {Phys. Rev. Lett.}\ }\textbf {\bibinfo {volume} {107}},\
  \bibinfo {pages} {113602} (\bibinfo {year} {2011})}\BibitemShut {NoStop}%
\bibitem [{\citenamefont {Ciuti}\ and\ \citenamefont
  {Nataf}(2012)}]{Ciuti:2012}%
  \BibitemOpen
  \bibfield  {author} {\bibinfo {author} {\bibfnamefont {C.}~\bibnamefont
  {Ciuti}}\ and\ \bibinfo {author} {\bibfnamefont {P.}~\bibnamefont {Nataf}},\
  }\href {\doibase 10.1103/PhysRevLett.109.179301} {\bibfield  {journal}
  {\bibinfo  {journal} {Phys. Rev. Lett.}\ }\textbf {\bibinfo {volume} {109}},\
  \bibinfo {pages} {179301} (\bibinfo {year} {2012})}\BibitemShut {NoStop}%
\bibitem [{\citenamefont {Lambert}\ \emph {et~al.}(2016)\citenamefont
  {Lambert}, \citenamefont {Matsuzaki}, \citenamefont {Kakuyanagi},
  \citenamefont {Ishida}, \citenamefont {Saito},\ and\ \citenamefont
  {Nori}}]{Lambert:2016}%
  \BibitemOpen
  \bibfield  {author} {\bibinfo {author} {\bibfnamefont {N.}~\bibnamefont
  {Lambert}}, \bibinfo {author} {\bibfnamefont {Y.}~\bibnamefont {Matsuzaki}},
  \bibinfo {author} {\bibfnamefont {K.}~\bibnamefont {Kakuyanagi}}, \bibinfo
  {author} {\bibfnamefont {N.}~\bibnamefont {Ishida}}, \bibinfo {author}
  {\bibfnamefont {S.}~\bibnamefont {Saito}}, \ and\ \bibinfo {author}
  {\bibfnamefont {F.}~\bibnamefont {Nori}},\ }\href {\doibase
  10.1103/PhysRevB.94.224510} {\bibfield  {journal} {\bibinfo  {journal} {Phys.
  Rev. B}\ }\textbf {\bibinfo {volume} {94}},\ \bibinfo {pages} {224510}
  (\bibinfo {year} {2016})}\BibitemShut {NoStop}%
\bibitem [{\citenamefont {Jaako}\ \emph {et~al.}(2016)\citenamefont {Jaako},
  \citenamefont {Xiang}, \citenamefont {Garcia-Ripoll},\ and\ \citenamefont
  {Rabl}}]{Jaako:2016}%
  \BibitemOpen
  \bibfield  {author} {\bibinfo {author} {\bibfnamefont {T.}~\bibnamefont
  {Jaako}}, \bibinfo {author} {\bibfnamefont {Z.-L.}\ \bibnamefont {Xiang}},
  \bibinfo {author} {\bibfnamefont {J.~J.}\ \bibnamefont {Garcia-Ripoll}}, \
  and\ \bibinfo {author} {\bibfnamefont {P.}~\bibnamefont {Rabl}},\ }\href
  {\doibase 10.1103/PhysRevA.94.033850} {\bibfield  {journal} {\bibinfo
  {journal} {Phys. Rev. A}\ }\textbf {\bibinfo {volume} {94}},\ \bibinfo
  {pages} {033850} (\bibinfo {year} {2016})}\BibitemShut {NoStop}%
\bibitem [{\citenamefont {Bamba}\ and\ \citenamefont
  {Imoto}(2017)}]{Bamba:2017}%
  \BibitemOpen
  \bibfield  {author} {\bibinfo {author} {\bibfnamefont {M.}~\bibnamefont
  {Bamba}}\ and\ \bibinfo {author} {\bibfnamefont {N.}~\bibnamefont {Imoto}},\
  }\href {\doibase 10.1103/PhysRevA.96.053857} {\bibfield  {journal} {\bibinfo
  {journal} {Phys. Rev. A}\ }\textbf {\bibinfo {volume} {96}},\ \bibinfo
  {pages} {053857} (\bibinfo {year} {2017})}\BibitemShut {NoStop}%
\bibitem [{\citenamefont {Kirton}\ \emph {et~al.}(2019)\citenamefont {Kirton},
  \citenamefont {Roses}, \citenamefont {Keeling},\ and\ \citenamefont
  {Torre}}]{Kirton:2019}%
  \BibitemOpen
  \bibfield  {author} {\bibinfo {author} {\bibfnamefont {P.}~\bibnamefont
  {Kirton}}, \bibinfo {author} {\bibfnamefont {M.~M.}\ \bibnamefont {Roses}},
  \bibinfo {author} {\bibfnamefont {J.}~\bibnamefont {Keeling}}, \ and\
  \bibinfo {author} {\bibfnamefont {E.~G.~D.}\ \bibnamefont {Torre}},\
  }\href@noop {} {\bibfield  {journal} {\bibinfo  {journal} {Adv. Quantum
  Technol.}\ }\textbf {\bibinfo {volume} {2}},\ \bibinfo {pages} {1800043}
  (\bibinfo {year} {2019})}\BibitemShut {NoStop}%
\bibitem [{\citenamefont {Andolina}\ \emph {et~al.}(2019)\citenamefont
  {Andolina}, \citenamefont {Pellegrino}, \citenamefont {Giovannetti},
  \citenamefont {MacDonald},\ and\ \citenamefont {Polini}}]{Andolina:2019}%
  \BibitemOpen
  \bibfield  {author} {\bibinfo {author} {\bibfnamefont {G.~M.}\ \bibnamefont
  {Andolina}}, \bibinfo {author} {\bibfnamefont {F.~M.~D.}\ \bibnamefont
  {Pellegrino}}, \bibinfo {author} {\bibfnamefont {V.}~\bibnamefont
  {Giovannetti}}, \bibinfo {author} {\bibfnamefont {A.~H.}\ \bibnamefont
  {MacDonald}}, \ and\ \bibinfo {author} {\bibfnamefont {M.}~\bibnamefont
  {Polini}},\ }\href {\doibase 10.1103/PhysRevB.100.121109} {\bibfield
  {journal} {\bibinfo  {journal} {Phys. Rev. B}\ }\textbf {\bibinfo {volume}
  {100}},\ \bibinfo {pages} {121109} (\bibinfo {year} {2019})}\BibitemShut
  {NoStop}%
\bibitem [{\citenamefont {Vukics}\ \emph {et~al.}(2014)\citenamefont {Vukics},
  \citenamefont {Grie\ss{}er},\ and\ \citenamefont {Domokos}}]{Vukics:2014}%
  \BibitemOpen
  \bibfield  {author} {\bibinfo {author} {\bibfnamefont {A.}~\bibnamefont
  {Vukics}}, \bibinfo {author} {\bibfnamefont {T.}~\bibnamefont {Grie\ss{}er}},
  \ and\ \bibinfo {author} {\bibfnamefont {P.}~\bibnamefont {Domokos}},\ }\href
  {\doibase 10.1103/PhysRevLett.112.073601} {\bibfield  {journal} {\bibinfo
  {journal} {Phys. Rev. Lett.}\ }\textbf {\bibinfo {volume} {112}},\ \bibinfo
  {pages} {073601} (\bibinfo {year} {2014})}\BibitemShut {NoStop}%
\bibitem [{\citenamefont {De~Bernardis}\ \emph
  {et~al.}(2018{\natexlab{a}})\citenamefont {De~Bernardis}, \citenamefont
  {Pilar}, \citenamefont {Jaako}, \citenamefont {De~Liberato},\ and\
  \citenamefont {Rabl}}]{DeBernardis:2018b}%
  \BibitemOpen
  \bibfield  {author} {\bibinfo {author} {\bibfnamefont {D.}~\bibnamefont
  {De~Bernardis}}, \bibinfo {author} {\bibfnamefont {P.}~\bibnamefont {Pilar}},
  \bibinfo {author} {\bibfnamefont {T.}~\bibnamefont {Jaako}}, \bibinfo
  {author} {\bibfnamefont {S.}~\bibnamefont {De~Liberato}}, \ and\ \bibinfo
  {author} {\bibfnamefont {P.}~\bibnamefont {Rabl}},\ }\href {\doibase
  10.1103/PhysRevA.98.053819} {\bibfield  {journal} {\bibinfo  {journal} {Phys.
  Rev. A}\ }\textbf {\bibinfo {volume} {98}},\ \bibinfo {pages} {053819}
  (\bibinfo {year} {2018}{\natexlab{a}})}\BibitemShut {NoStop}%
\bibitem [{\citenamefont {Stokes}\ and\ \citenamefont
  {Nazir}(2019{\natexlab{a}})}]{Stokes:2019a}%
  \BibitemOpen
  \bibfield  {author} {\bibinfo {author} {\bibfnamefont {A.}~\bibnamefont
  {Stokes}}\ and\ \bibinfo {author} {\bibfnamefont {A.}~\bibnamefont {Nazir}},\
  }\href {\doibase 10.1038/s41467-018-08101-0} {\bibfield  {journal} {\bibinfo
  {journal} {Nature Communications}\ }\textbf {\bibinfo {volume} {10}},\
  \bibinfo {pages} {499} (\bibinfo {year} {2019}{\natexlab{a}})}\BibitemShut
  {NoStop}%
\bibitem [{\citenamefont {Stokes}\ and\ \citenamefont
  {Nazir}(2019{\natexlab{b}})}]{Stokes:2019b}%
  \BibitemOpen
  \bibfield  {author} {\bibinfo {author} {\bibfnamefont {A.}~\bibnamefont
  {Stokes}}\ and\ \bibinfo {author} {\bibfnamefont {A.}~\bibnamefont {Nazir}},\
  }\href@noop {} {\bibfield  {journal} {\bibinfo  {journal} {arXiv preprint
  arXiv:1902.05160}\ } (\bibinfo {year} {2019}{\natexlab{b}})}\BibitemShut
  {NoStop}%
\bibitem [{\citenamefont {Garziano}\ \emph {et~al.}(2020)\citenamefont
  {Garziano}, \citenamefont {Settineri}, \citenamefont {Stefano}, \citenamefont
  {Savasta},\ and\ \citenamefont {Nori}}]{Garziano:2020}%
  \BibitemOpen
  \bibfield  {author} {\bibinfo {author} {\bibfnamefont {L.}~\bibnamefont
  {Garziano}}, \bibinfo {author} {\bibfnamefont {A.}~\bibnamefont {Settineri}},
  \bibinfo {author} {\bibfnamefont {O.~D.}\ \bibnamefont {Stefano}}, \bibinfo
  {author} {\bibfnamefont {S.}~\bibnamefont {Savasta}}, \ and\ \bibinfo
  {author} {\bibfnamefont {F.}~\bibnamefont {Nori}},\ }\href@noop {} {\enquote
  {\bibinfo {title} {Gauge invariance of the dicke and hopfield models},}\ }
  (\bibinfo {year} {2020}),\ \Eprint {http://arxiv.org/abs/2002.04241}
  {arXiv:2002.04241 [quant-ph]} \BibitemShut {NoStop}%
\bibitem [{\citenamefont {Todorov}\ and\ \citenamefont
  {Sirtori}(2012)}]{Todorov:2012}%
  \BibitemOpen
  \bibfield  {author} {\bibinfo {author} {\bibfnamefont {Y.}~\bibnamefont
  {Todorov}}\ and\ \bibinfo {author} {\bibfnamefont {C.}~\bibnamefont
  {Sirtori}},\ }\href {\doibase 10.1103/PhysRevB.85.045304} {\bibfield
  {journal} {\bibinfo  {journal} {Phys. Rev. B}\ }\textbf {\bibinfo {volume}
  {85}},\ \bibinfo {pages} {045304} (\bibinfo {year} {2012})}\BibitemShut
  {NoStop}%
\bibitem [{\citenamefont {Todorov}(2014)}]{Todorov:2014a}%
  \BibitemOpen
  \bibfield  {author} {\bibinfo {author} {\bibfnamefont {Y.}~\bibnamefont
  {Todorov}},\ }\href {\doibase 10.1103/PhysRevB.89.075115} {\bibfield
  {journal} {\bibinfo  {journal} {Phys. Rev. B}\ }\textbf {\bibinfo {volume}
  {89}},\ \bibinfo {pages} {075115} (\bibinfo {year} {2014})}\BibitemShut
  {NoStop}%
\bibitem [{\citenamefont {Todorov}\ and\ \citenamefont
  {Sirtori}(2014)}]{Todorov:2014b}%
  \BibitemOpen
  \bibfield  {author} {\bibinfo {author} {\bibfnamefont {Y.}~\bibnamefont
  {Todorov}}\ and\ \bibinfo {author} {\bibfnamefont {C.}~\bibnamefont
  {Sirtori}},\ }\href {\doibase 10.1103/PhysRevX.4.041031} {\bibfield
  {journal} {\bibinfo  {journal} {Phys. Rev. X}\ }\textbf {\bibinfo {volume}
  {4}},\ \bibinfo {pages} {041031} (\bibinfo {year} {2014})}\BibitemShut
  {NoStop}%
\bibitem [{\citenamefont {Vukics}\ \emph {et~al.}(2015)\citenamefont {Vukics},
  \citenamefont {Grie\ss{}er},\ and\ \citenamefont {Domokos}}]{Vukics:2015}%
  \BibitemOpen
  \bibfield  {author} {\bibinfo {author} {\bibfnamefont {A.}~\bibnamefont
  {Vukics}}, \bibinfo {author} {\bibfnamefont {T.}~\bibnamefont {Grie\ss{}er}},
  \ and\ \bibinfo {author} {\bibfnamefont {P.}~\bibnamefont {Domokos}},\ }\href
  {\doibase 10.1103/PhysRevA.92.043835} {\bibfield  {journal} {\bibinfo
  {journal} {Phys. Rev. A}\ }\textbf {\bibinfo {volume} {92}},\ \bibinfo
  {pages} {043835} (\bibinfo {year} {2015})}\BibitemShut {NoStop}%
\bibitem [{\citenamefont {De~Bernardis}\ \emph
  {et~al.}(2018{\natexlab{b}})\citenamefont {De~Bernardis}, \citenamefont
  {Jaako},\ and\ \citenamefont {Rabl}}]{DeBernardis:2018a}%
  \BibitemOpen
  \bibfield  {author} {\bibinfo {author} {\bibfnamefont {D.}~\bibnamefont
  {De~Bernardis}}, \bibinfo {author} {\bibfnamefont {T.}~\bibnamefont {Jaako}},
  \ and\ \bibinfo {author} {\bibfnamefont {P.}~\bibnamefont {Rabl}},\ }\href
  {\doibase 10.1103/PhysRevA.97.043820} {\bibfield  {journal} {\bibinfo
  {journal} {Phys. Rev. A}\ }\textbf {\bibinfo {volume} {97}},\ \bibinfo
  {pages} {043820} (\bibinfo {year} {2018}{\natexlab{b}})}\BibitemShut
  {NoStop}%
\bibitem [{\citenamefont {Schir\'o}\ \emph {et~al.}(2012)\citenamefont
  {Schir\'o}, \citenamefont {Bordyuh}, \citenamefont {\"Oztop},\ and\
  \citenamefont {T\"ureci}}]{Schiro:2012}%
  \BibitemOpen
  \bibfield  {author} {\bibinfo {author} {\bibfnamefont {M.}~\bibnamefont
  {Schir\'o}}, \bibinfo {author} {\bibfnamefont {M.}~\bibnamefont {Bordyuh}},
  \bibinfo {author} {\bibfnamefont {B.}~\bibnamefont {\"Oztop}}, \ and\
  \bibinfo {author} {\bibfnamefont {H.~E.}\ \bibnamefont {T\"ureci}},\ }\href
  {\doibase 10.1103/PhysRevLett.109.053601} {\bibfield  {journal} {\bibinfo
  {journal} {Phys. Rev. Lett.}\ }\textbf {\bibinfo {volume} {109}},\ \bibinfo
  {pages} {053601} (\bibinfo {year} {2012})}\BibitemShut {NoStop}%
\bibitem [{\citenamefont {Garbe}\ \emph {et~al.}(2017)\citenamefont {Garbe},
  \citenamefont {Egusquiza}, \citenamefont {Solano}, \citenamefont {Ciuti},
  \citenamefont {Coudreau}, \citenamefont {Milman},\ and\ \citenamefont
  {Felicetti}}]{Garbe:2017}%
  \BibitemOpen
  \bibfield  {author} {\bibinfo {author} {\bibfnamefont {L.}~\bibnamefont
  {Garbe}}, \bibinfo {author} {\bibfnamefont {I.}~\bibnamefont {Egusquiza}},
  \bibinfo {author} {\bibfnamefont {E.}~\bibnamefont {Solano}}, \bibinfo
  {author} {\bibfnamefont {C.}~\bibnamefont {Ciuti}}, \bibinfo {author}
  {\bibfnamefont {T.}~\bibnamefont {Coudreau}}, \bibinfo {author}
  {\bibfnamefont {P.}~\bibnamefont {Milman}}, \ and\ \bibinfo {author}
  {\bibfnamefont {S.}~\bibnamefont {Felicetti}},\ }\href@noop {} {\bibfield
  {journal} {\bibinfo  {journal} {Phys. Rev. A}\ }\textbf {\bibinfo {volume}
  {95}},\ \bibinfo {pages} {053854} (\bibinfo {year} {2017})}\BibitemShut
  {NoStop}%
\bibitem [{\citenamefont {Cui}\ \emph {et~al.}(2019)\citenamefont {Cui},
  \citenamefont {H\'ebert}, \citenamefont {Gr\'emaud}, \citenamefont
  {Rousseau}, \citenamefont {Guo},\ and\ \citenamefont {Batrouni}}]{Cui:2019}%
  \BibitemOpen
  \bibfield  {author} {\bibinfo {author} {\bibfnamefont {S.}~\bibnamefont
  {Cui}}, \bibinfo {author} {\bibfnamefont {F.}~\bibnamefont {H\'ebert}},
  \bibinfo {author} {\bibfnamefont {B.}~\bibnamefont {Gr\'emaud}}, \bibinfo
  {author} {\bibfnamefont {V.~G.}\ \bibnamefont {Rousseau}}, \bibinfo {author}
  {\bibfnamefont {W.}~\bibnamefont {Guo}}, \ and\ \bibinfo {author}
  {\bibfnamefont {G.~G.}\ \bibnamefont {Batrouni}},\ }\href {\doibase
  10.1103/PhysRevA.100.033608} {\bibfield  {journal} {\bibinfo  {journal}
  {Phys. Rev. A}\ }\textbf {\bibinfo {volume} {100}},\ \bibinfo {pages}
  {033608} (\bibinfo {year} {2019})}\BibitemShut {NoStop}%
\bibitem [{\citenamefont {Carusotto}\ and\ \citenamefont
  {Ciuti}(2013)}]{Carusotto:2013}%
  \BibitemOpen
  \bibfield  {author} {\bibinfo {author} {\bibfnamefont {I.}~\bibnamefont
  {Carusotto}}\ and\ \bibinfo {author} {\bibfnamefont {C.}~\bibnamefont
  {Ciuti}},\ }\href {\doibase 10.1103/RevModPhys.85.299} {\bibfield  {journal}
  {\bibinfo  {journal} {Rev. Mod. Phys.}\ }\textbf {\bibinfo {volume} {85}},\
  \bibinfo {pages} {299} (\bibinfo {year} {2013})}\BibitemShut {NoStop}%
\bibitem [{\citenamefont {Schmidt}\ and\ \citenamefont
  {Koch}(2013)}]{Schmidt:2013}%
  \BibitemOpen
  \bibfield  {author} {\bibinfo {author} {\bibfnamefont {S.}~\bibnamefont
  {Schmidt}}\ and\ \bibinfo {author} {\bibfnamefont {J.}~\bibnamefont {Koch}},\
  }\href {\doibase 10.1002/andp.201200261} {\bibfield  {journal} {\bibinfo
  {journal} {Ann. Phys. (Berlin)}\ }\textbf {\bibinfo {volume} {525}},\
  \bibinfo {pages} {395} (\bibinfo {year} {2013})}\BibitemShut {NoStop}%
\bibitem [{\citenamefont {Noh}\ and\ \citenamefont
  {Angelakis}(2016)}]{Noh:2016}%
  \BibitemOpen
  \bibfield  {author} {\bibinfo {author} {\bibfnamefont {C.}~\bibnamefont
  {Noh}}\ and\ \bibinfo {author} {\bibfnamefont {D.~G.}\ \bibnamefont
  {Angelakis}},\ }\href {\doibase 10.1088/0034-4885/80/1/016401} {\bibfield
  {journal} {\bibinfo  {journal} {Rep. Prog. Phys.}\ }\textbf {\bibinfo
  {volume} {80}},\ \bibinfo {pages} {016401} (\bibinfo {year}
  {2016})}\BibitemShut {NoStop}%
\end{thebibliography}%

\end{document}